\documentclass[12pt]{article}  
\usepackage{graphicx}
\usepackage{epsfig}
\usepackage{hyperref}
\usepackage {amsmath}
\usepackage {amssymb}
\usepackage {bm}  
\usepackage[usenames,dvipsnames]{color}
\usepackage{xspace} 
\newcommand{\pt}{\mbox{$p_T$}\xspace}

\newcommand{\Npart}{\mbox{$N_{\rm part}$}\xspace}
\newcommand{\Ncoll}{\mbox{$N_{\rm coll}$}\xspace}
\newcommand{\Nch}{\mbox{$N_{\rm ch}$}\xspace}
\newcommand{\Et}{\mbox{${\rm E}_T$}\xspace}

\newcommand{\sqs}{\mbox{$\sqrt{s}$}\xspace}
\newcommand{\sqsn}{\mbox{$\sqrt{s_{_{NN}}}$}\xspace}

\newcommand{\Nqp}{\mbox{$N_{qp}$}\xspace}
\def\lsim{\raise0.3ex\hbox{$<$\kern-0.75em\raise-1.1ex\hbox{$\sim$}}}
\def\gsim{\raise0.3ex\hbox{$>$\kern-0.75em\raise-1.1ex\hbox{$\sim$}}}

\def\mean#1{\left<#1\right>}
\def\Journal#1#2#3#4{{#1}{\bf #2} (#4) #3}
\def\IJMPA{{Int. J. Mod. Phys. A}}
\def\JPG{{J. Phys. G}}

\def\NIMA{{Nucl. Instrum. Methods A}}
\def\NPA{{Nucl. Phys. A}}
\def\NPB{{Nucl. Phys. B}}
\def\PLB{{Phys. Lett. B}}

\def\PL{Phys. Lett.\ }
\def\PRL{Phys. Rev. Lett.\ }
\def\PRD{{Phys. Rev. D}}
\def\PRC{{Phys. Rev. C}}

\def\ZPC{{Z. Phys. C}}
\def\ARNPS{{Ann. Rev. Nucl. Part. Sci.\ }} 
\def\RMP{Rev. Mod. Phys.\ }

\def\QGP{{\color{Red} Q}{\color{Blue} G}{\color{Green} P}} 
\def\QCD{{\color{Red} Q}{\color{Green} C}{\color{Blue} D}}

\normalsize
\begin{document}

\title{Highlights from BNL and RHIC 2014}
\author{M.~J.~Tannenbaum
\thanks{Research supported by U.~S.~Department of Energy, DE-AC02-98CH10886.}
\\Physics Department, 510c,\\
Brookhaven National Laboratory,\\
Upton, NY 11973-5000, USA\\
mjt@bnl.gov} 
\date{}
\maketitle
\vspace*{-2pc}
\section{Introduction}\label{sec:introduction}
\begin{figure}[!htb]
\begin{center}
\includegraphics[width=0.9\textwidth]{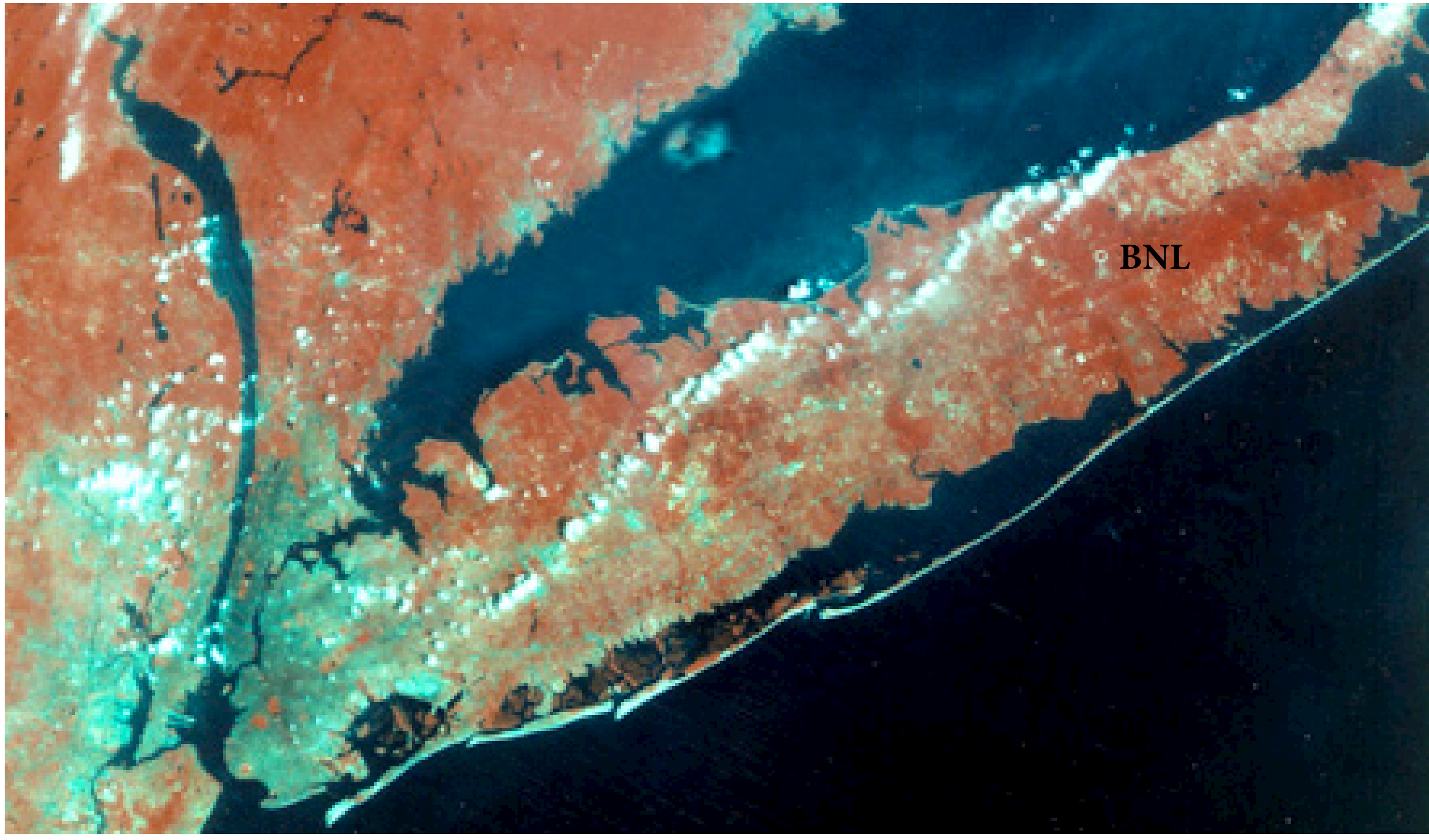}
\end{center}
\caption[]{\footnotesize NASA infra-red photo of Long Island and the New York Metro Region from space. RHIC is the white circle to the left of the word BNL. Manhattan Island in New York City, $\sim$100 km west of BNL, is also clearly visible on the left side of the photo.    }
\label{fig:NASA}
\end{figure}

The Relativistic Heavy Ion Collider (RHIC) at Brookhaven National Laboratory (BNL) is one of the two remaining operating hadron colliders, the other being the LHC at CERN. Unlike the LHC, which is buried in a deep underground tunnel, RHIC is built in an enclosure on the surface which is covered by an earth berm for shielding which can be seen from outer space (Fig.~\ref{fig:NASA}).

\begin{figure}[!htb]
\begin{center}
\includegraphics[width=0.9\textwidth]{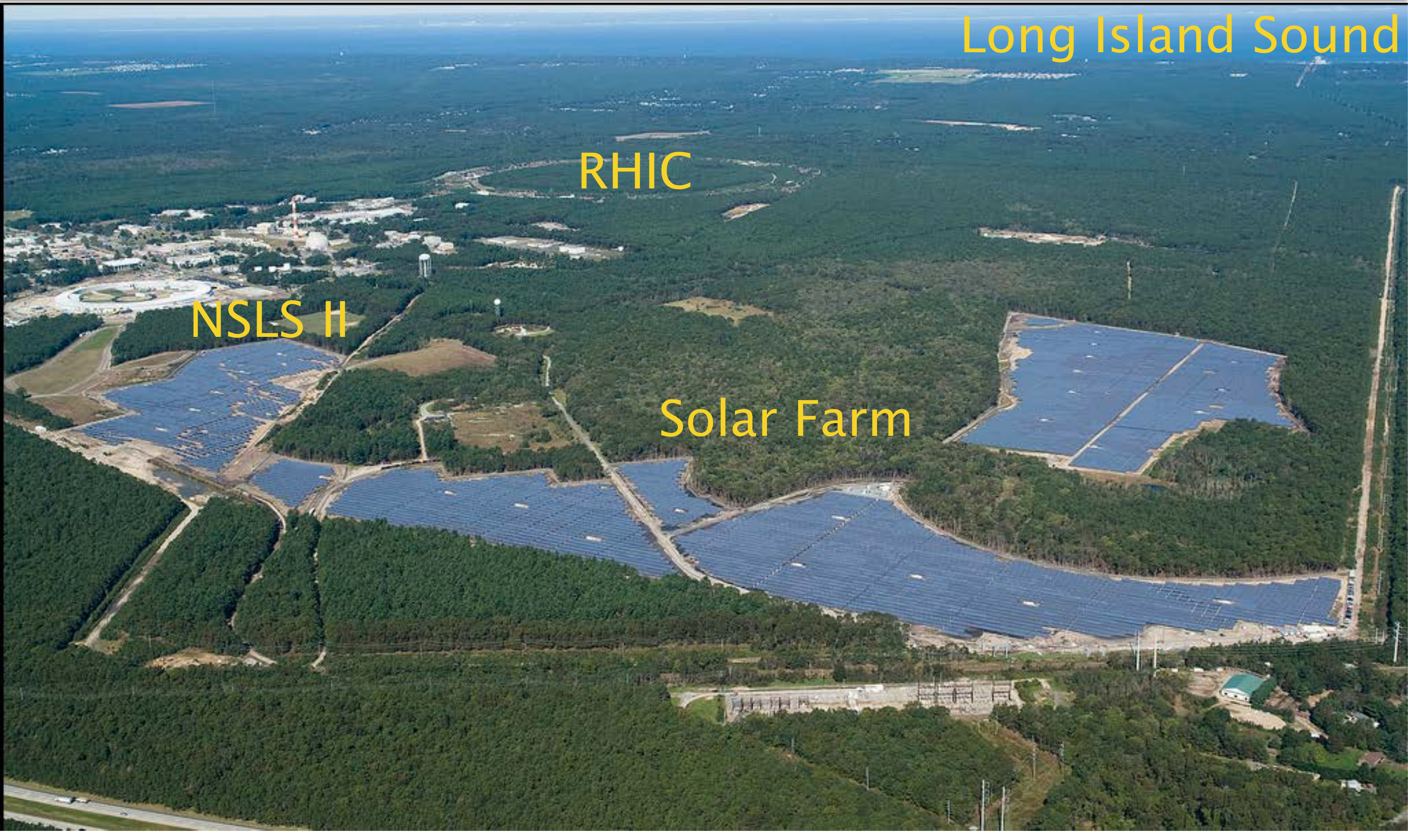}
\end{center}\vspace*{-1.5pc}
\caption[]{\footnotesize Aerial view of BNL with NSLS II, RHIC and the Solar Farm indicated.   }
\label{fig:BNLother}\vspace*{-2.0pc}
\end{figure}

BNL is a multipurpose laboratory, quite different in scope from Fermilab and CERN, with many ``cutting edge'' major research facilities in addition to RHIC. Figure \ref{fig:BNLother} shows the two newest facilities: the National Synchrotron Light Source II (NSLS II) to come on-line October 1, 2014; and the Long Island Solar Farm which has an experimental section as well as supplying 32MW peak power to nearby homes in partnership with the local electric company.  
\section{News from BNL since ISSP2013}
Although Fiscal Year 2014 started on October 1, 2013 with the U.~S.~Government shut down for the first 16 days due to the lack of an approved budget, the rest of the FY2014 turned out very well for BNL and RHIC. At the administrative level, Professor Robert Tribble of Texas A\&M University, well known for both his physics research and leadership of two important U.~S. Department of Energy (DoE) Panels in 2005 and 2013, was appointed Deputy Director for Science \& Technology, effective February 2014. Not long after that, in March, the DoE issued a Request For Proposals for a ``management and operating (M\&O) contractor'' for BNL, which is owned by the U.S. Government, but run by an M\&O contractor. The present contractor, BSA, is a partnership of Battelle Memorial Institute, a private non-profit science and technology development company, headquartered in Columbus, Ohio, and Stony Brook University. BSA has been the M\&O contractor at BNL for the past 15 years (out of the BNL's 67 year existence), with the ``engagement'' of six of the world's leading research universities (Columbia, Cornell, Harvard, MIT, Princeton and Yale) who were among the universities that formed the founding M\&O contractor, Associated Universities Incorporated, along with Johns Hopkins, and the Universities of Pennsylvania and Rochester. The new contract starts on January 1, 2015, preceded by a maximum 2 month transition phase-in period so should be awarded near or soon after November 1, 2014~\cite{DoESolicitation}. 

The U. S. High Energy Physics bureaucracy was not idle during this period, with the release of the ``Particle Physics Project Prioritization Panel (P5)'' Report to the High energy Physics Advisory Panel (HEPAP) on May 21, 2014. The charge of the panel was ``to develop an updated strategic plan for U.S. high energy physics that
can be executed over a 10 year timescale, in the context of a 20 year global vision for the field.'' Their reasonable top priority for constrained budget scenarios was to ``Use the Higgs boson as a new tool for discovery'' which is good news for the U. S. HEP groups working at the LHC at CERN; but lots of internal U. S. activities were ``redirected''~\cite{P5report}. For the unconstrained budget scenario, all they could come up with was:
\begin{itemize}\vspace*{-0.5pc}
\item Develop a greatly expanded accelerator R\&D program that would emphasize the ability to build very high-energy accelerators beyond the High-Luminosity LHC (HL-LHC) and ILC at dramatically lower cost. \vspace*{-0.5pc}
\item Play a world-leading role in the ILC experimental program and provide critical expertise and components to the accelerator, {\it should this exciting scientific opportunity be realized in Japan.}\vspace*{-0.5pc}
\end{itemize}
which IMHO lacks the imagination and drive of previous generations of U.S High Energy Physicists who had proposed and were constructing a 40 TeV $p$$+$$p$ collider for completion in 1995 if not for \ldots~\cite{StantheMan}.  

Not to be outdone, the new Long Range Planning exercise for U.~S.~ Nuclear Physics was initiated in April 2014. \vspace*{-1pc}
\section{RHIC Operations and accelerator future plans}\vspace*{-1pc}
Since beginning operation in the year 2000, RHIC, which can collide any species with any other species including polarized protons, has provided collisions at 14 different values of nucleon-nucleon c.m. energy, $\sqrt{s_{NN}}$, and ten different species combinations including Au+Au, d+Au, Cu+Cu, Cu+Au, U+U, and in 2014 He$^3$+Au, if differently polarized protons are counted as different species.  The performance history of RHIC with A+A 
\begin{figure}[!bh]
\begin{center}
\includegraphics[width=0.355\textwidth]{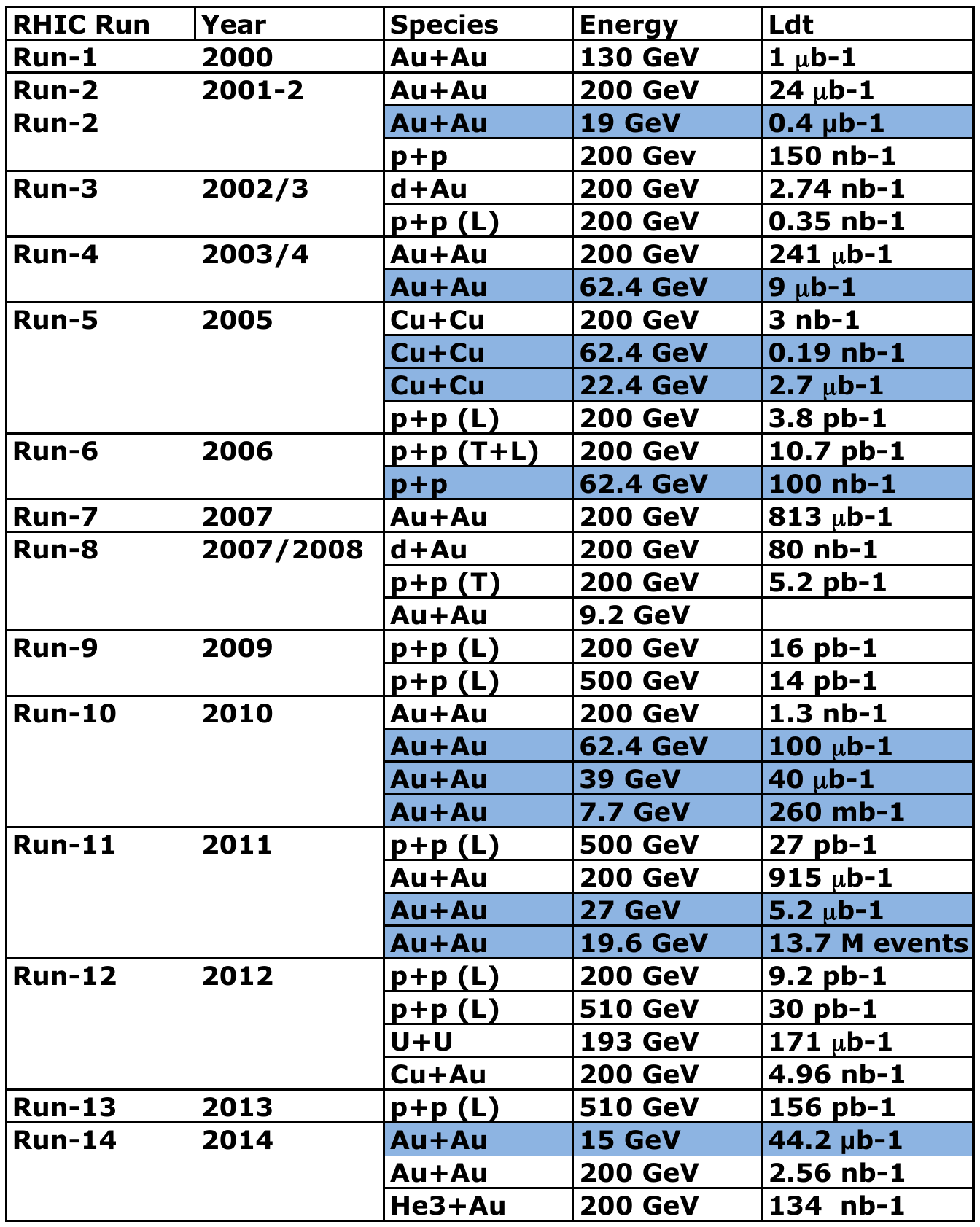}
\includegraphics[width=0.63\textwidth]{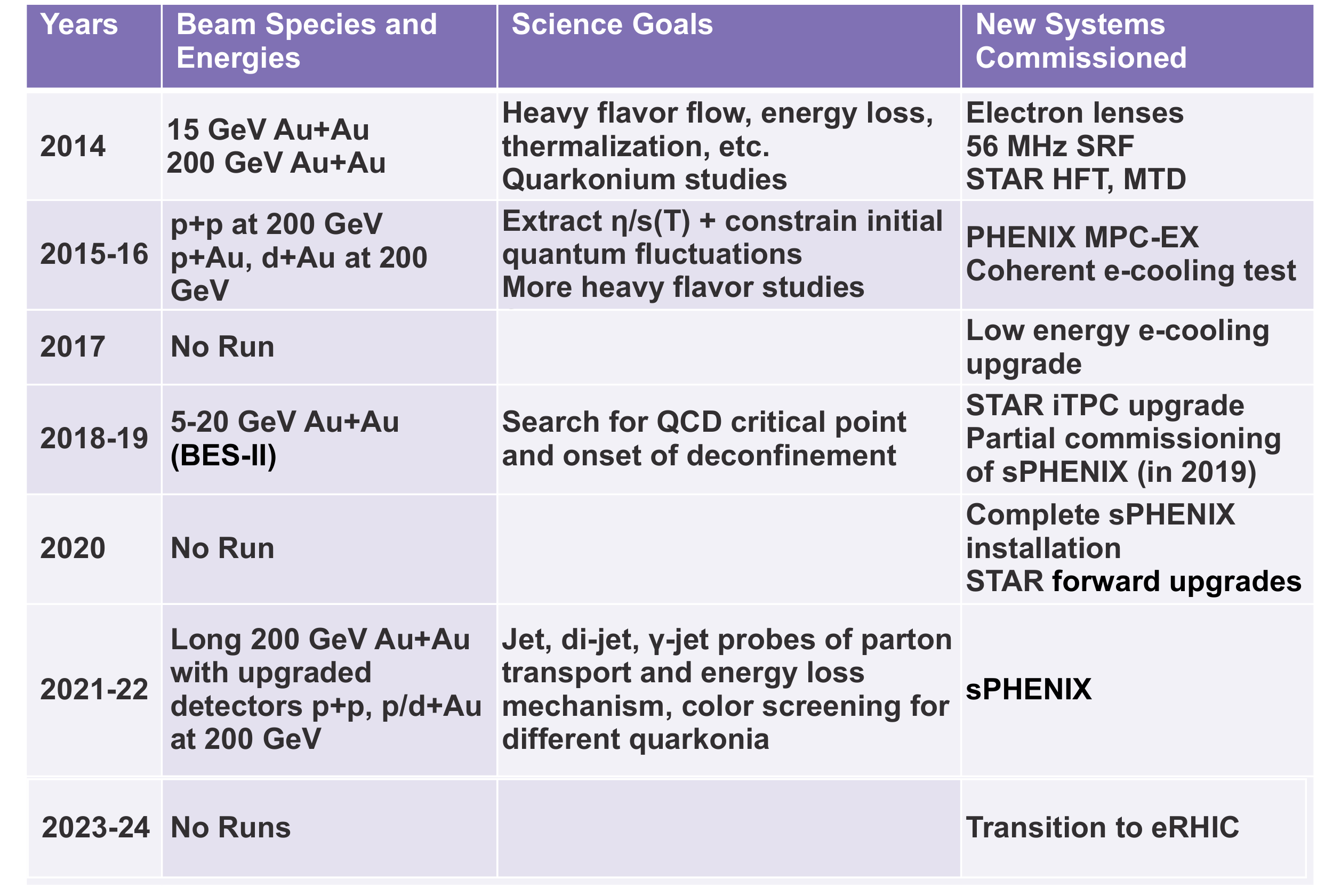}
\end{center}\vspace*{-1.5pc}
\caption[]{\footnotesize a)(left) Year, species and proton polarization ({\bf L}ongitudinal or {\bf T}ransverse), \sqsn, integrated luminosity of RHIC runs. b) (right) Future run schedule and new equipment.}
\label{fig:RHICperf}\vspace*{-0.5pc}
\end{figure}
and polarized $p$$+$$p$ collisions is shown in Fig.~\ref{fig:RHICperf}a; and in Fig.~\ref{fig:RHICperf}b, the plans for future runs are shown. 
 
\begin{figure}[!htb]
\begin{center}
\includegraphics[width=0.7\textwidth]{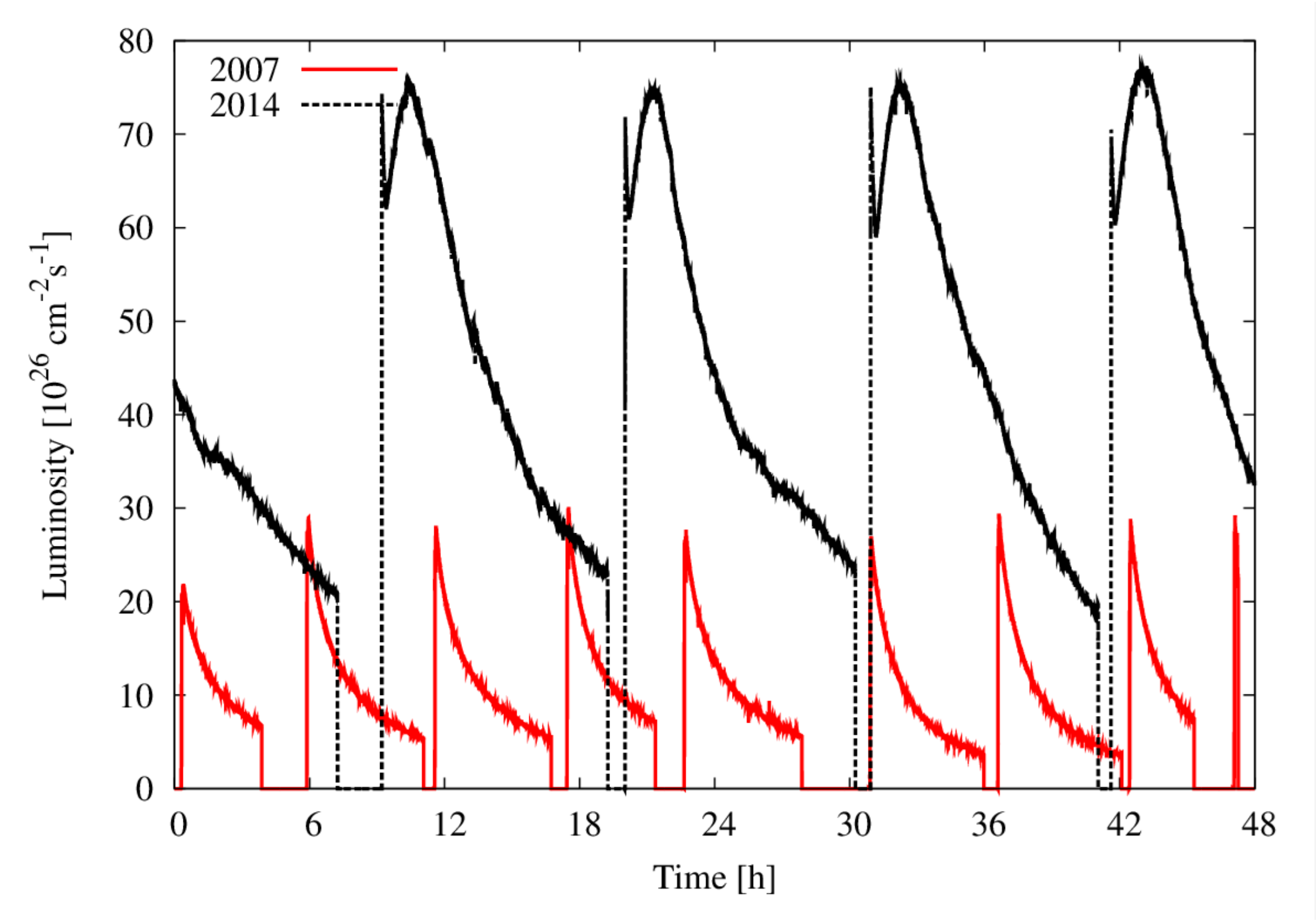}
\end{center}\vspace*{-1.9pc}
\caption[]{\footnotesize Run-14 luminosity vs. storage time compared to Run-7, courtesy Wolfram Fischer. }
\vspace*{-1.0pc}
\label{fig:RHICLumvst2014}
\end{figure}

For this year's run (2014) the full 3 dimensional cooling including electron lenses for partial compensation of the beam-beam tune shift and 56 MhZ storage r.f. for stronger longitudinal focusing were implemented which led to a higher initial luminosity and much longer lifetime of the beam with a more level luminosity load due to the 3d stochatic cooling (Fig.~\ref{fig:RHICLumvst2014}). The luminosity performance of RHIC with A+A and polarized $p$$+$$p$ collisions is shown in Fig.~\ref{fig:RHICperf}. Notably, the Au$+$Au $\int\!{\cal L}\ dt$ in 2014 exceeds all previous Au+Au runs combined as did the p$+$p $\int\!{\cal L}\ dt$ in 2013. 
\begin{figure}[!h]
\begin{center}
\includegraphics[width=0.49\textwidth]{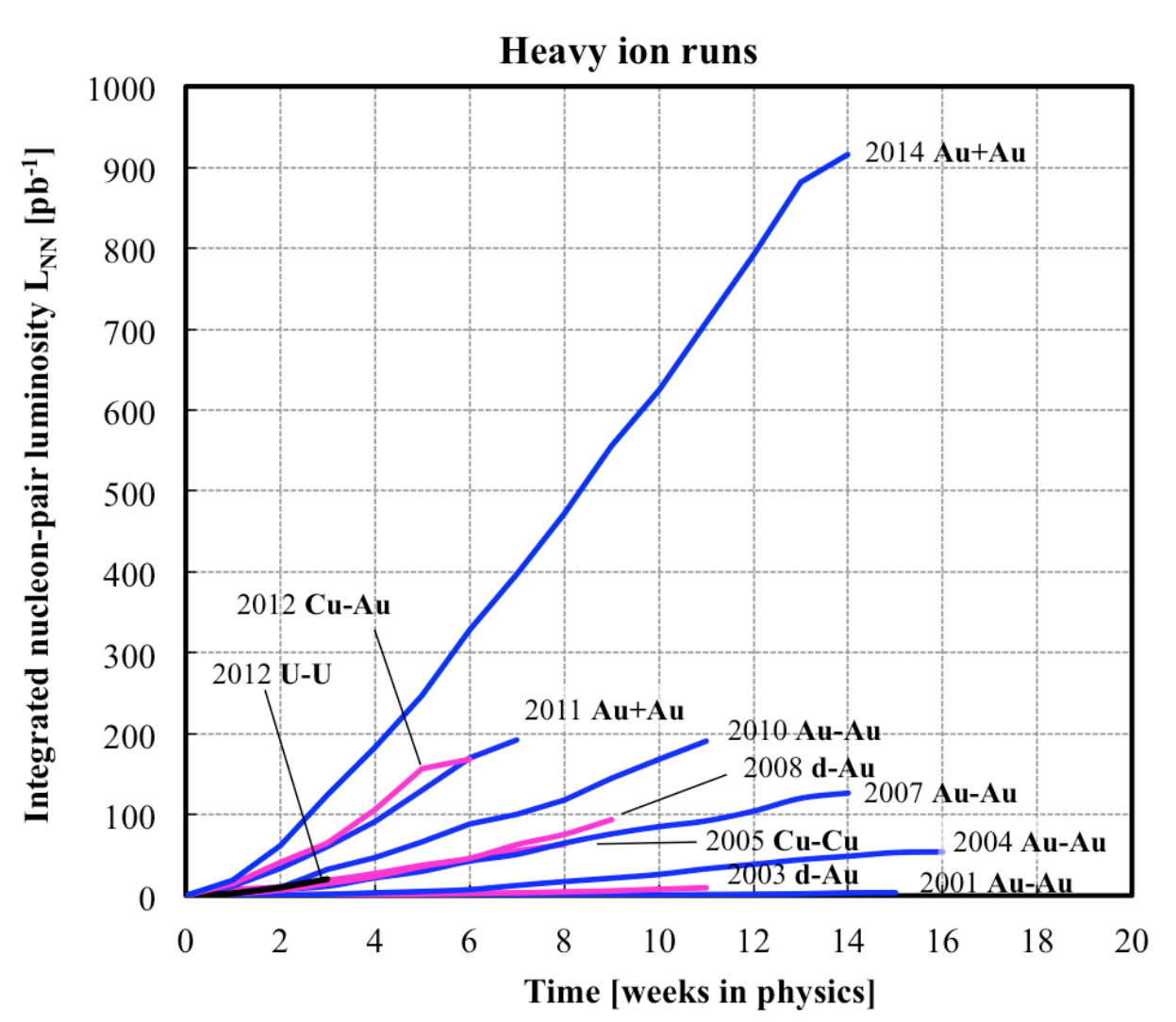}
\includegraphics[width=0.49\textwidth]{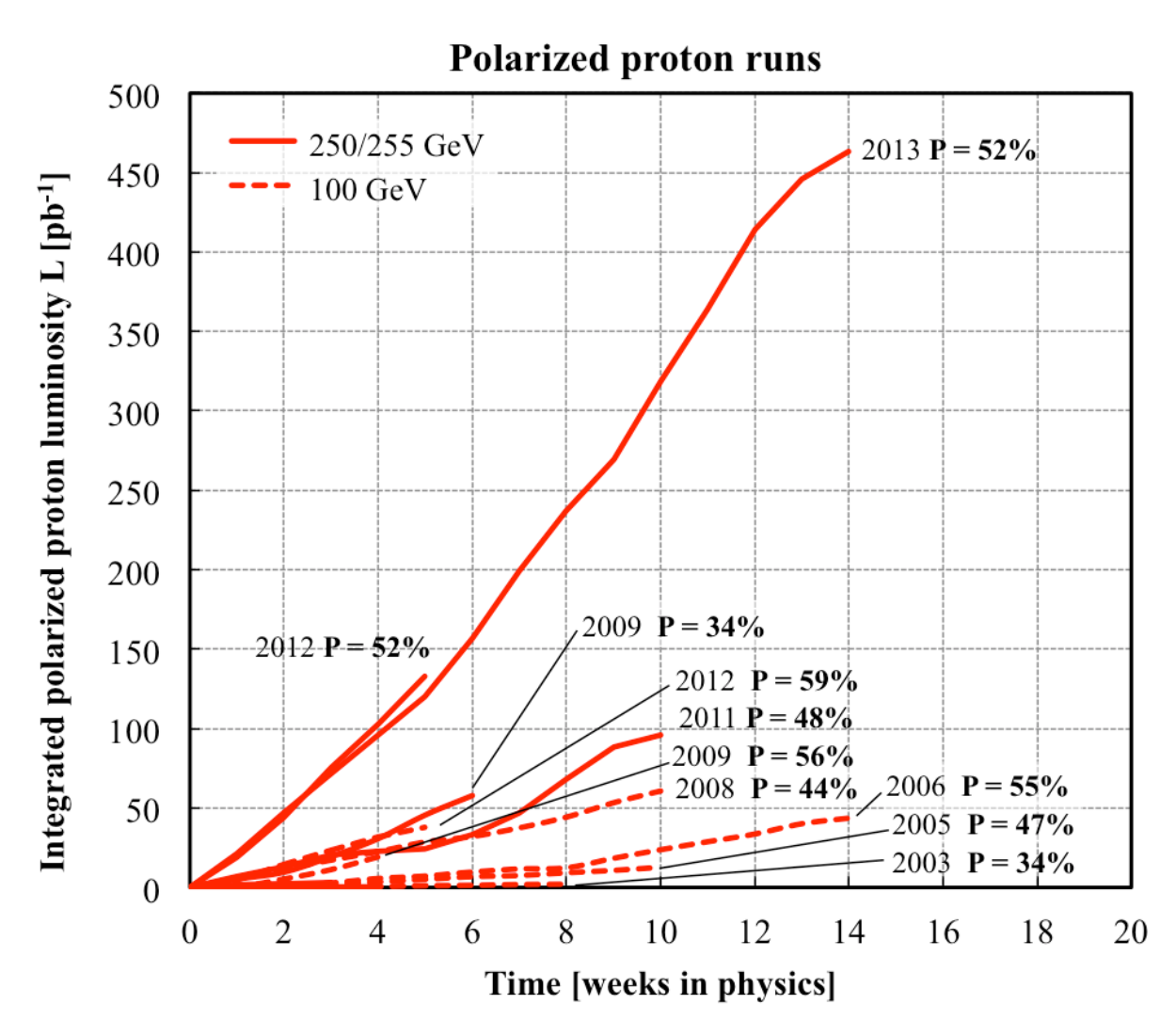}
\end{center}\vspace*{-1.5pc}
\caption[]{\footnotesize a)(left) Au+Au performance, where the nucleon-pair luminosity is defined as $L_{\rm NN}=A\times B\times L$, where $L$ is the luminosity and $A$, $B$ are the number of nucleons in the colliding species. b) (right) Polarized $p$$+$$p$ performance. Courtesy Wolfram Fischer.}
\label{fig:RHICperf}
\end{figure}
  
\begin{figure}[!htb]
\begin{center}
\includegraphics[width=0.8\textwidth]{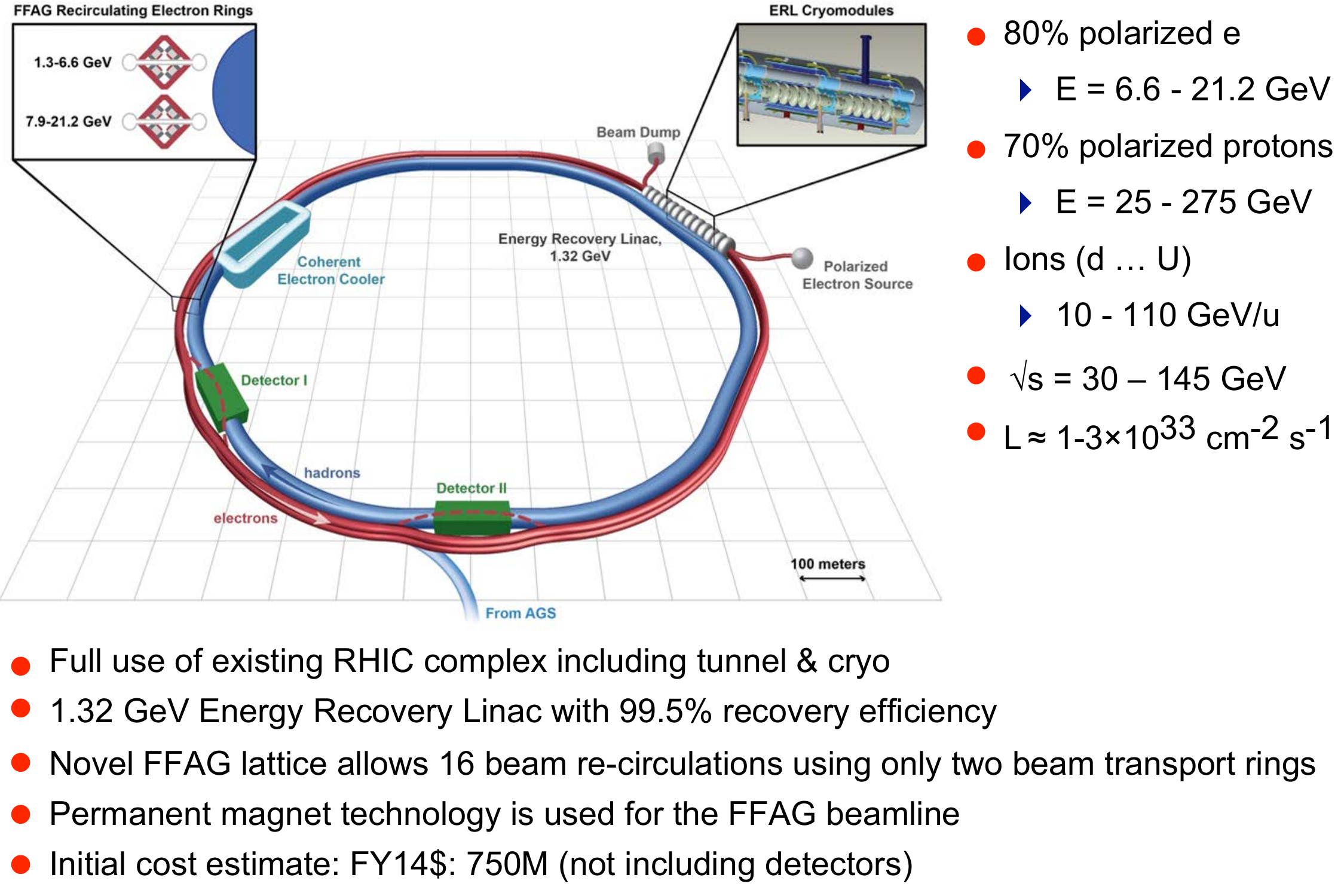}
\end{center}\vspace*{-1.95pc}
\caption[]{\footnotesize New BNL design for eRHIC with annotations. }
\label{fig:eRHIC}\vspace*{-1.0pc}
\end{figure}

The major future plan for accelerators in Nuclear Physics concerns an electron-ion collider, which if located at BNL will be called eRHIC. A new highly innovative and cost-effective design of eRHIC was proposed this year based on a Fixed Focus Alternating Gradient (FFAG) electron accelerator and an Energy Recovery Linac (Fig.~\ref{fig:eRHIC}). 
\vspace*{-1pc}
\section{Detector issues in A$+$A compared to p$+$p collisions}
\begin{figure}[!h]
\begin{center}
\begin{tabular}{cc}
\psfig{file=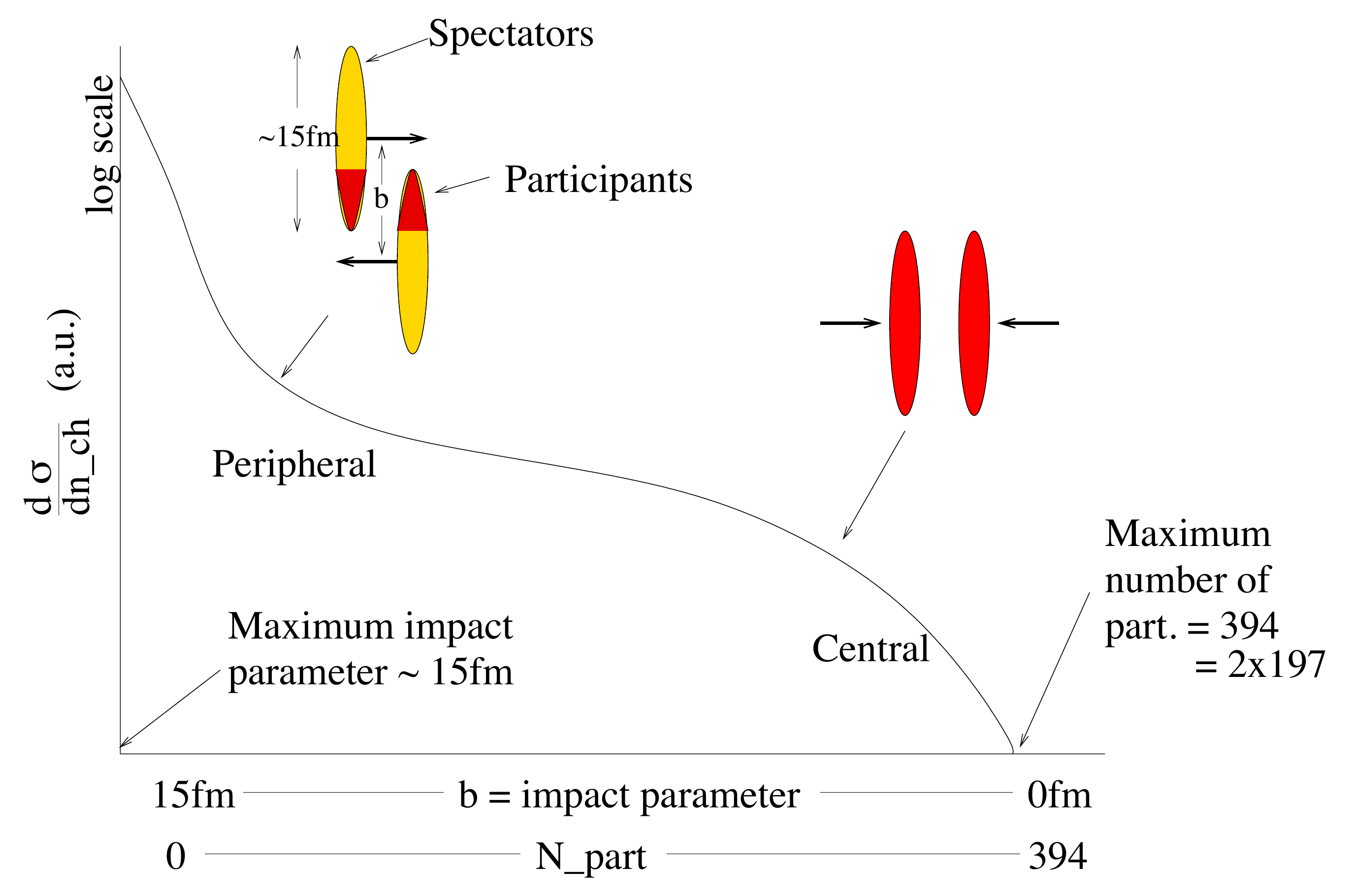,width=0.50\linewidth}\hspace*{-1pc}
\psfig{file=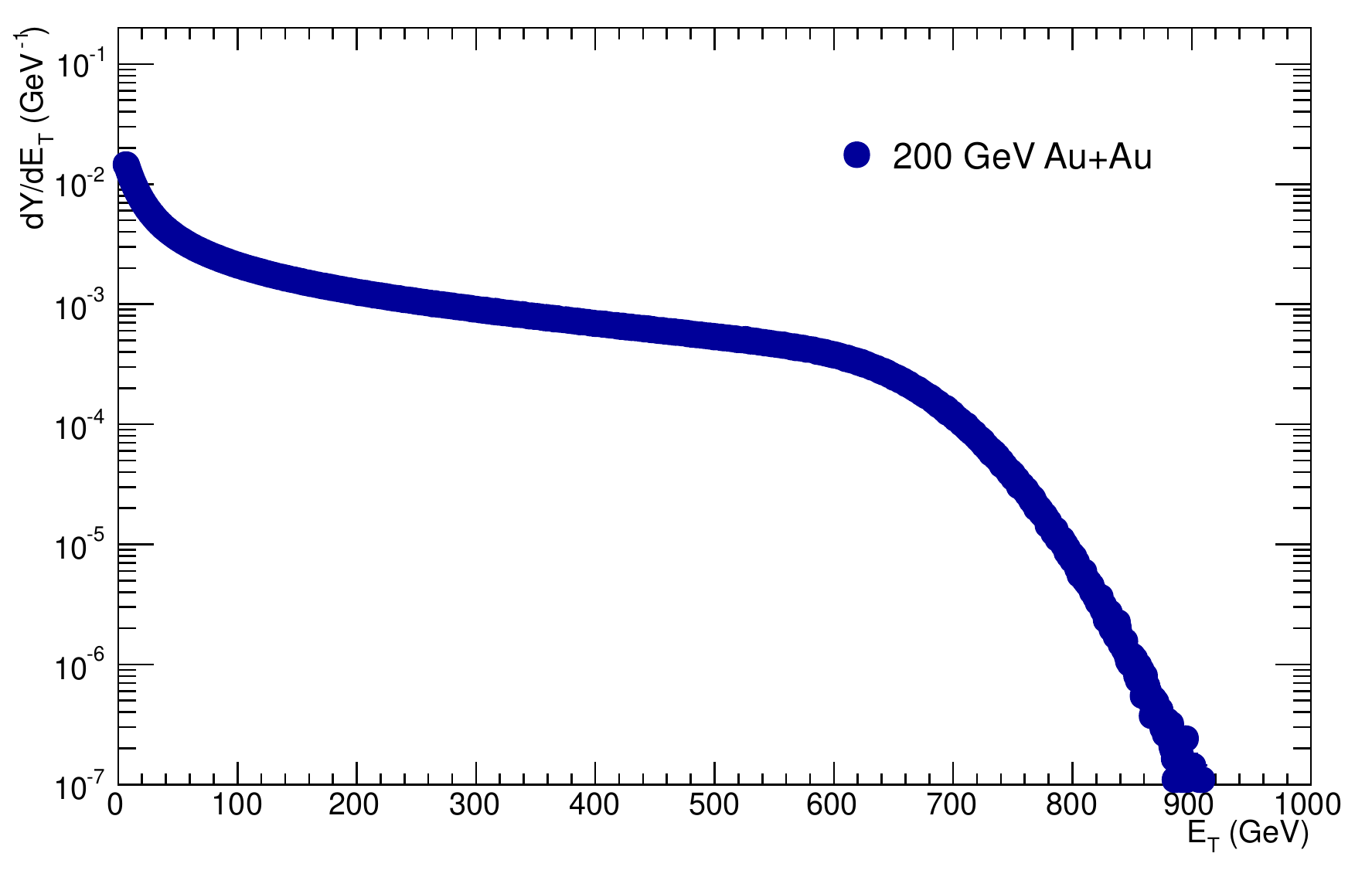,width=0.52\linewidth,angle=0}\end{tabular}
\end{center}\vspace*{-1.5pc}
\caption[]{\footnotesize a) (left) Schematic of collision in the $N$-$N$ c.m. system of two Lorentz contracted nuclei with radius $R$ and impact parameter $b$. The curve with the ordinate labeled $d\sigma/d n_{\rm ch}$ represents the relative probability of charged particle  multiplicity $n_{\rm ch}$ which is directly proportional to the number of participating nucleons, $N_{\rm part}$. b)(right) \Et distribution in Au+Au at  $\sqrt{s_{NN}}=200$ GeV from PHENIX~\cite{ppg100}. \label{fig:nuclcoll}}\vspace*{-0.5pc}
\end{figure}

\begin{figure}[!h]
\begin{center}
\begin{tabular}{cc}
\psfig{file=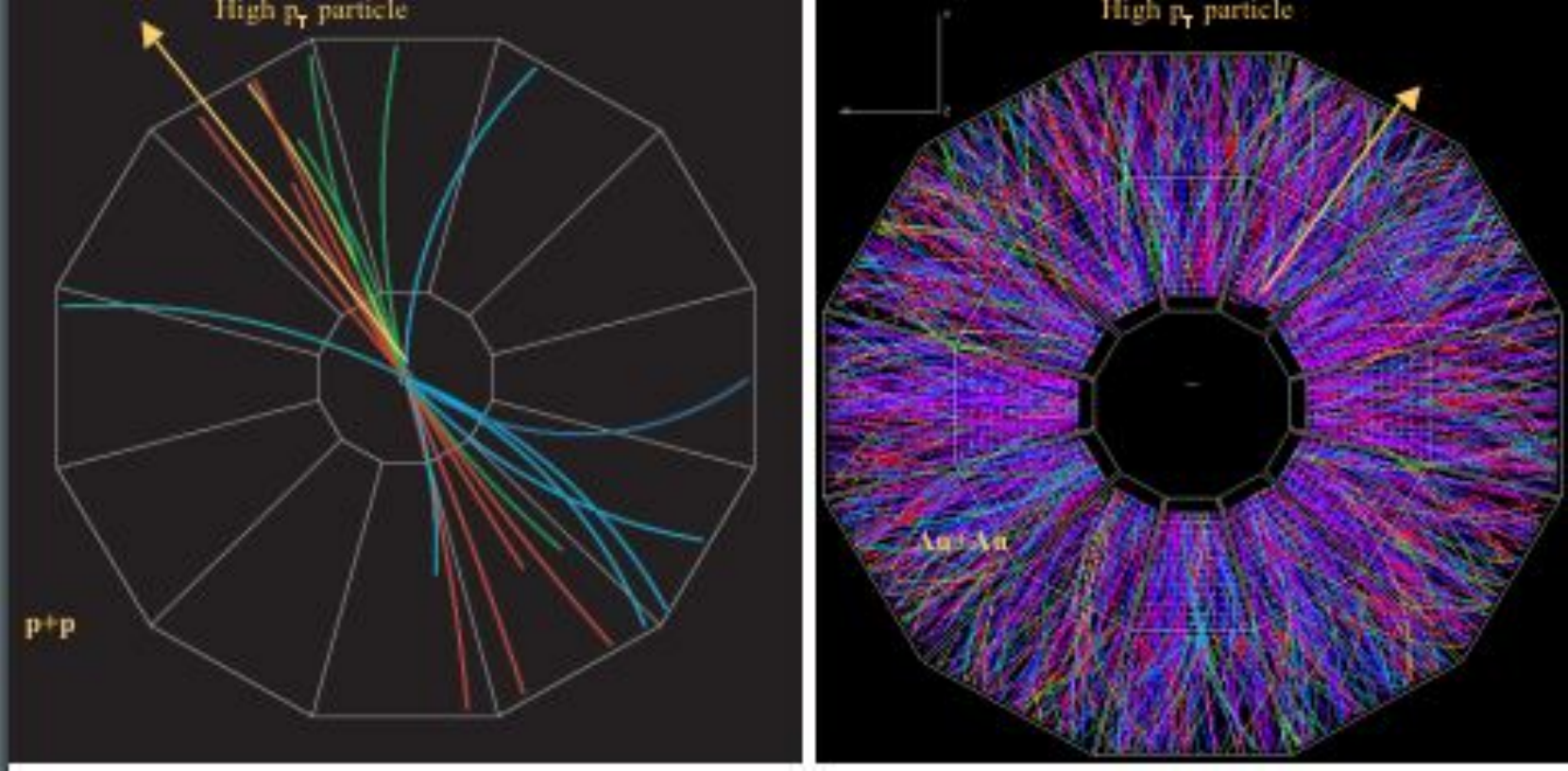,width=0.64\linewidth}&\hspace*{-0.025\linewidth}
\psfig{file=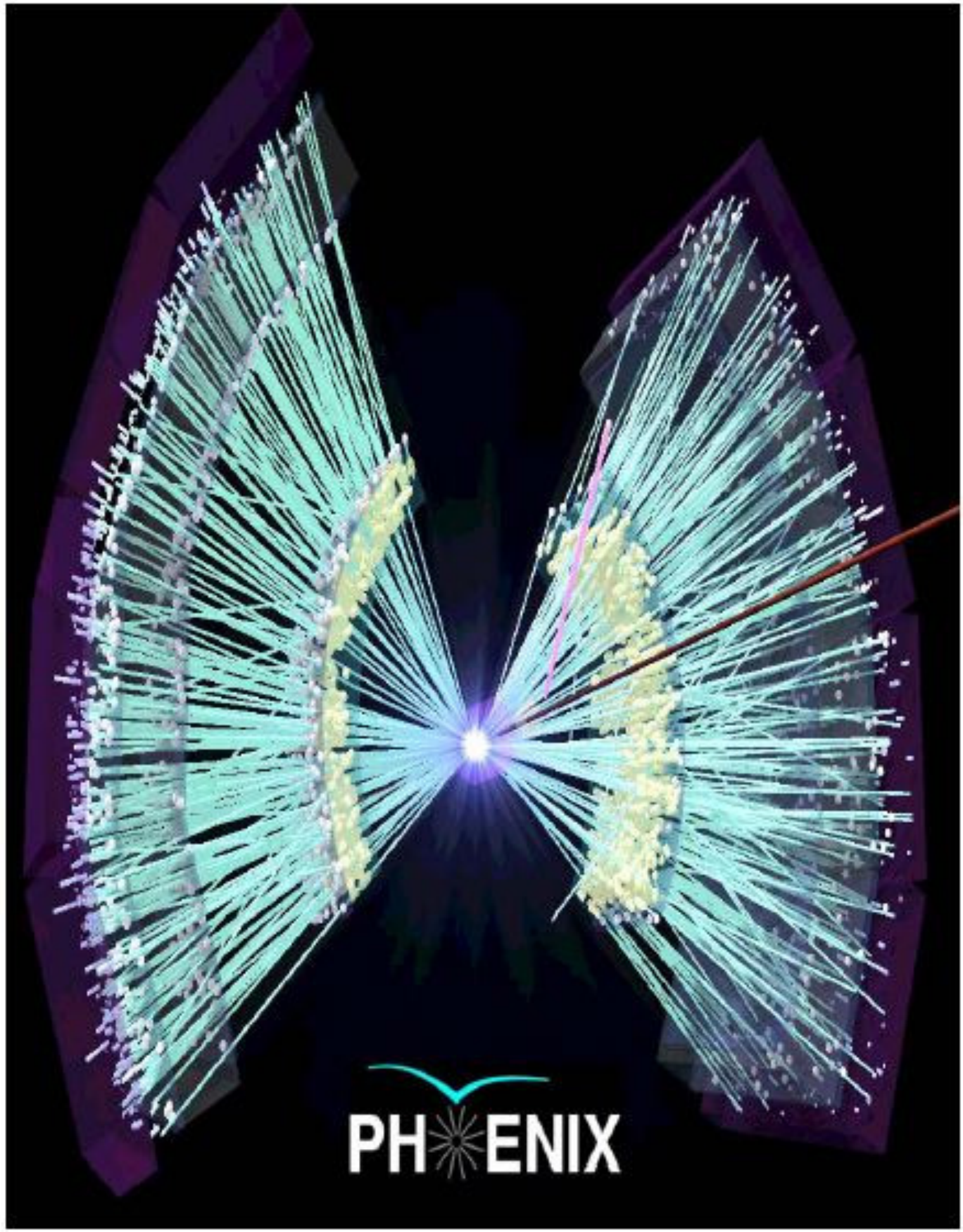,width=0.315\linewidth,height=0.315\linewidth} 
\end{tabular}
\end{center}\vspace*{-1.5pc}
\caption[]{\footnotesize a) (left) A $p$$+$$p$ collision in the STAR detector viewed along the collision axis; b) (center) Au+Au central collision at $\sqsn=200$ GeV in STAR;  c) (right) Au+Au central collision at $\sqsn=200$ GeV in PHENIX.  
\label{fig:collstar}}\vspace*{-0.5pc}
\end{figure}

 	A main concern of experimental design in RHI collisions is the huge multiplicity in A+A central collisions compared to  $p$$+$$p$ collisions. 
A schematic drawing of a collision of two relativistic Au nuclei is shown in Fig.~\ref{fig:nuclcoll}a. 
In the center of mass system of the nucleus-nucleus collision, the two Lorentz-contracted nuclei of radius $R$ approach each other with impact parameter $b$. In the region of overlap, the ``participating" nucleons interact with each other, while in the non-overlap region, the ``spectator" nucleons simply continue on their original trajectories and can be measured in Zero Degree Calorimeters (ZDC), in fixed target experiments, so that the number of spectators can be measured from which the number of participants (\Npart) can be determined for symmetric A+A collisions.  The degree of overlap is called the centrality of the collision, with $b\sim 0$, being the most central and $b\sim 2R$, the most peripheral. The maximum time of overlap is $\tau_\circ=2R/\gamma\,c$ where $\gamma$ is the Lorentz factor and $c$ is the speed of light in vacuum.  
The energy of the inelastic collision is predominantly dissipated by multiple particle production, where \Nch, the number of charged particles produced, or \Et, the energy emitted transverse to the beam direction, is directly proportional~\cite{PXWP} to \Npart  as sketched on Fig.~\ref{fig:nuclcoll}a. Thus, \Nch and \Et  in central Au+Au collisions are roughly $A$ times larger than in a $p$$+$$p$ collision, as shown in actual events from the STAR and PHENIX detectors at RHIC (Fig.~\ref{fig:collstar}). 

At colliders, the impact parameter $b$ can not be measured directly because charged spectators are swept away from zero degrees by the collider magnets. Instead, the centrality of a collision is defined in terms of the upper percentile e.g. top 10\%-ile, upper 10$-$20\%-ile, of \Nch or \Et distributions as in Fig.~\ref{fig:nuclcoll}b. Unfortunately the ``upper'' and ``-ile'' are usually not mentioned which sometimes confuses the uninitiated. Also a model is required to derive \Npart from the measurement so that the derived value of \Npart at a collider or the number of binary nucleon-nucleon collisions (\Ncoll) is model dependent and may have biases.

\begin{figure}[!h]
\begin{center}
a) \includegraphics[width=0.79\textwidth]{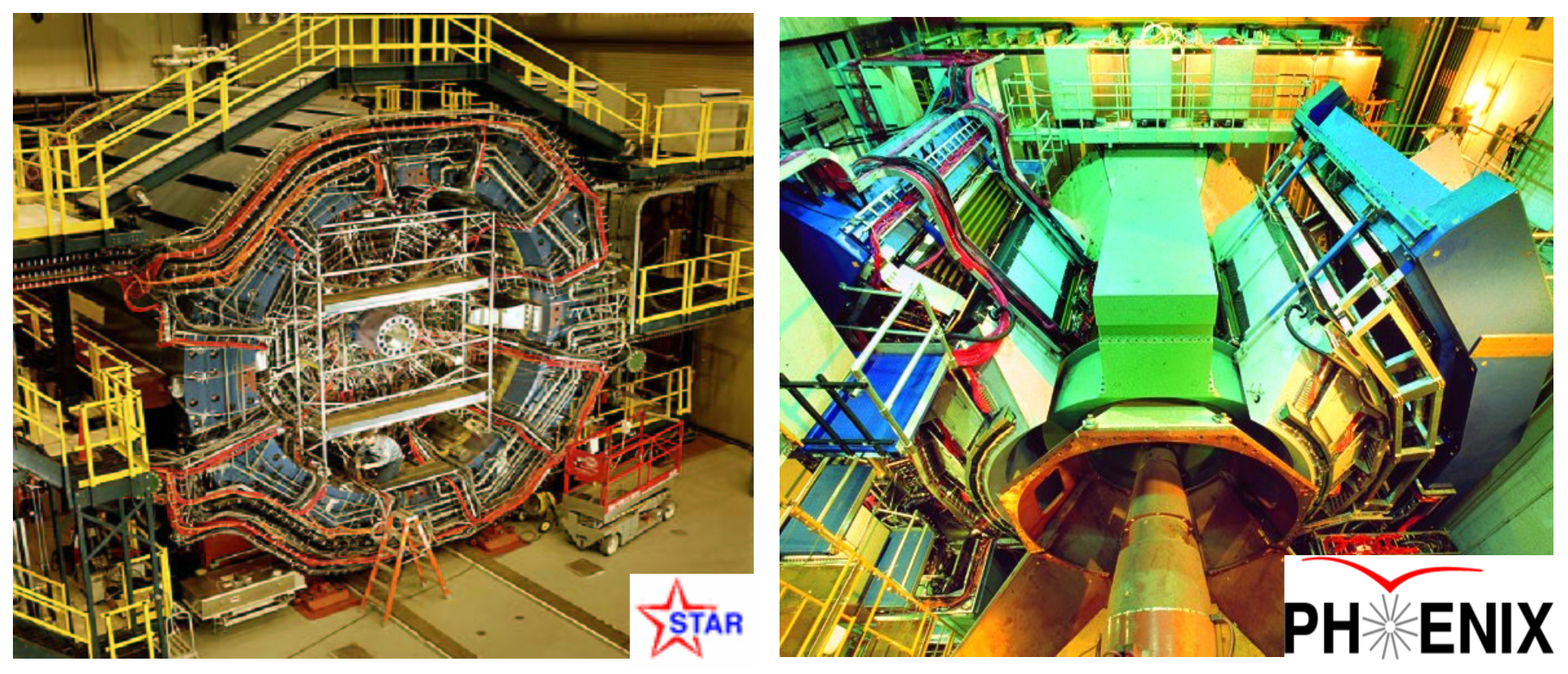} b)
\end{center}\vspace*{-1.5pc}
\caption[]{\footnotesize Actual STAR (a)  and PHENIX (b) detectors, compare with Figs~\ref{fig:collstar}b,c. The direction of the beam is along the axis of the STAR solenoid; and in (b) between the two spectrometer arms (perpendicular to the caption). }
\label{fig:DetectorPIX}\vspace*{-1.0pc}
\end{figure}
	Since it is a huge task to reconstruct the momenta and identity of all the particles produced in these events, the initial detectors at RHIC~\cite{RHICNIM} concentrated on the measurement of single-particle or multi-particle inclusive variables to analyze RHI collisions, with inspiration from the CERN ISR which emphasized those techniques before the era of jet reconstruction (see, for example, Refs.~\cite{MJTIJMPA2011} and \cite{RATCUP}). There are at present two major detectors in operation at RHIC, STAR and PHENIX, and there were also two smaller detectors, BRAHMS and PHOBOS, which have completed their program. As may be surmised from Figs.~\ref{fig:collstar}a,b and \ref{fig:DetectorPIX}a, STAR, which emphasizes hadron physics, is most like a conventional general purpose collider detector, a Time Projection Chamber to detect all charged particles over the full azimuth ($\Delta\phi=2\pi$) and  $\pm 1$ units of pseudo-rapidity ($\eta$); while PHENIX (Figs.~\ref{fig:collstar}c and \ref{fig:DetectorPIX}b), is a very high granularity high resolution special purpose detector: a two-arm spectrometer at mid-rapidity, with each arm covering solid angle $|\eta|\leq 0.35, \Delta\phi=90^\circ$, together with two full-azimuth muon detectors at forward and backward rapidity ($1.1\leq|\eta|\leq 2.3$).\footnote{The detector is so non-conventional that it made the cover of \href{http://www.phenix.bnl.gov/phenix/WWW/docs/covers/phystoday/2003oct/phystoday03oct.jpg}{Physics Today, October 2003}.} For the present runs, both STAR and PHENIX have excellent particle identification (PID) capability with electromagnetic calorimeters (EMcal) for photon and electron detection and Time of Flight for charged hadrons. PHENIX has a Ring Imaging CHerenkov counter for enhanced electron detection and triggering and small but full azimuth EM calorimeters (MPC) just before each muon arm covering $3.1\leq|\eta|\leq 3.7$, while STAR obtains enhanced hadron identification using dE/dx in the TPC. For the 2014 run, both PHENIX (VTX, FVTX) and STAR (HFT) are equipped with micro-vertex detectors for tagging Heavy-Flavor $c$ and $b$ quarks via displaced vertices.  

The main objectives of buliding RHIC were i) to discover the Quark Gluon Plasma (\QGP), which was achieved as I have discussed in detail in review articles based on previous ISSP proceedings~\cite{MJTIJMPA2011,MJTIJMPA2014}; ii) to measure its properties, which were much different than expected, namely a ``perfect  fluid" of quarks and gluons with their color charges exposed rather than a gas. The latest measurements from RHIC continue to be very interesting. \vspace*{-1.0pc}
\section{\Nch, \Et distributions and constituent-quarks as the fundamental elements of particle production}
\begin{figure}[!t] 
      \begin{center}
      \begin{tabular}{cc}
      a)\includegraphics[width=0.59\textwidth,angle=0.2]{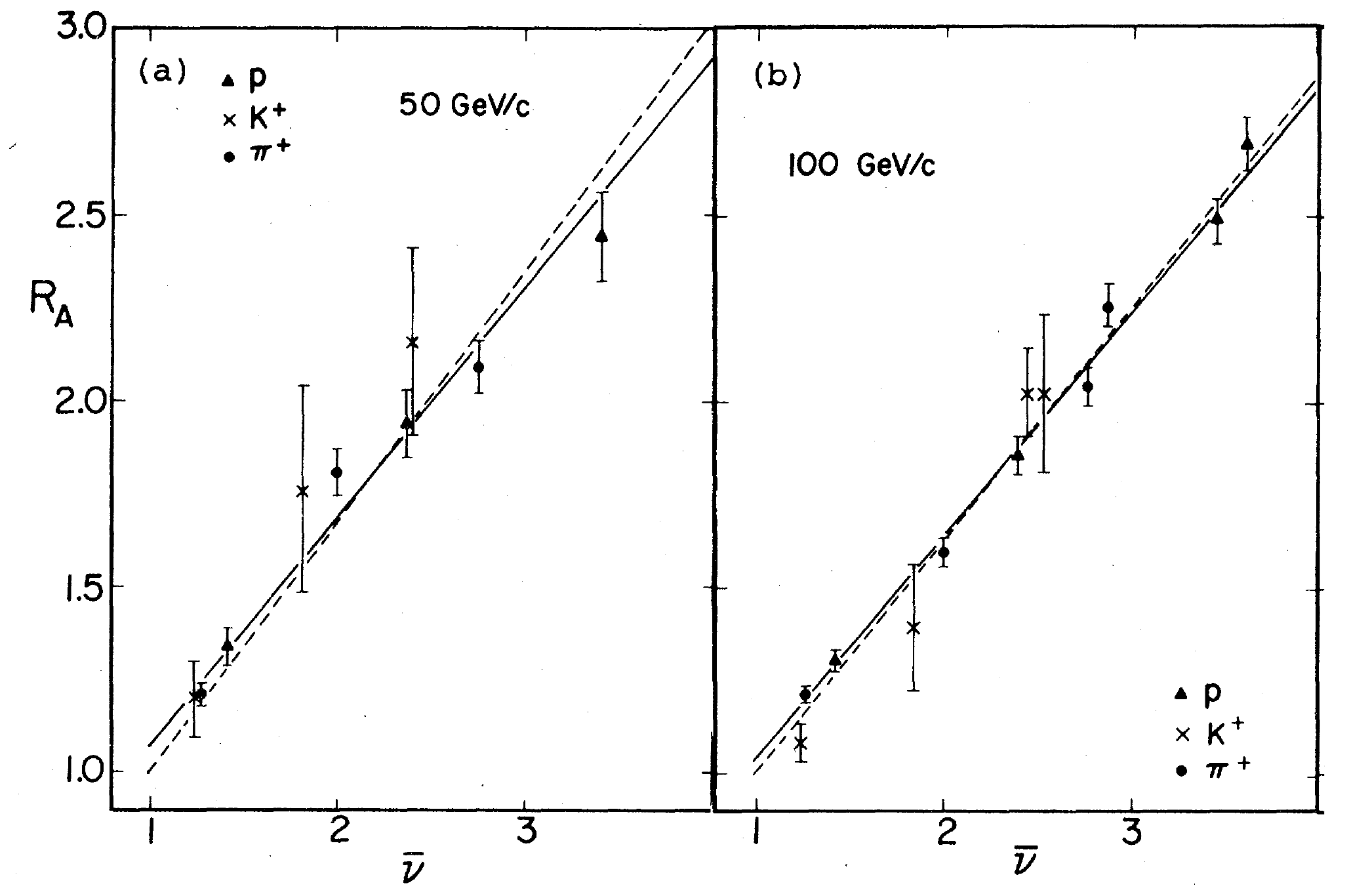}&\hspace*{0.2pc}
      b)\includegraphics[width=0.34\textwidth,angle=-1]{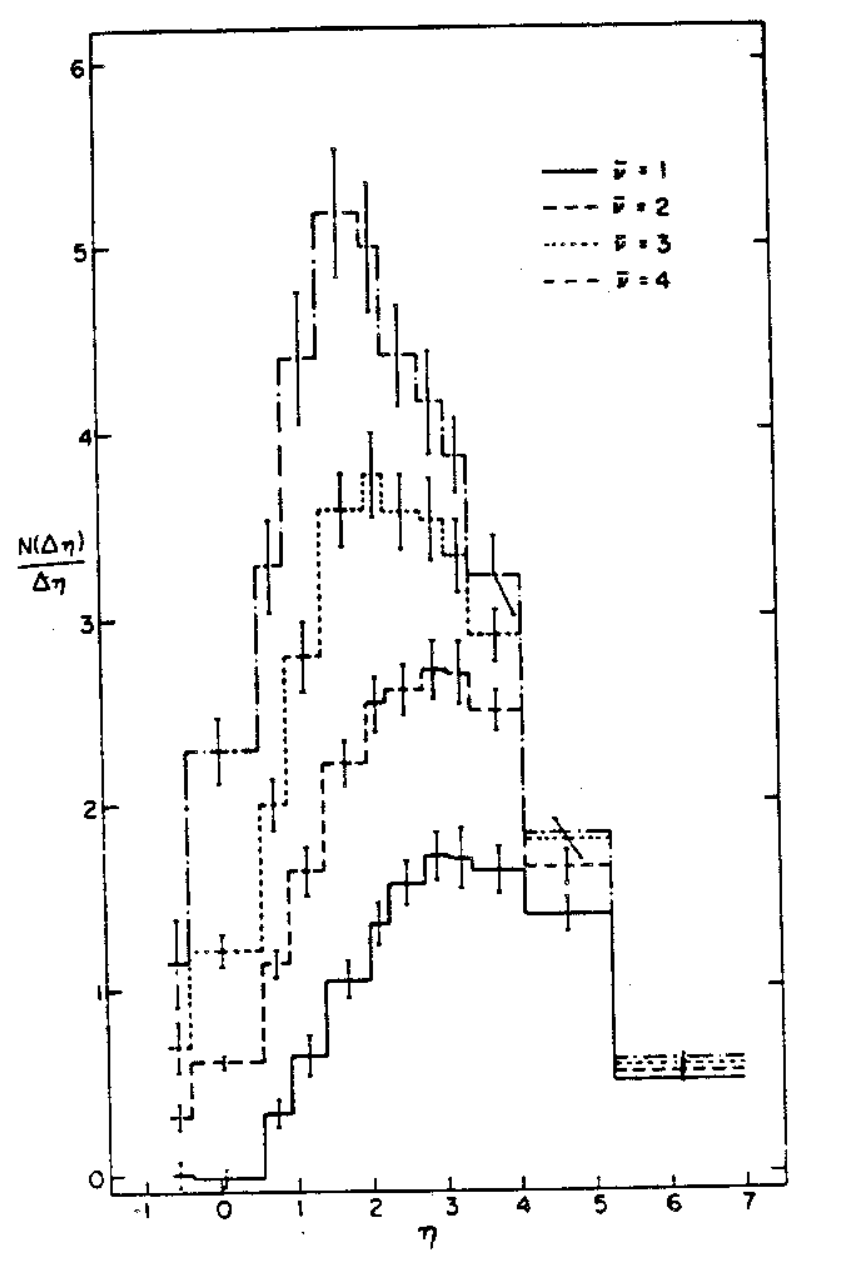}      
      \end{tabular}
      \end{center}\vspace*{-1.0pc}
      \caption[]{\footnotesize a) $R_A=\mean{\Nch}_{hA}/\mean{\Nch}_{hp}$ as a function of the average thickness of each nucleus given in terms of the mean free path, $\overline{\nu}=\mean{\Ncoll}$~\cite{EliasPRD22} for 50 and 100 GeV/c h+A  collisions; b)  Charged particle multiplicity density, $d\Nch/d\eta$, as a function of $A$ (represented by $\overline{\nu}$) for 200 GeV/c p+A collisions~\cite{HalliwellPRL39}.}
      \label{fig:pAdists}\vspace*{-1.0pc}
      \end{figure}

     The first experiment specifically designed to measure the dependence of the charged particle multiplicity in high energy  p+A collisions as a function of the nuclear size was performed by Wit Busza and collaborators at Fermilab using beams of $\sim 50-200$ GeV/c hadrons colliding with various fixed nuclear targets. They found the extraordinary result~\cite{BuszaPRL34} that the average charged particle multiplicity $\mean{\Nch}_{hA}$ in hadron+nucleus (h+A) interactions was not simply proportional to the number of collisions (absorption-mean-free-paths), $\mean{\Ncoll}=\overline{\nu}$, but increased much more slowly, proportional to the number of participants $\mean{\Npart}$ . Thus, relative to h+p collisions (Fig.~\ref{fig:pAdists}a)~\cite{EliasPRD22}: 
         \begin{equation}
         R_A=\mean{\Nch}_{hA}/\mean{\Nch}_{hp}=\mean{\Npart}_{hA}/\mean{\Npart}_{hp}=(1+\overline{\nu})/2 \label{eq:npartscaling}\qquad.
         \end{equation} 
Since the different projectiles, $h=\pi^+, K^+, p$ in Fig.~\ref{fig:pAdists}a have different mean free paths, the fit to the same straight line in terms of $\overline{\nu}$ is convincing.
      
      The other striking observation (Fig.~\ref{fig:pAdists}b)~\cite{HalliwellPRL39} was that a relativistic incident proton could pass through e.g. $\nu=4$ absorption-mean-free-paths of a target nucleus and emerge from the other side; and furthermore there was no intra-nuclear cascade of produced particles 
(a stark difference from what would happen to the same proton in a macroscopic 4 mean-free-path hadron calorimeter). 
In the forward fragmentation region of 200 GeV/c p+A collisions, within 1 unit of rapidity from the beam, $y^{\rm beam}=6.06$, there was essentially no change in $d\Nch/d\eta$ as a function of $A$, while at mid-rapidity ($\eta\approx y^{\rm cm}_{_{NN}}=3.03$), $d\Nch/d\eta$ increased with $A$ together with a small backward shift of the peak of the distribution  resulting in a huge relative increase of multiplicity in the target fragmentation region, $\eta<1$ in the laboratory system. 
These striking features of the $\sim 200$ GeV/c fixed target hadron-nucleus data ($\sqsn\sim 19.4$ GeV) showed the importance of taking into account the time and distance scales of the soft multi-particle production process including quantum mechanical effects. \vspace*{-1.0pc}
\subsection{The Wounded Nucleon Model} 
The observations in Fig.~\ref{fig:pAdists} had clearly shown that the target nucleus was rather transparent so that a relativistic incident nucleon 
could make many successive collisions while passing through the nucleus, and emerge intact. 
Immediately after a  relativistic nucleon interacts inside a nucleus, the only thing that can happen consistent with relativity and quantum mechanics is for it to become an excited nucleon with roughly the same energy and reduced longitudinal momentum and rapidity. It remains in that state inside the nucleus because the uncertainty principle and time dilation prevent it from fragmenting into particles until it is well outside the nucleus. This feature immediately eliminates the possibility of 
a cascade in the nucleus from the {rescattering} of the secondary products. 
If one makes the further assumptions that an excited nucleon interacts with the same 
cross section as an unexcited nucleon and that the successive collisions 
of the excited nucleon do not affect the excited state or its eventual 
fragmentation products~\cite{midFrankel}, this leads to the conclusion (c. 1977) that the elementary process for particle 
production in nuclear collisions is the excited nucleon, and to the prediction 
that the multiplicity in nuclear interactions should be proportional to 
the total number of projectile and target participants, rather than to the 
total number of collisions, or $R_A=\mean{\Nch}_{hA}/\mean{\Nch}_{hp}=\mean{\Npart}_{hA}/\mean{\Npart}_{hp}=(1+\overline{\nu})/2$, as observed.  This is called the Wounded Nucleon Model (WNM)~\cite{WNM} and, in the common usage, Wounded Nucleons (WN) are called participants. In a later model from the early 1980's, the Additive Quark Model, AQM~\cite{AQMPRD25}, constituent-quark participants were introduced; but the AQM is actually a model of particle production by color-strings in which only one color-string can be attached to a constituent-quark participant, effectively a projectile quark participant model.  \vspace*{-2.0pc}
\subsection{Extreme Independent Models}
\label{sec:ExtInd}
The models mentioned above are examples of Extreme Independent Models in which the effect of the nuclear geometry of the interaction can be calculated independently of the dynamics of particle production which can be taken directly from experimental measurements. The nuclear geometry is represented by the relative probability, $w_n$ per A$+$B interaction for a given number $n$ of fundamental elements, in the present case, number of collisions (\Ncoll), number of nucleon participants (wounded nucleon model-WNM~\cite{WNM}), number of constituent-quark participants (\Nqp), number of color strings (AQM). The dynamics of particle production, the \Nch or \Et distribution of the fundamental element, is taken from the measured $p$$+$$p$ data in the same detector: e.g. the measured \Nch distribution for a $p$$+$$p$ collision represents: 1 collision; 2 participants (WNM); a predictable convolution of constituent-quark-participants (NQP), or projectile-quark-participants (AQM). Glauber calculations of the nuclear geometry ($w_n$) together with the $p$$+$$p$ measurement provide a prediction for the A$+$B measurement in the same detector as the result of particle production by multiple independent fundamental elements.  

I became acquainted with these models in my first talk at a Quark Matter conference (QM1984) where   
\begin{figure}[!h] 
      \centering
      \small
a)\hspace*{-0.1pc}\raisebox{-1.3pc}{\includegraphics[width=0.31\linewidth,height=0.36\linewidth]{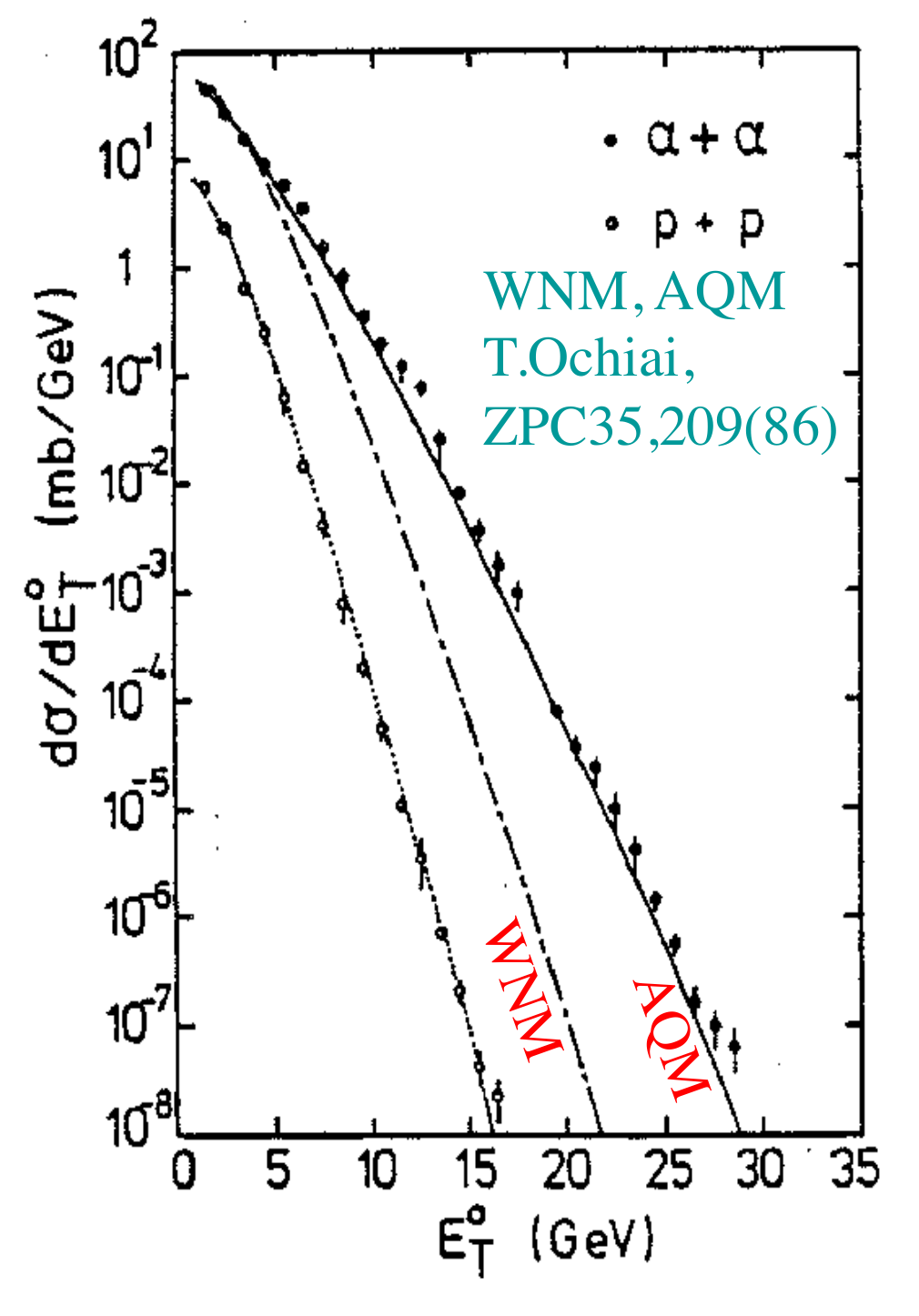}}b)\hspace*{-0.3pc}
\raisebox{-1pc}{\includegraphics[width=0.31\linewidth,height=0.35\linewidth,angle=1]{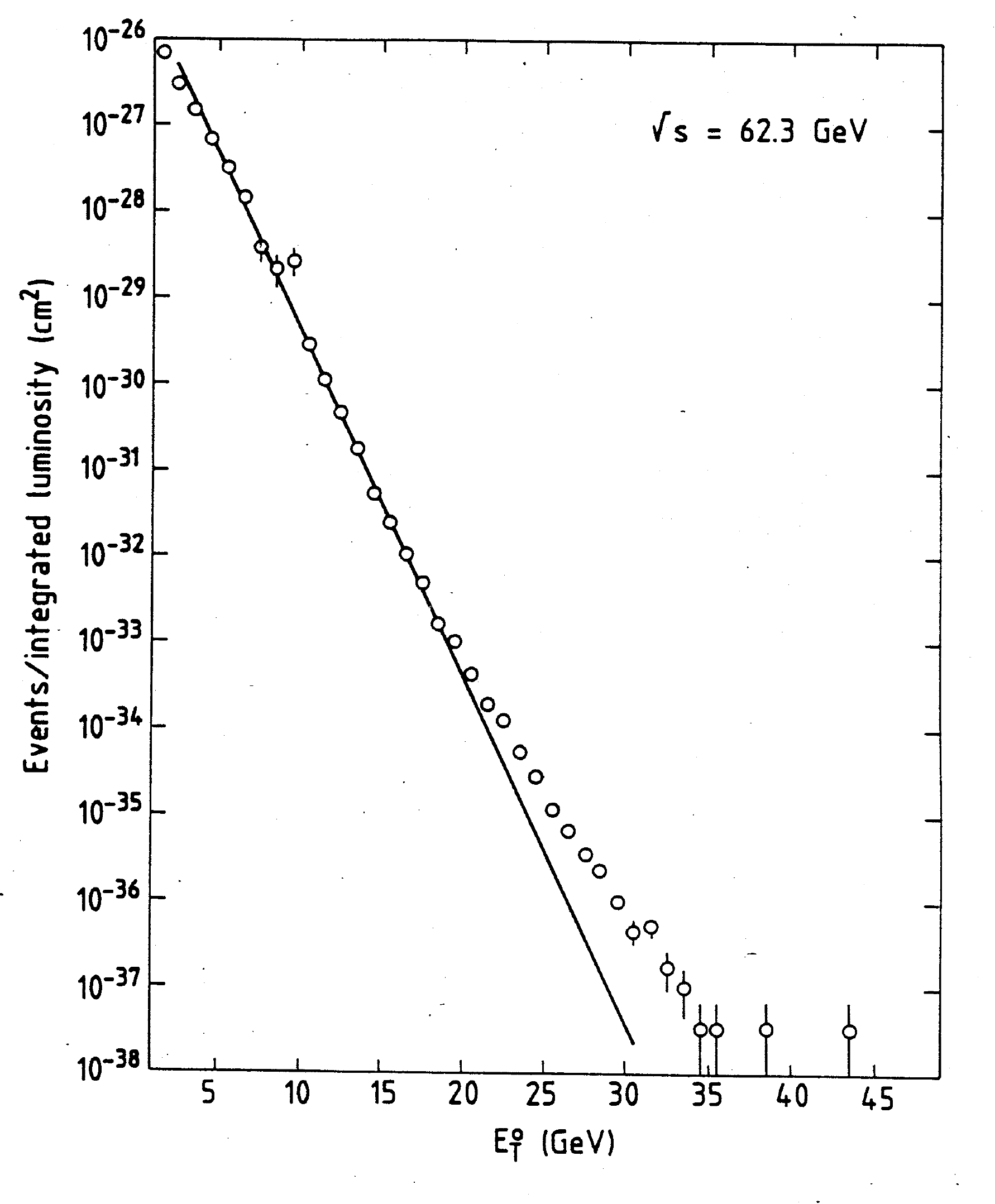}}\hspace*{-0.3pc}
c)\raisebox{-0.2pc}{\includegraphics[width=0.31\linewidth,height=0.33\linewidth,angle=-0.0]{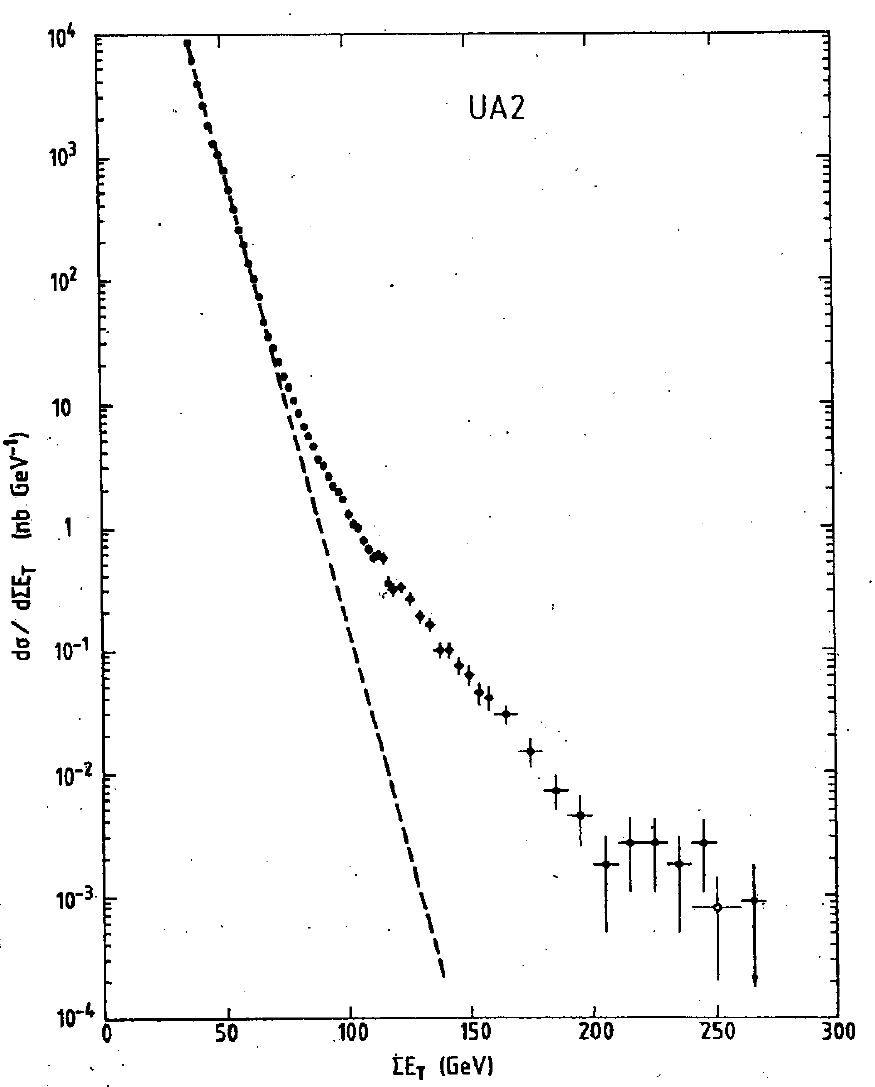}}\vspace*{-1.0pc}
     \caption[]{\footnotesize (a) \Et distributions in $p$$+$$p$, $\alpha$$+$$\alpha$~\cite{BCMOR-alfalfa} at \sqsn=31 GeV, with AQM and WNM calculations~\cite{OchiaiZPC35}. (b),(c) \Et distributions with breaks indicating jets: (b) $p$$+$$p$ \sqs=62.3 GeV~\cite{CMORNPB244}; (c) $d\sigma/d\Et$ (nb/GeV) vs. \Et for $\bar{p}$$+$$p$ \sqs=540 GeV~\cite{DiLellaARNPS85}.}
      \label{fig:3plots}\vspace*{-1.0pc}
   \end{figure}
I presented measurements of transverse energy distributions from $p$$+$$p$ and $\alpha$$+$$\alpha$ interactions at \sqsn=31 GeV at the CERN-ISR (Fig.~\ref{fig:3plots}a~\cite{BCMOR-alfalfa,MJTQM84proc}). The transverse energy, \Et, is a multiparticle variable defined as the sum
\begin{equation}
   \Et=\sum_i E_i\,\sin\theta_i \qquad d\Et(\eta)/d\eta=\sin\theta(\eta)\, dE(\eta)/d\eta, 
\label{eq:ETdef}
\end{equation}
where $\theta$ is the polar angle, $\eta=-\ln \tan\theta/2$ is the 
pseudorapidity, $E_i$ is by convention taken as the kinetic energy for 
baryons, the kinetic energy + 2 $m_N$ for antibaryons, and the total 
energy for all other particles, and the sum is taken over all particles 
emitted into a fixed solid angle for each event.  

The transverse energy was introduced by high energy physicists~\cite{WillisISAproc72,BjorkenPRD8} 
as an improved method to detect and study the jets from hard-scattering compared to high \pt single particle spectra by which hard-scattering was discovered in $p$$+$$p$ collisions and used as a hard-probe in Au$+$Au collisions at RHIC. However, it didn't work as expected: \Et distributions, like \Nch distributions, are dominated by soft particles near $\mean{\pt}$ (e.g. see Ref.~\cite{RATCUP} for details).
Nevertheless, it was claimed at the conference~\cite{CallenQM84}, in comments to my talk, that the deviation from the WNM in Fig.~\ref{fig:3plots}a was due to jets, but in both proceedings~\cite{MJTQM84proc,CallenQM84} 
it was demonstrated that~\cite{CallenQM84} ``there is no \ldots\ sign of jets. This indicates that soft processes are still dominant, and that we are still legimately testing the WNM at these high values of \Et.'' As shown in Fig.~\ref{fig:3plots}a, the the AQM~\cite{OchiaiZPC35}, rather than the WNM, followed the data.  Jets do appear in \Et distributions as a break $\lsim10^{-5}$ down in cross section (Figs.~\ref{fig:3plots}b,c). \vspace*{-1.0pc}  
\subsection{\Et and \Nch distributions cut on centrality}
\begin{figure}[!h] 
      \centering
      \small
a)\raisebox{0pc}{\includegraphics[width=0.29\linewidth]{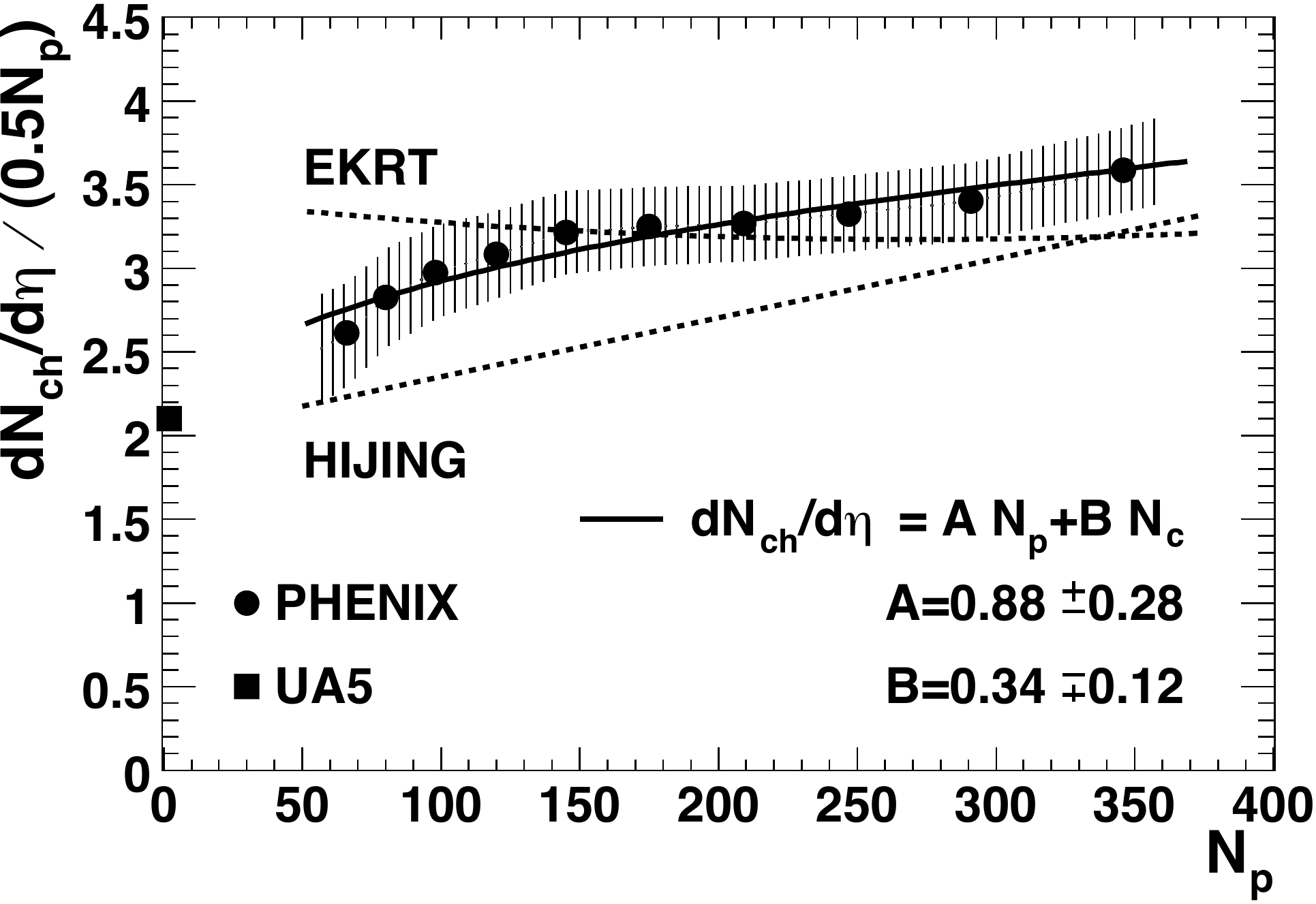}}\hspace*{1pc} 
b)\raisebox{0pc}{\includegraphics[width=0.29\linewidth,height=0.20\linewidth,angle=0]{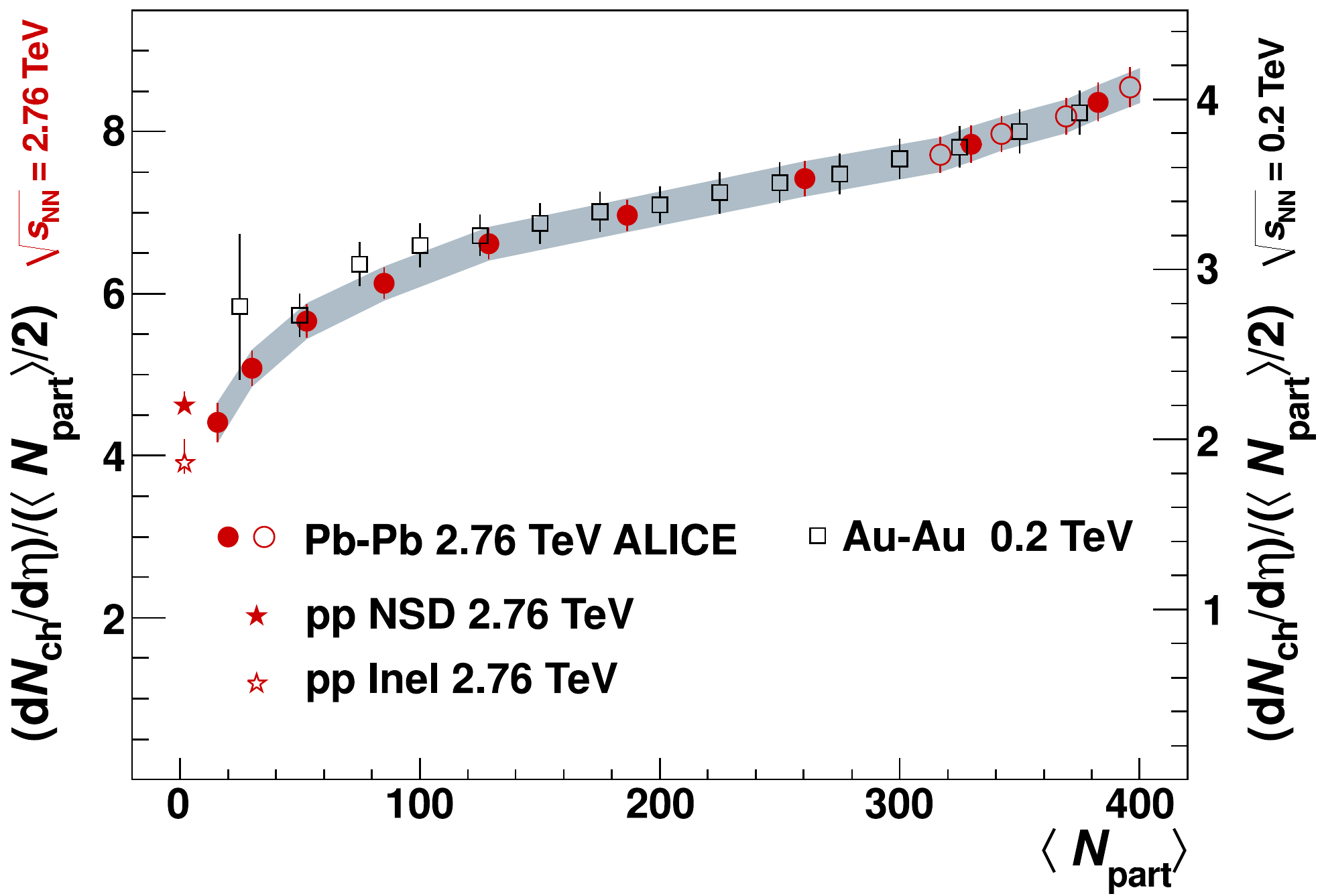}}\hspace*{0.5pc}
c)\raisebox{0pc}{\includegraphics[width=0.29\linewidth,height=0.21\linewidth,angle=-0.0]{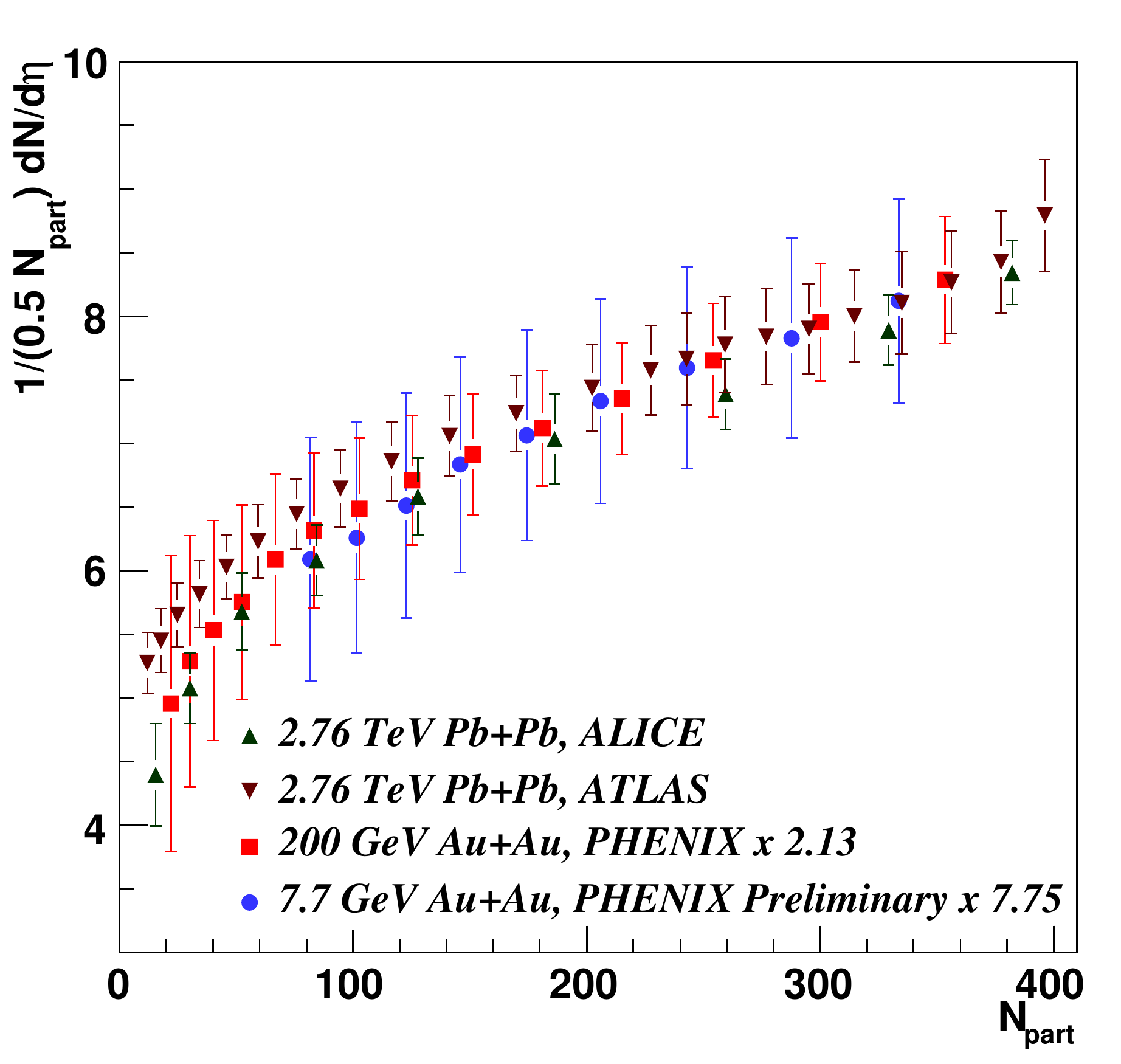}}
\normalsize
     \caption[]{\footnotesize (a) PHENIX, Au$+$Au, \sqsn=130 GeV~\cite{Adcox:2000sp}; (b) ALICE, Pb+Pb, \sqsn=2.76 TeV~\cite{ALICEPRL106}; (c) PHENIX preliminary Au$+$Au, \sqsn=7.7 GeV compared to the data at larger \sqsn~\cite{JTMCPOD8}. }
      \label{fig:3RHICplots}
   \end{figure}

At RHIC, following the style of the CERN SpS rather than the BNL-AGS fixed target heavy ion program, \Et and \Nch distributions were not generally shown. The measurements were presented cut in centrality in the form  $\mean{d\Nch^{\rm AA}/d\eta}/(\mean{\Npart}/2)$ vs. $\mean{\Npart}$ (Fig.~\ref{fig:3RHICplots}), 
which would be a constant equal to $\mean{d\Nch^{\rm pp}/d\eta}$ if the WNM worked. The measurements clearly deviate from the WNM (Fig.~\ref{fig:3RHICplots}a)~\cite{Adcox:2000sp}; so the PHENIX collaboration, inspired by    
the preceding article in the journal~\cite{WangGyulassyPRL86}, fit their data to the two-component model:
\begin{equation} 
{d\Nch^{AA}/d\eta}=({d\Nch^{pp}/d\eta})\ [(1-x)\,\mean{\Npart}/2 +x\,\mean{\Ncoll} ] \label{eq:ansatz}
\end{equation}
where the \Ncoll term implied a hard-scattering component for \Et and \Nch, known to be absent in $p$$+$$p$ collisions~\footnote{It was noted in Ref.~\cite{Adcox:2000sp} that hard-scattering was not a unique interpretation.  
The shape of the centrality dependences of $d\Et^{AA}/d\eta$ parameterized as $\Npart^\alpha$ were the same for Pb+Pb at \sqsn=17.6 GeV at the CERN SpS and Au$+$Au at \sqsn=130 GeV, with $\alpha=1.1$ and $\alpha=1.16\pm 0.04$, respectively. The LHC data 10 years later~\cite{ALICEPRL106} gave $\alpha=1.19\pm 0.02$ for Pb+Pb at \sqsn=2760 GeV, again the same shape.\label{fn:2}}
 (recall Fig.~\ref{fig:3plots}). A decade later, the first measurement from Pb+Pb collisions with \sqsn=2.76 TeV at the LHC Fig.~\ref{fig:3RHICplots}b~\cite{ALICEPRL106}, showed exactly the same shape  vs. \Npart as the RHIC Au$+$Au data at \sqsn=200 GeV, although $\mean{\Ncoll}$ is a factor of 1.6 larger and the hard-scattering cross section is more than a factor of 20 larger. This strongly argued against a hard-scattering component and for a nuclear geometrical effect, which was reinforced by a PHENIX preliminary measurement in Au$+$Au at \sqsn=7.7 GeV (Fig.~\ref{fig:3RHICplots}c)~\cite{JTMCPOD8} which also showed the same shape for the evolution of $\mean{d\Nch^{\rm AA}/d\eta}/(\mean{\Npart}/2)$ with \Npart as the \sqsn=200 and 2760 GeV measurements. 
It had previously been proposed that the number of constituent-quark participants provided the nuclear geometry that could explain the RHIC Au$+$Au data without the need to introduce a hard-scattering component~\cite{EreminVoloshinPRC67}.  However an asymmetric system is necessary in order to distinguish the NQP model from the AQM so the two models were applied to the RHIC $d$$+$Au data. 
\subsection{The number of constituent-quark participants model (NQP)}
\begin{figure}[!h]\vspace*{-0.5pc} 
      \centering
      \small
a)\raisebox{0pc}{\includegraphics[width=0.32\linewidth]{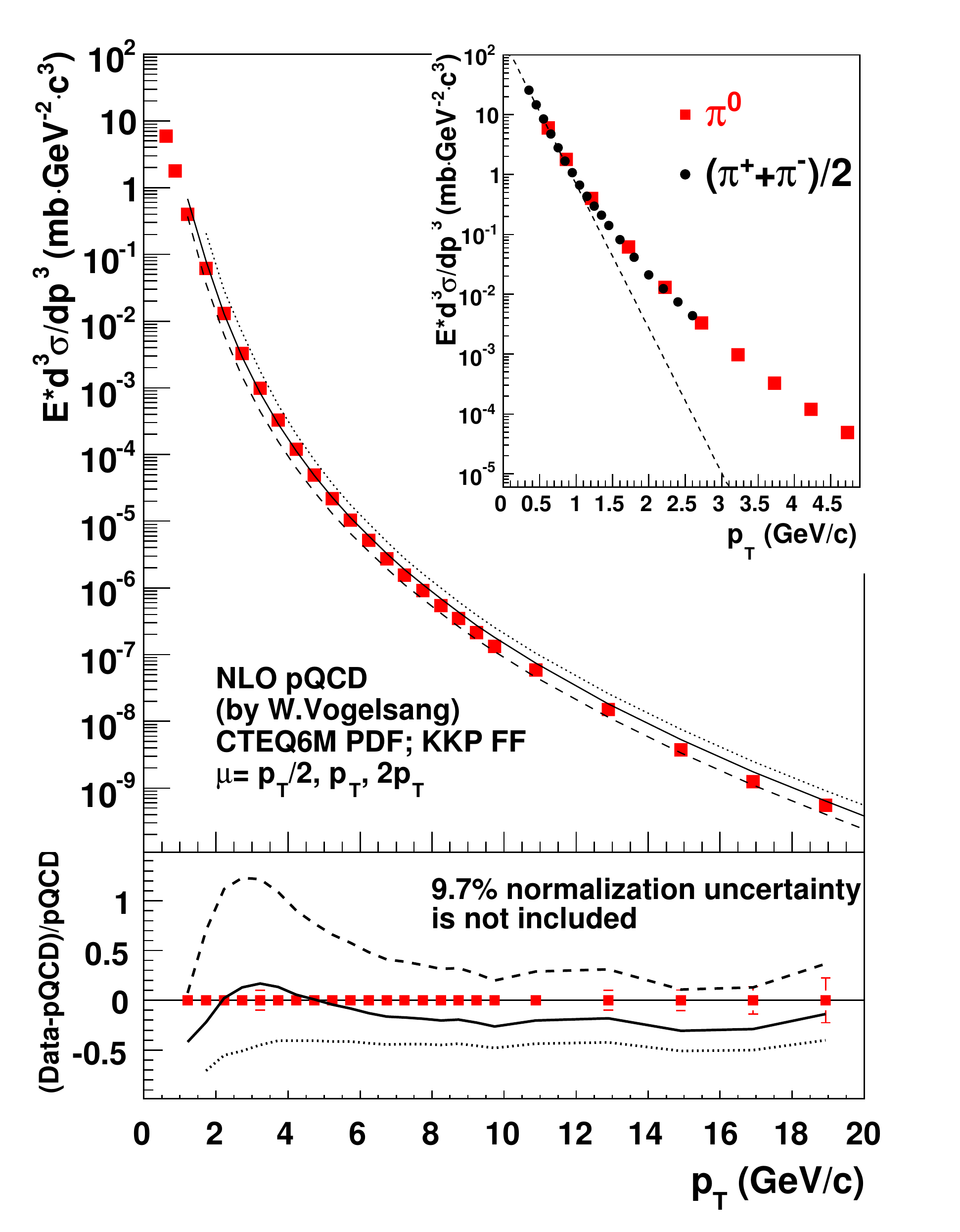}}\hspace*{4pc} 
\raisebox{0pc}{\includegraphics[width=0.52\linewidth,angle=0]{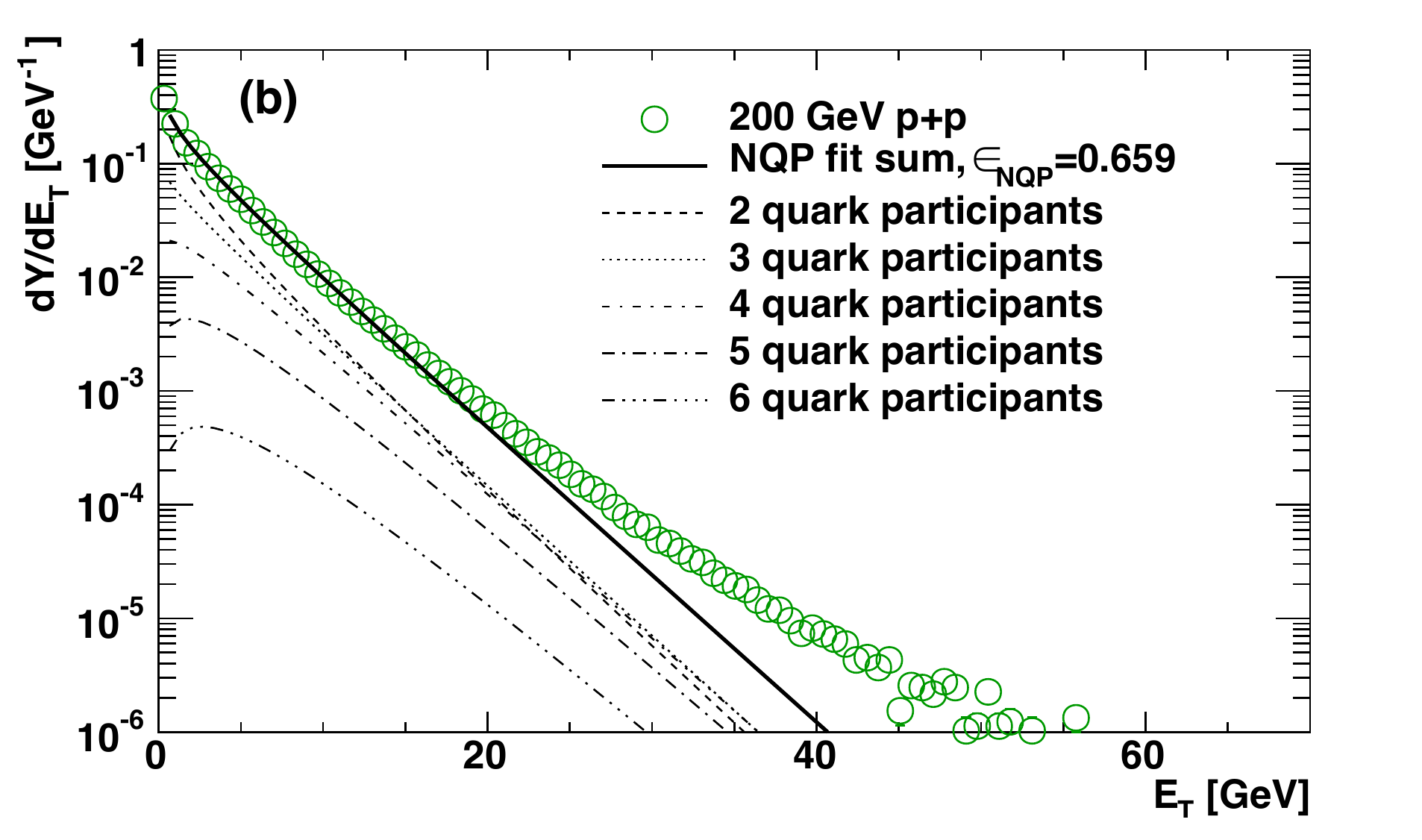}}\hspace*{0.5pc}
\normalsize
     \caption[]{\footnotesize (a) Invariant cross section of $\pi^0$ vs. $p_T$ at mid-rapidity in $p$$+$$p$ collisions at \sqs=200 GeV~\cite{PXpi0PRD}. The inset shows the transition from an exponental to a power-law in the range $1<\pt<2$ GeV/c (b) PHENIX deconvolution of $p$$+$$p$ \Et distribution at \sqs=200 GeV~\cite{ppg100}}
      \label{fig:pi0pT}
   \end{figure}

The massive constituent-quarks~\cite{MGMquark}, which form mesons and nucleons (e.g. a proton=$uud$), are relevant for static properties and soft physics with $p_T\lsim1.4$ GeV/c. They are complex objects or quasiparticles~\cite{ShuryakNPB203} made of the massless partons (valence quarks, gluons and sea quarks) of DIS~\cite{DIS2} such that the valence quarks acquire masses $\approx 1/3$ the nucleon mass with radii $\approx 0.3$ fm when bound in the nucleon. With  smaller resolution one can see inside the bag to resolve the massless partons which can scatter at large angles according to QCD. At RHIC, hard-scattering starts to be visible as a power law above soft (exponential) particle production only for $p_T>$ 1.4 GeV/c at mid-rapidity (Fig~\ref{fig:pi0pT}a), where $Q^2=2p_T^2=4$ (GeV/c)$^2$ which corresponds to a distance scale (resolution) $<0.1$ fm.\footnote{Shuryak and collaborators recently made similar arguments about resolution in separating hard from soft processes although their mechanism for soft particle and \QGP\ production was 2 color strings per wounded nucleon from Pomeron exchange~\cite{ColorSpaghettiPRC90}.}

A standard Monte Carlo Glauber calculation is used to assemble the initial positions of all the nucleons. Then three quarks are distributed around the center of each nucleon according to the proton charge distribution \mbox{$\rho(\vec{r})\propto e^{-{\sqrt{12}\,r/r_{\rm rms}}}$}, where $r_{\rm rms}=0.81$ fm is the rms charge radius of the proton~\cite{HofstadterRMP}. The $q$--$q$ inelastic scattering cross section is adjusted to 9.36 mb  at \sqs=200 GeV to give the correct $p$$+$$p$ inelastic cross section (42 mb)  and then used in the A$+$B calculations. 
\begin{figure}[!h]\vspace*{-0.5pc}  
      \centering 
\raisebox{0pc}{\includegraphics[width=0.48\linewidth,angle=0]{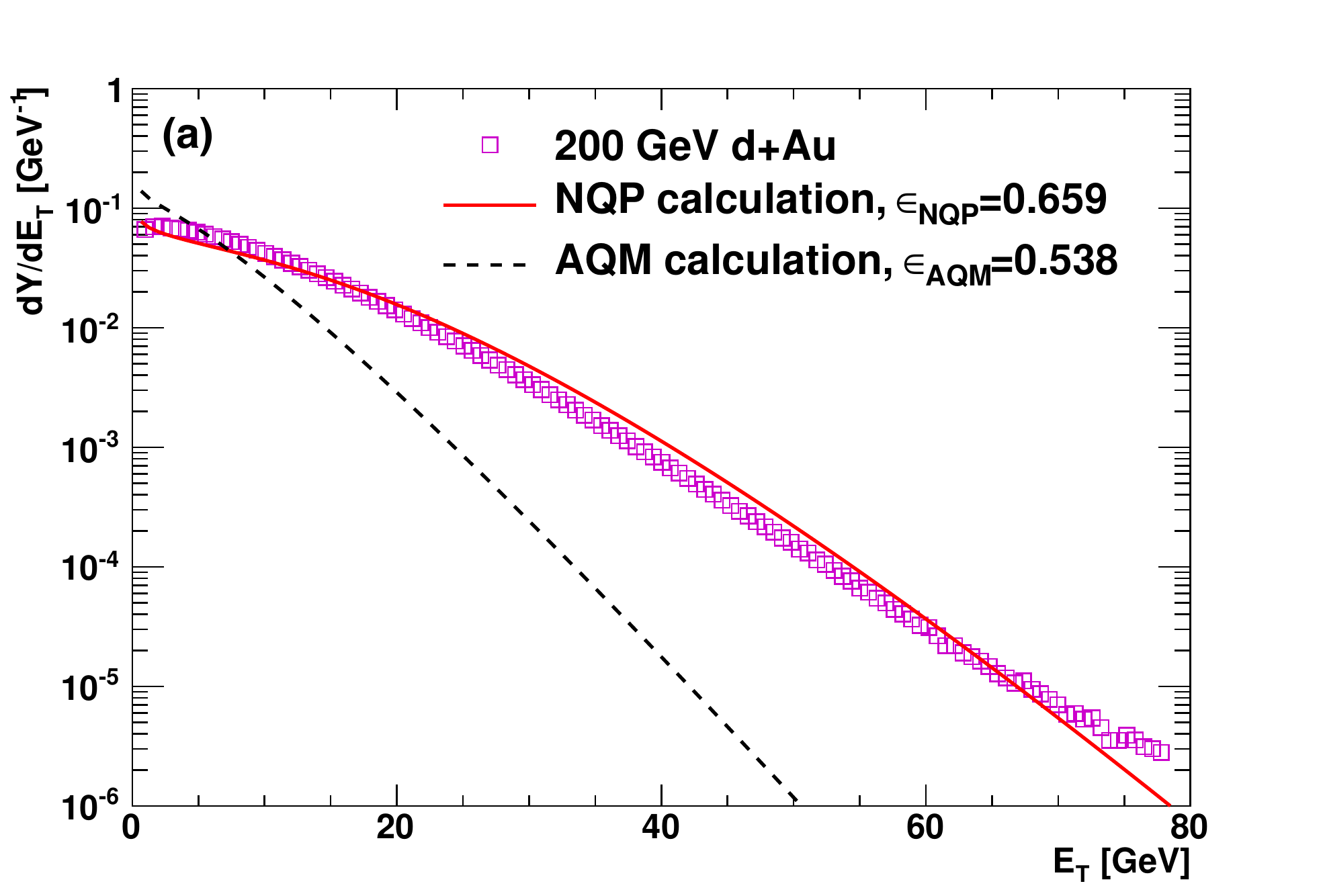}}\hspace*{0.5pc}
\raisebox{0pc}{\includegraphics[width=0.48\linewidth,angle=-0.0]{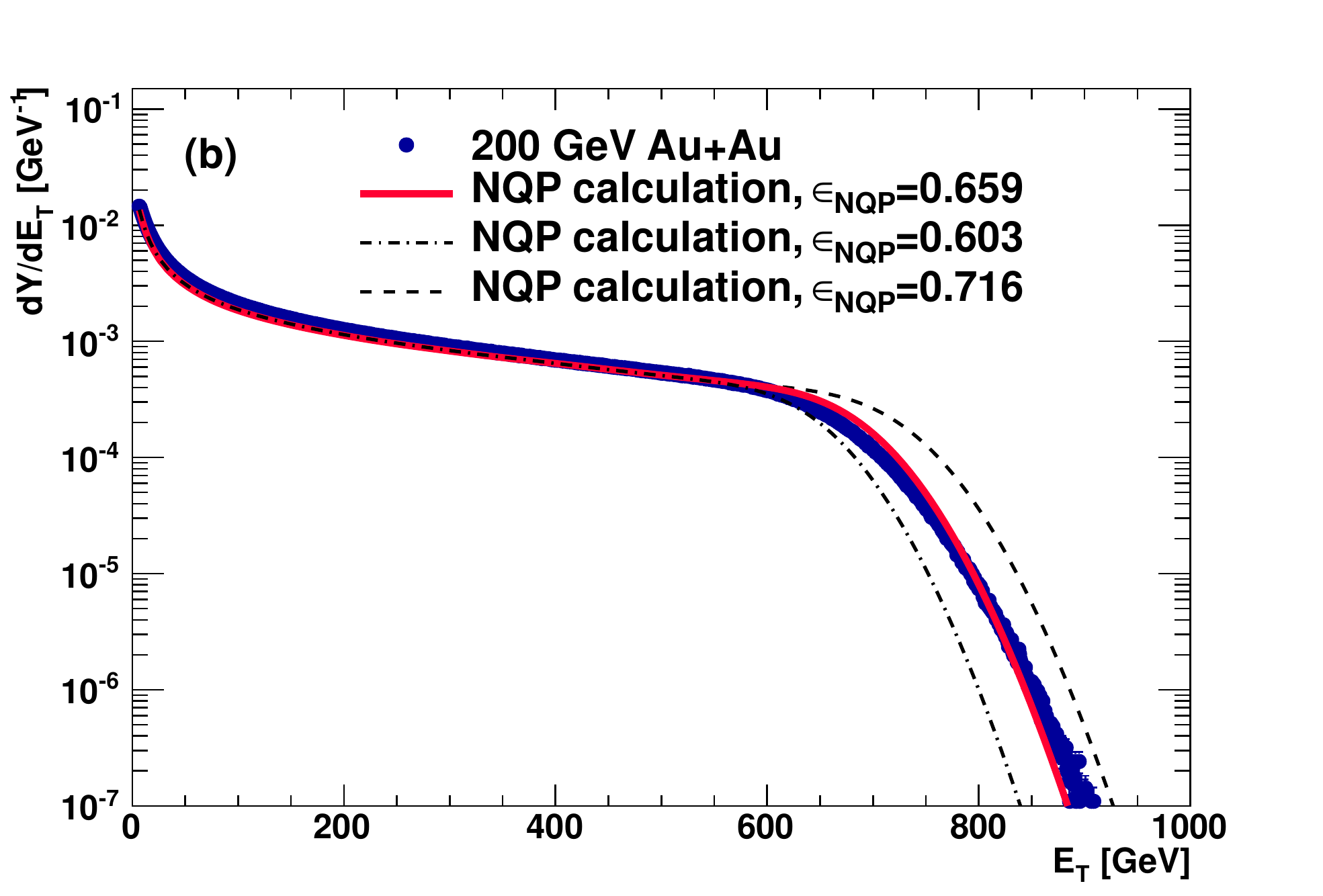}}
\normalsize \vspace*{-0.25pc}
     \caption[]{\footnotesize PHENIX NQP calculations~\cite{ppg100} based on the \Et distribution of a constituent-quark participant from Fig.~\ref{fig:pi0pT}b for: (a) $d$$+$Au (also AQM), (b) Au$+$Au \Et distributions at \sqsn=200 GeV. }
      \label{fig:NQPcalc}
   \end{figure}

\noindent Fig.~\ref{fig:pi0pT}b shows the deconvolution of the $p$$+$$p$ \Et distribution to the sum of 2--6 constituent-quark participants from which the \Et distribution of a constituent-quark is determined and applied to $d$$+$Au (Fig.~\ref{fig:NQPcalc}a) and Au$+$Au (Fig.~\ref{fig:NQPcalc}b) reactions in the same detector. 

   The NQP calculations closely follow the measured $d$$+$Au and Au$+$Au \Et distributions in shape and magnitude over a range of more than 1000 in cross section. A complete calculation was also done for the AQM which fails to describe the $d$$+$Au data (Fig.~\ref{fig:NQPcalc}a). The conclusion is that the number of constituent-quark participants determines the \Nch and \Et distributions and that the AQM calculation which describes the $\alpha$--$\alpha$ data at \sqsn=31 GeV (Fig.~\ref{fig:3plots}a) is equivalent to the NQP in the symmetric system. 

\begin{figure}[!h] \vspace*{-0.6pc}
      \centering
            \small 
a)\raisebox{0.0pc}{\includegraphics[width=0.49\linewidth]{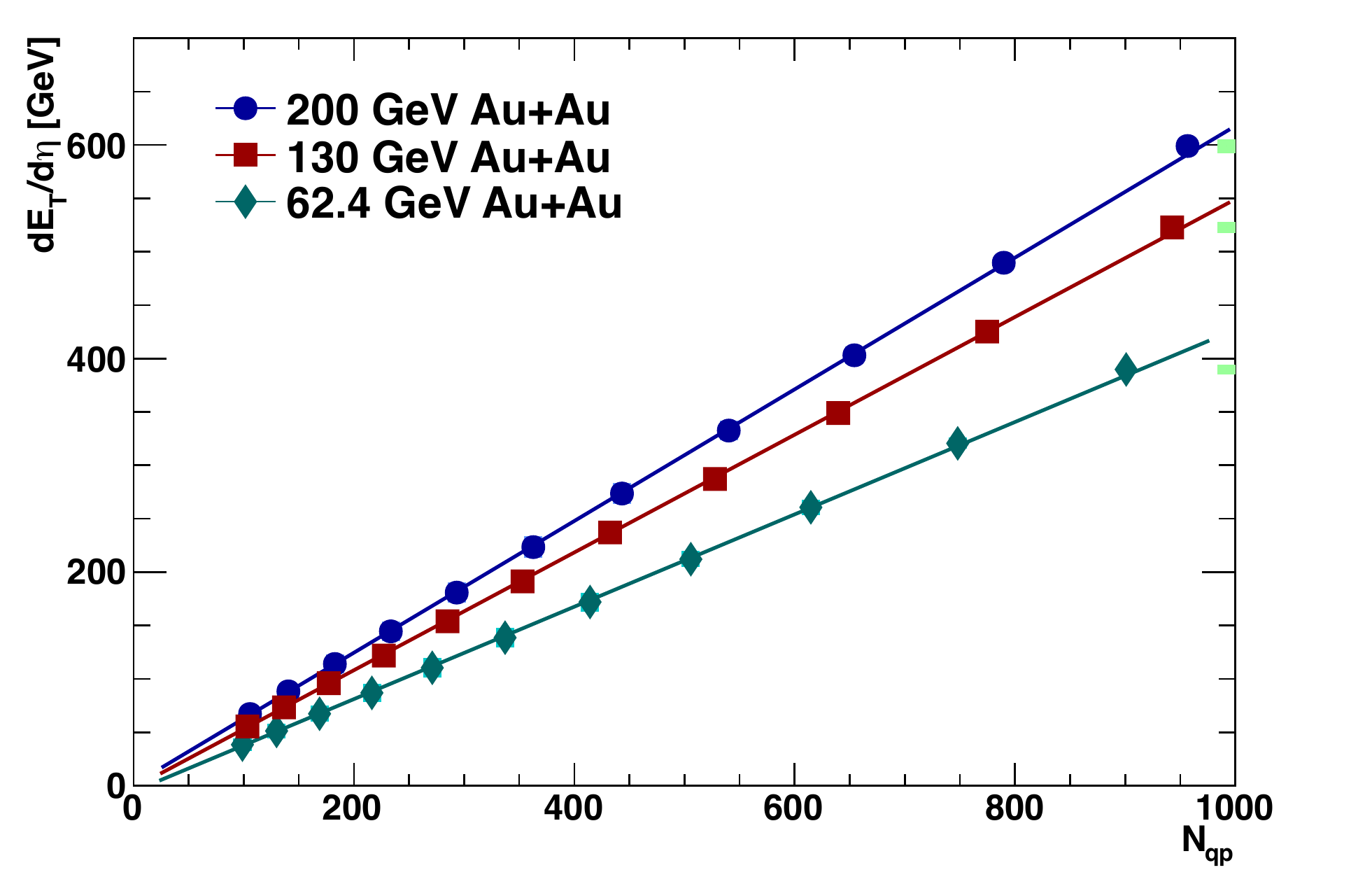}}
b)\raisebox{0pc}{\includegraphics[width=0.49\linewidth,angle=0]{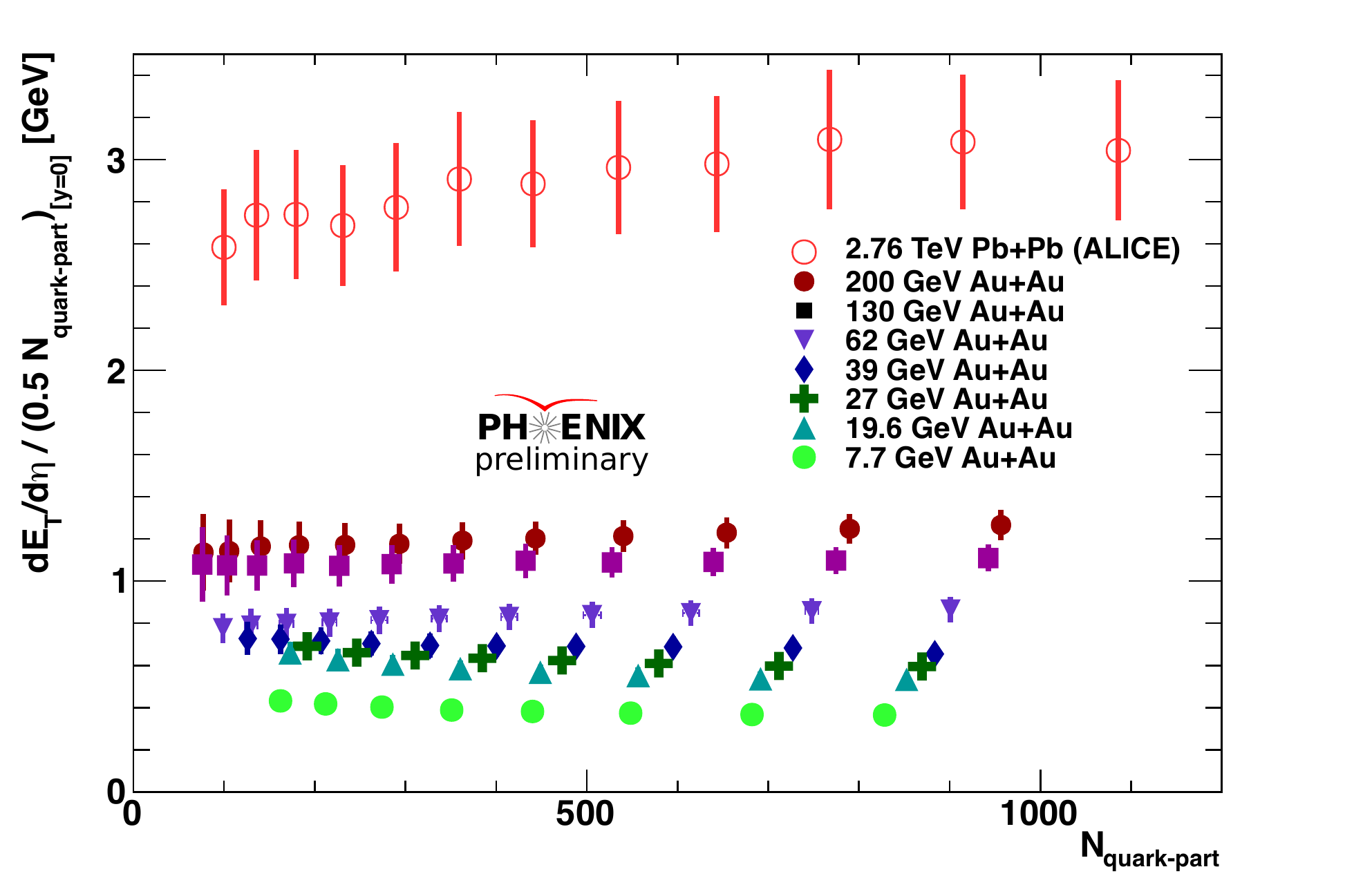}}\hspace*{-0.2pc}
\normalsize\vspace*{-0.6pc}
     \caption[]{\footnotesize PHENIX~\cite{ppg100}: (a) $d\Et/d\eta$ vs. $N_{qp}$ ; (b) $d\Et/d\eta/(\Nqp/2)$ vs. $N_{qp}$ }
      \label{fig:NQPcent}
   \end{figure}

The NQP model was also applied to the centrality-cut PHENIX data by making a plot of $d\Et/d\eta$ for a given centrality bin as a function of the number of constituent-quark participants \Nqp in the bin for Au$+$Au collisions at $\sqsn=62.4$, 130 and 200 GeV (Fig.~\ref{fig:NQPcent}a).
 
\vspace*{-0.2pc}
      
\noindent The data for each \sqsn are well described by a straight line and all are consistent with a zero intercept. This means that $d\Et/d\eta$ is strictly proportional to \Nqp (Fig~\ref{fig:NQPcent}a) so that $d\Et/d\eta/(\Nqp/2)$ vs. \Nqp is a constant for $\sqsn > 39$ GeV (Fig~\ref{fig:NQPcent}b) even up to the LHC $\sqsn=2.76$ TeV. This brought up a very interesting question, with a very important and fundamental answer.

Most experiments at RHIC, starting with PHENIX (Fig.~\ref{fig:3RHICplots})~\cite{Adcox:2000sp} had successfully fit their measurements of $d\Et/d\eta$ or $d\Nch/d\eta$ as a function of centrality (represented by \Npart) to the two-component model (Eq.~\ref{eq:ansatz}). Also, both the ATLAS~\cite{ATLASPLB107} and ALICE~\cite{ALICEPRC88} experiments at the LHC computed the ansatz, $[(1-x) \mean{\Npart}/2 + x\, \mean{\Ncoll}]$, in event-by-event Glauber Monte Carlo calculations which fit their forward \Et measurements used to define centrality in Pb+Pb collisions. ALICE realized that the combination of the two components \Npart and \Ncoll in the ansatz represented the number of emitting sources of particles, which they named ``ancestors''.  Since the number of constituent-quarks \Nqp also represents the number of emitting sources in a simple linear relationship, Bill Zajc of PHENIX suggested that ``the success of the two component model is not because there are some contributions proportional to \Npart and some going as \Ncoll, but because a particular linear combination of \Npart and \Ncoll turns out to be an empirical proxy for the number of contsituent-quarks.'' 
We had a nice table of $\mean{\Npart}$, $\mean{\Nqp}$, $\mean{\Ncoll}$  as a function of centrality in Au$+$Au collisions at \sqsn=200 GeV, so it did not take very long for me to verify the striking result that indeed it was true: with $x=0.08$, the ratio $\mean{\Nqp}/[(1-x) \mean{\Npart}/2 + x\, \mean{\Ncoll}]=3.88$ on the average  and varies by less than  1\% over the entire centrality range in 5\% bins except for the most peripheral bin where it is 5\% low (Table~\ref{table:ansatzratio}).  This result clearly demonstrates that the ansatz works because the particular linear combination of \Npart and \Ncoll turns out to be an empirical proxy for \Nqp and not because the \Ncoll term implies hard-scattering.
\begin{table}[!b]
\centering
\includegraphics[width=0.71\linewidth]{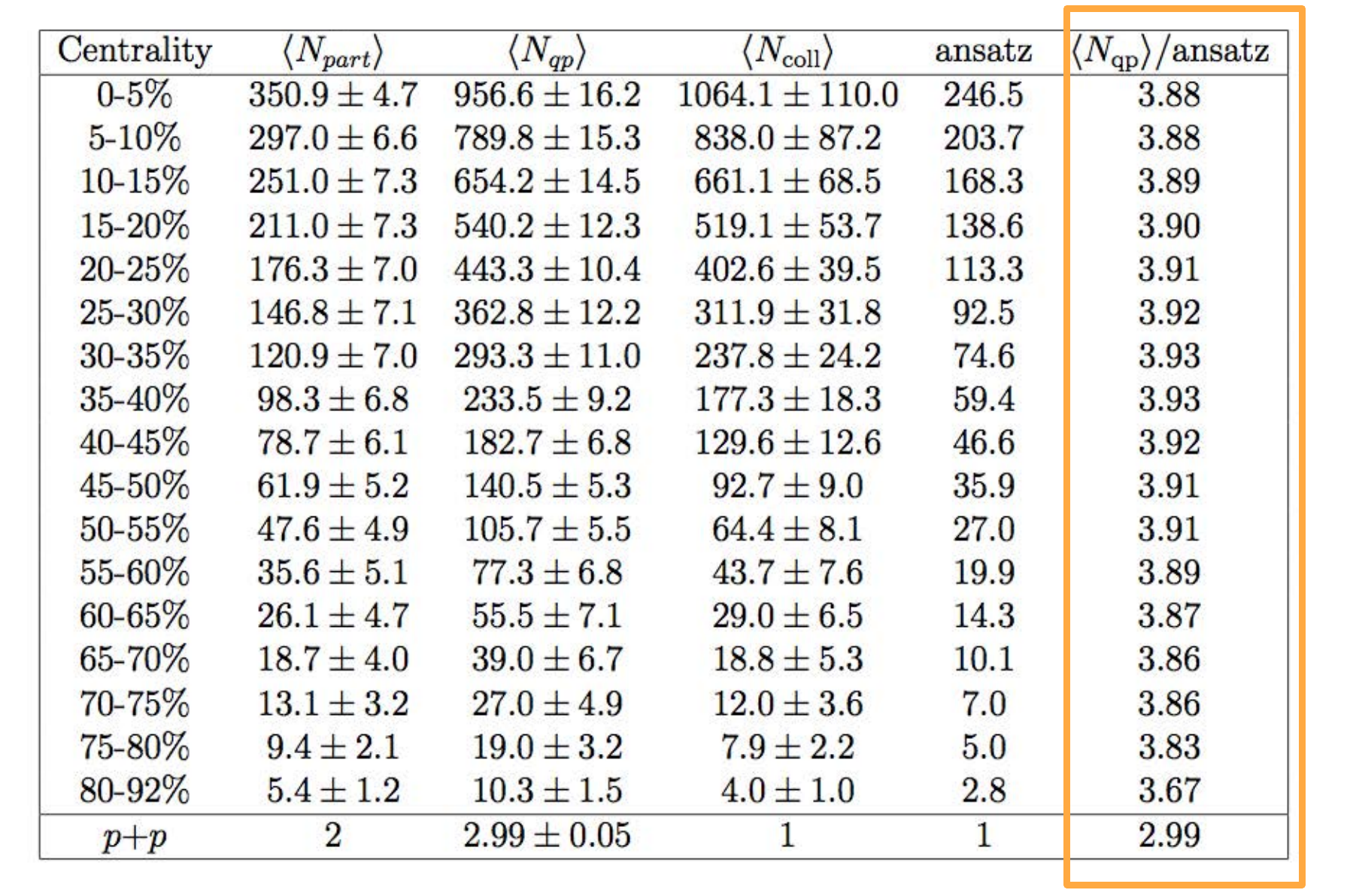}
\caption[]{\footnotesize Verification that the ansatz, $[(1-x)\,\mean{\Npart}/2 +x\,\mean{\Ncoll} ]$, from Eq.~\ref{eq:ansatz}, with $x=0.08$, is a proxy for \Nqp. The errors quoted on $\mean{\Npart}$, $\mean{\Nqp}$,  $\mean{\Ncoll}$ are correlated and largely cancel in the $\mean{\Nqp}$/ansatz ratio. For $x=0.09$ the average $\mean{\Nqp}/{\rm ansatz}$=3.81, the maximum variation is less than 1.6\%, but 4\% low in the most peripheral bin. }
\label{table:ansatzratio}
\end{table}

  The fact that the $\mean{\Nqp}$/ansatz ratio drops from an average of 3.88 for Au$+$Au collisions to 2.99 for $p$$+$$p$ collisions is also interesting. This is consistent with the PHOBOS~\cite{PHOBOSPRC70} result that a fit of Eq.~\ref{eq:ansatz} to
$\mean{d\Nch^{AA}/d\eta}$ with $x=0.09$ leaving $\mean{d\Nch^{pp}/d\eta}$ as a free parameter gives the result $\mean{d\Nch^{pp}/d\eta}=2.70$ which is above the measured inelastic value of 2.29. The lower value of $\mean{\Nqp}$/ansatz for $p$$+$$p$ would then give a value of $2.70\times2.99/3.88=2.08$ (2.12 for $x=0.09$) for $\mean{d\Nch^{pp}/d\eta}$,  much closer to the measured  value. In that same paper, PHOBOS also noted that their data were consistent with a constant value of $x$ from \sqsn=19.6 to 200 GeV (more recently extended to \sqsn=2.76 TeV~\cite{GSFSPRC90}) which indicated that the fraction of hard-processes contributing to multiplicity did not increase over a huge range of \sqs even though the hard-scattering cross section greatly increased over this same range. 
\subsection{Constituent-quark participants resolve several outstanding puzzles}
\begin{figure}[!th]
       \begin{center}
\includegraphics[width=0.40\textwidth]{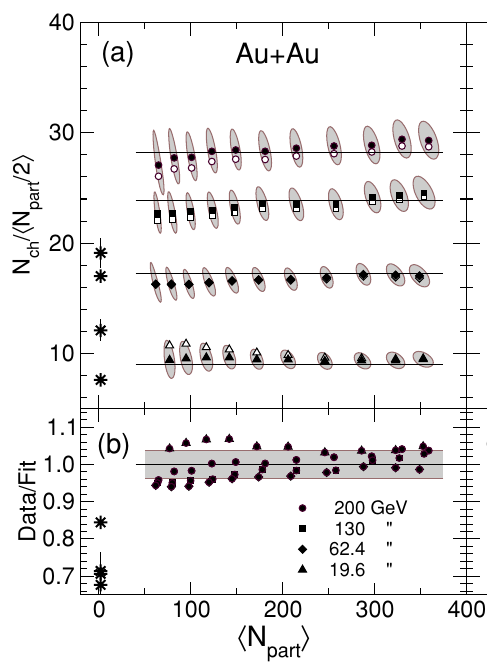}\hspace{1.0pc}
b)\includegraphics[width=0.46\textwidth]{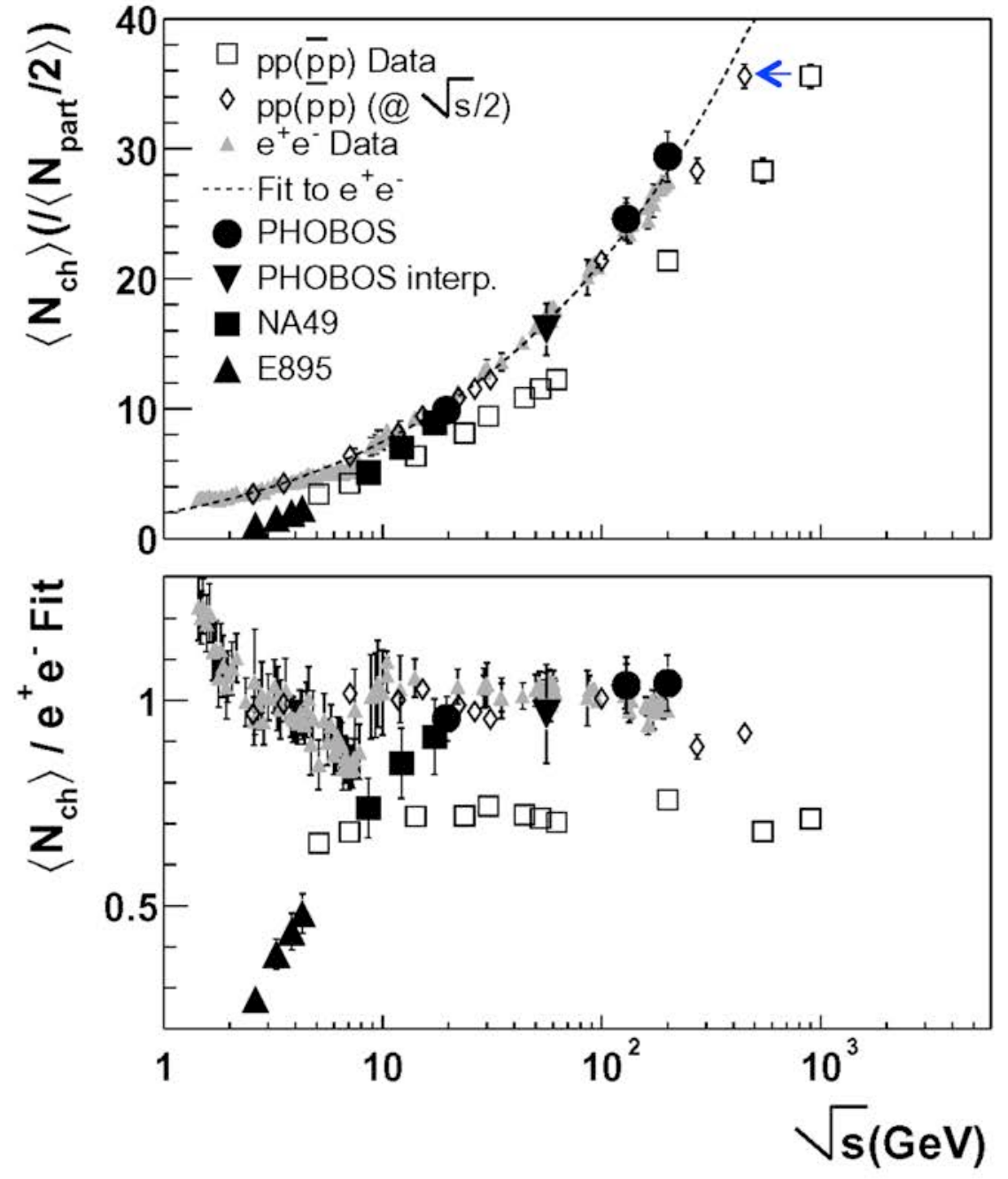}
\end{center}\vspace*{-1.25pc}
\caption[]{\footnotesize a) Total charged multiplicity per nucleon pair $\Nch/\mean{\Npart/2}$ vs. centrality, \Npart, for the \sqsn indicated.  Open points are the measured $\Nch|_{|\eta|<5.4}/(\mean{\Npart/2}$; solid points are extrapolated to $|\eta|\leq y_{\rm beam}$. b)~Total charged multiplicity per nucleon pair in $p$$+$$p$ and A+A collisions as a function of c.m. energy $\sqrt{s}$ compared to $e^+ + e^-$ collisions~\cite{PHOBOSPRC74}.
\label{fig:PHOBOSWNM?}}
\end{figure}

PHOBOS also made some very nice measurements of the charged particle multiplicity over the full rapidity range, not just mid-rapidity. The total charged multiplicity $\Nch|_{|\eta|<5.4}$ was measured and $\Nch|_{|\eta|<5.4}/\mean{\Npart/2}$ plotted as a function of centrality for \sqsn=19.6, 62.4,130, and 200 GeV (Fig.~\ref{fig:PHOBOSWNM?}a)~\cite{PHOBOSPRC83}. At first glance the data appear to follow the WNM because the multiplicity/per nucleon pair $\Nch|_{|\eta|<5.4}/\mean{\Npart/2}$ appears to be constant in Au$+$Au collisions. However, the true believers, e.g. Ref.~\cite{Bialkowska2006}, claim that the WNM does not work because the value in Au$+$Au collisions is much larger than the $p$$+$$p$ value shown; but ``still the proportionality of these multiplicities to the number of participants holds''~\cite{Bialkowska2006}. 

In fact, the difference between the $p$$+$$p$ and Au$+$Au values may be related to another interesting observation by PHOBOS~\cite{PHOBOSPRC74} that the ``leading particle effect'' in $p$$+$$p$ collisions, as discovered by Zichichi and collaborators~\cite{BasilePLB95}---in which the total multiplicity at c.m. energy $\sqrt{s_{\rm pp}}$ is equal to that in $e^+ e^-$ collisions at $\sqrt{s_{\rm ee}}=\sqrt{s_{\rm pp}}/2$ (the ``effective energy'') (Fig.~\ref{fig:PHOBOSWNM?}b) because the leading protons carry away half the $p$$+$$p$ c.m. energy---is absent in A+A collisions where the leading protons can reinteract. This observation seems to contradict the WNM, in which the key assumption is that what counts is whether or not a nucleon was struck, not how many times it was struck.

Both these effects can be reconciled by constituent-quark participants.

In the NQP model (Table~\ref{table:ansatzratio}), the $\mean{\Nqp/\Npart}$ is 1.5 for a $p$$+$$p$ collision but rises to 2.27--2.73 (a factor of 1.51--1.82) for the more central (0-50\%, $\mean{\Npart}>60$) Au$+$Au collisions plotted in Fig.~\ref{fig:PHOBOSWNM?}a. This would correspond to an increase in $\Nch/\mean{\Npart/2}$ by a factor of $\sim 1.5$ as observed. It also might explain the slight rise of the open points with increasing \Npart. Similarly, the increase in ``effective energy'' for particle production shown by the increase in $\Nch/\mean{\Npart/2}$ from $p$$+$$p$ to Au+Au collisions is due to an increase in the number of (constituent-quark) participants, not because of additional collisions of a given nucleon-participant. Furthermore, the factor 1.5 decrease in $\Nch/\mean{\Npart/2}$ from Au$+$Au to $p$$+$$p$ corresponds to a reduction in \sqs for the observed $\Nch/\mean{\Npart/2}$ from 200 to 100 GeV $p$$+$$p$ collisions on Fig.~\ref{fig:PHOBOSWNM?}b, the same factor of 2 discussed in the original measurement~\cite{PHOBOSPRC74}. Thus, the NQP model rather than the WNM preserves the assumption in these ``extreme-independent'' participant models that successive collisions of a participant do not increase its particle emission while explaining these two interesting observations.   

Another argument against the \Npart, \Ncoll ansatz representing actual hard-collisions rather than simply being a proxy for constituent-quark participants concerns the measurement of elliptic flow in central U+U collisions~\cite{STARv2UU}.

	\section{Collective Flow} 
   For many years, since the days of the Bevalac~\cite{HansAkePRL52}, collective flow~\cite{Ollitrault} has been observed in A$+$A collisions over the full range of energies studied,  from incident kinetic energy of $100A$ MeV to c.m. energy of $\sqsn=2.76$ TeV,  and thought to be a distinguishing feature of A$+$A collisions compared to either $p$$+$$p$ or $p$$+$A collisions.  
Collective flow, or simply flow, is a collective effect which can not be obtained from a superposition of independent N$+$N collisions. I first present a short review (details can be found in previous ISSP proceedings~\cite{MJTIJMPA2011,MJTIJMPA2014}) and then move on to the newer results.

\begin{figure}[!t]
   \begin{center}
a)\raisebox{1.0pc}{\includegraphics[width=0.45\linewidth]{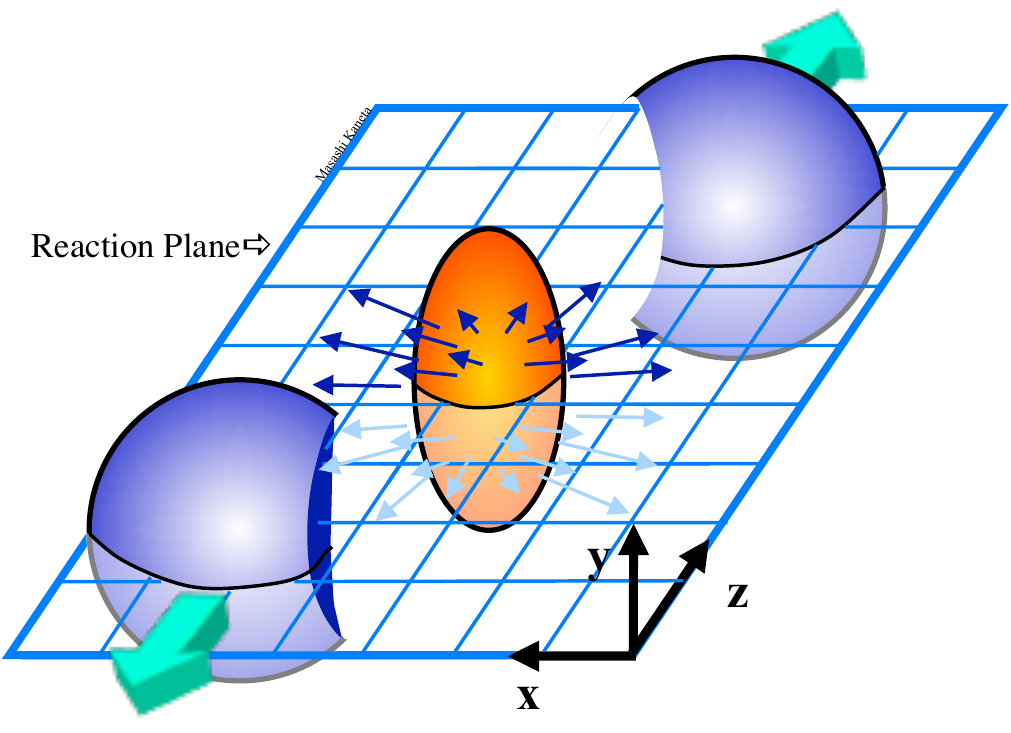}}\hspace*{1.0pc}
b)\raisebox{0.0pc}{\includegraphics[width=0.45\linewidth]{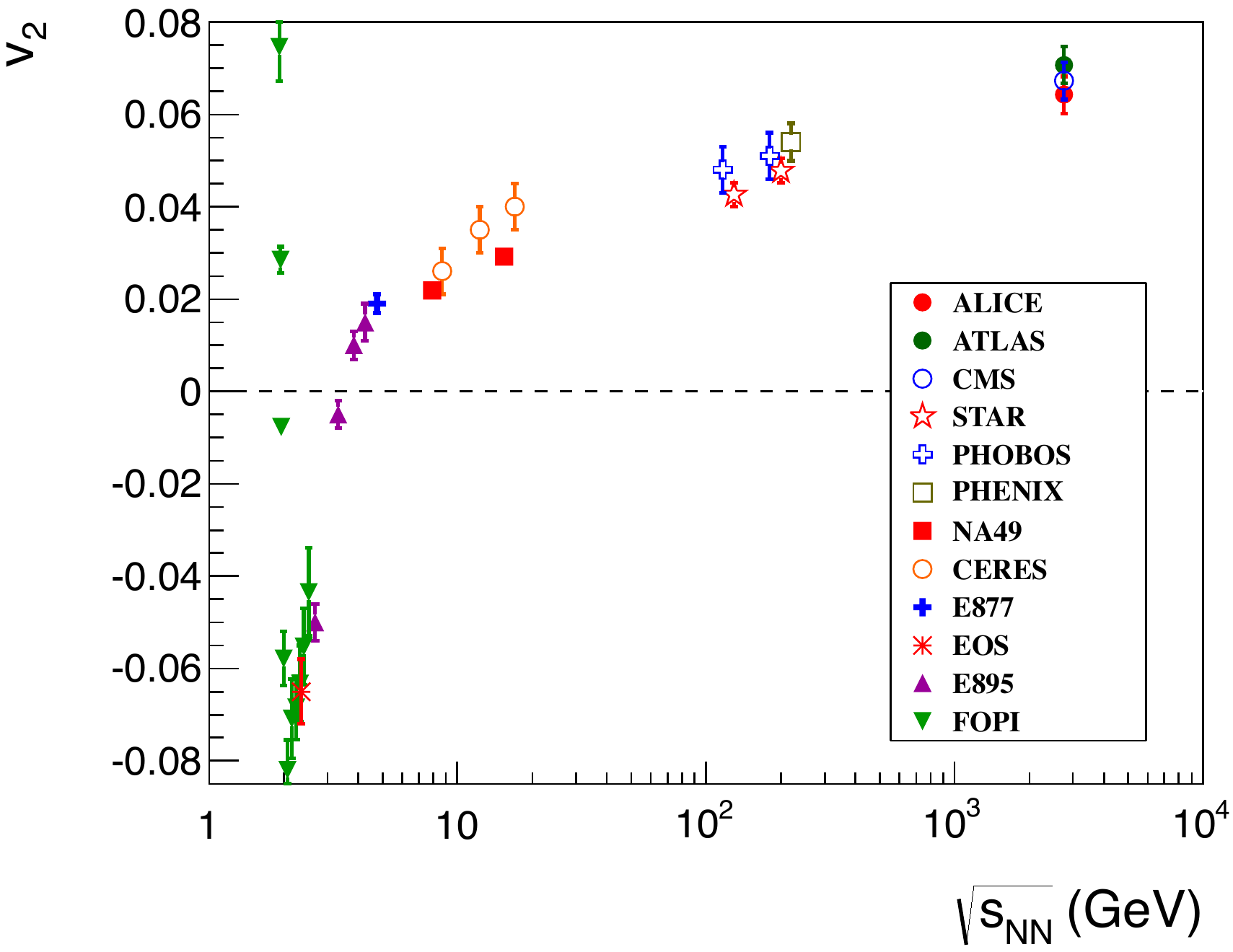}}
\end{center}\vspace*{-1.0pc}
\caption[]{\footnotesize (a) Almond shaped overlap zone generated just after an A$+$A collision where the incident nuclei are moving along the $\pm z$ axis. The reaction plane by definition contains the impact parameter vector (along the $x$ axis)~\cite{KanetaQM04}. (b) $v_2$ for charged particles integrated over \pt at \sqsn=2.76 TeV for 20--30\% centrality compared to the measurements at lower \sqsn at the same centrality~\cite{HeinzSnellingsARNPS}.  
\label{fig:MasashiFlow}}\vspace*{-1.0pc}
\end{figure}

Immediately after an A+A collision, the overlap region defined by the nuclear geometry is almond shaped (see Fig~\ref{fig:MasashiFlow}a) with the shortest axis along the impact parameter vector. The different pressure gradients along the short and long axes of the ellipse break the $\phi$ symmetry of the problem and create an azimuthal angular dependence of the semi-inclusive single particle spectrum with respect to the reaction plane, $\phi-\Phi_R$, which is represented  by an expansion in harmonics~\cite{VoloshinPoskanzerSnellings}, where the angle of the reaction plane $\Phi_R$ is defined to be along the impact parameter vector, the $x$ axis in Fig.~\ref{fig:MasashiFlow}a: 
  \begin{equation}
\frac{Ed^3 N}{dp^3}=\frac{d^3 N}{p_T dp_T dy d\phi}
=\frac{d^3 N}{2\pi\, p_T dp_T dy} \left[ 1+\sum_n 2 v_n \cos n(\phi-\Phi_R)\right] .
\label{eq:siginv2} 
\end{equation} 
The Fourier coefficient $v_2$, called elliptic flow, is predominant at mid-rapidity. The evolution of $v_2$ with  \sqsn (Fig.~\ref{fig:MasashiFlow}b)~\cite{HeinzSnellingsARNPS} is the result of competing processes. At very low \sqsn corresponding to values of $\sim 100A$ MeV~\cite{FOPINPA679} the main effect among many others is from nuclei bouncing off each other and breaking to fragments, which is sensitive to the equation of state of the nuclei---soft, like sponges, hard like billiard balls? The negative $v_2$ at larger \sqsn is produced by the effective ``squeeze-out'' (in the $y$ direction) of the produced particles by slow moving minimally Lorentz-contracted spectators (as in Fig.~\ref{fig:MasashiFlow}a) which block the particles emitted in the reaction plane. With increasing \sqsn, the spectators move faster and become  more contracted so the blocking stops. The increase of $v_2$ with \sqsn is generally described by hydrodynamics in the \QGP\ region, but is also described by hadron transport theories for $\sqsn\lsim 10$ GeV~\cite{LokeshFriendsPRC82}. 

Flow measurements contributed two of the most important results about the properties of the \QGP: i) the scaling of $v_2$ of identified particles at mid-rapidity with the number of constituent-quarks $n_q$ in the particle---$v_2/n_q$ scales with the transverse kinetic energy per constituent-quark, $KE_T/n_q$, because particles have not formed at the time flow develops; ii) the  persistence of flow for $p_T>1$ GeV/c which implied that the viscosity is small~\cite{TeaneyPRC68}, perhaps as small as a quantum viscosity bound from string theory~\cite{Kovtun05}, $\eta/s=1/(4\pi)$ where $\eta$ is the shear viscosity and $s$ the entropy density per unit volume.  This led to the description of the ``s\QGP'' produced at RHIC as ``the perfect fluid''. 

New insight came in 2013, when measurements in $p$$+$Pb at LHC and $d$$+$Au at RHIC  observed what looked very much like collective flow in these systems that were believed to be too small to support collective effects. This was the reason for the He$^3+$Au run at RHIC in 2014, to see whether triangular flow, $v_3$, would be more prominent with a 3 nucleon projectile. 
The improvement of the $p$$+$Au and $d$$+$Au measurements this year to identified pions and protons strengthened the case that the observed $v_2$ in these small systems is really hydrodynamic collective flow. 

\begin{figure}[!thb]
   \begin{center}
\raisebox{0.4pc}{\includegraphics[width=0.38\linewidth]{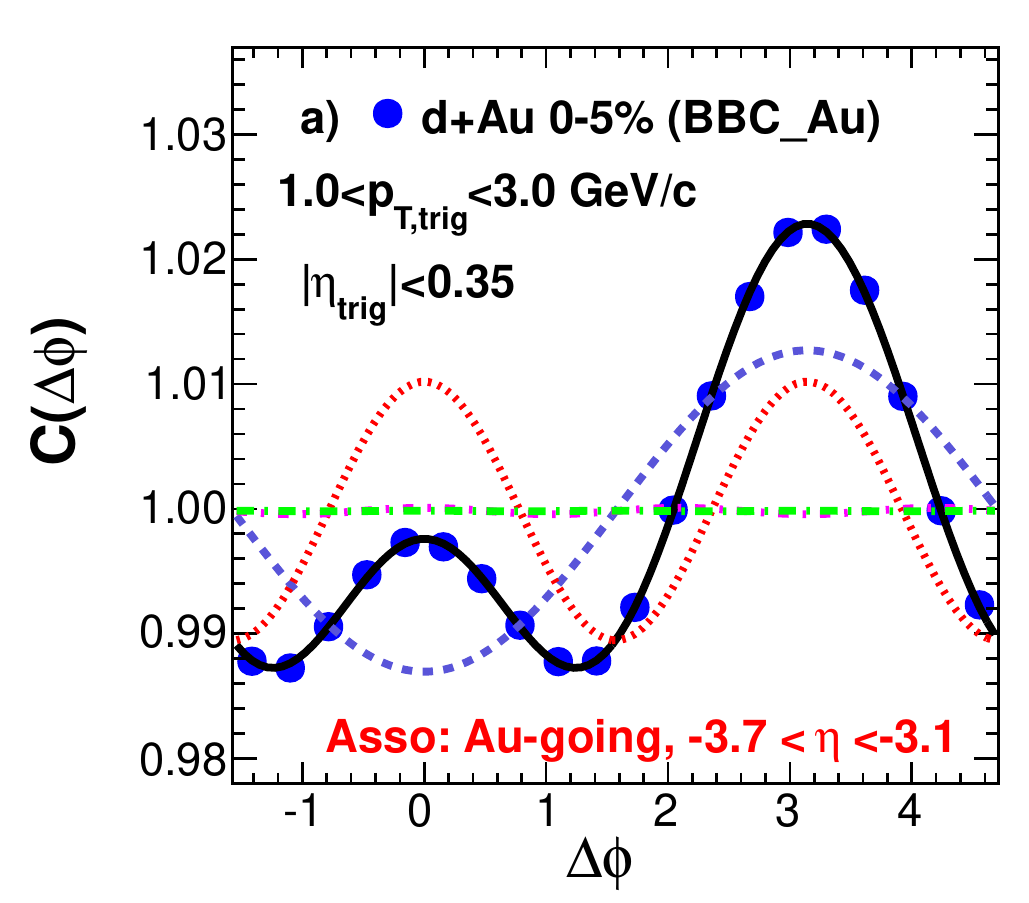}}\hspace*{1.0pc}
\raisebox{0.0pc}{\includegraphics[width=0.58\linewidth]{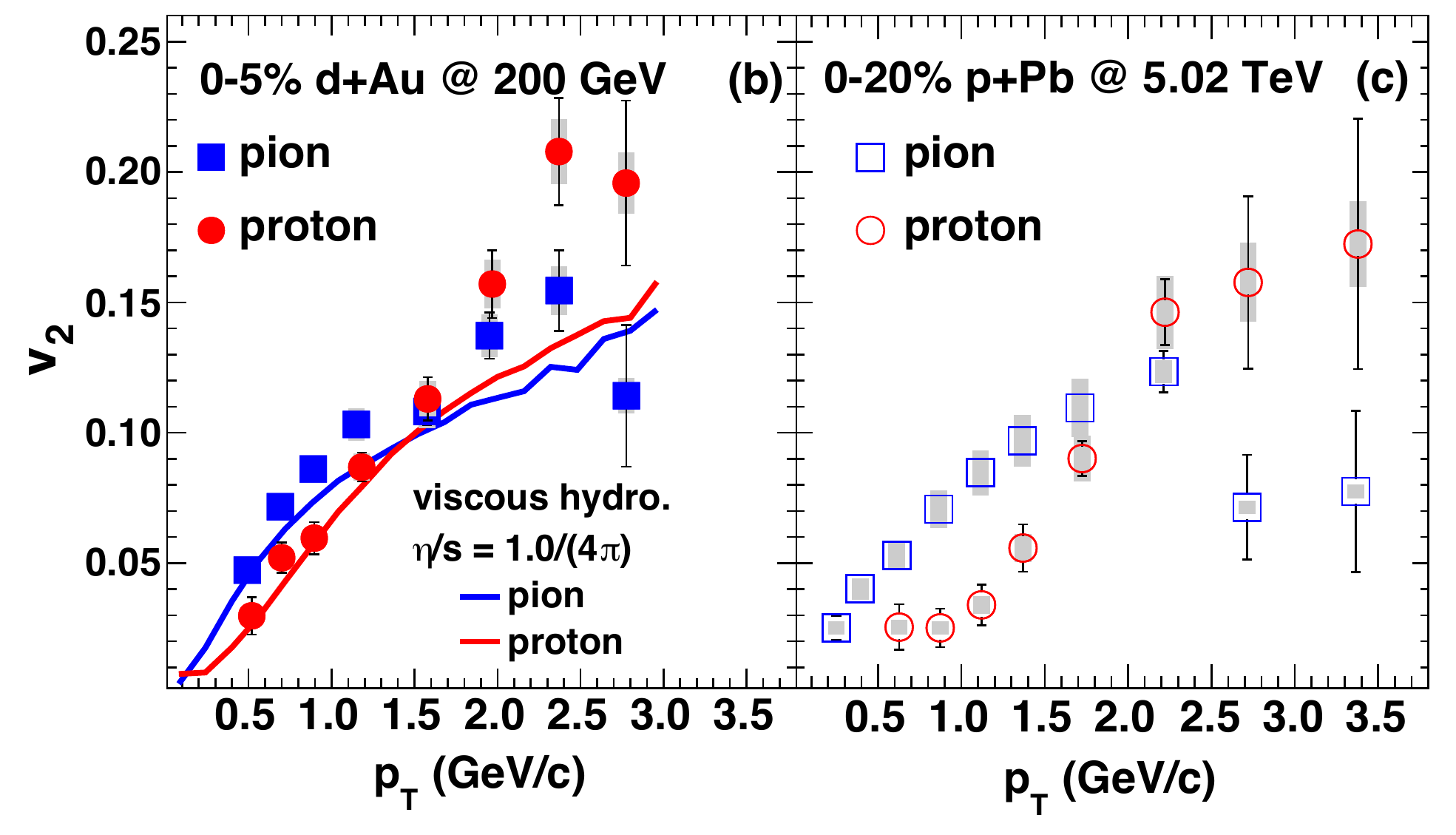}}
\end{center}\vspace*{-2.0pc}
\caption[]{\footnotesize (a) PHENIX~\cite{PXppg161} two-particle azimuthal correlation function (Eq.~\ref{eq:2partdphi}) for $1.0<p_{\rm T,trig}<3.0$ GeV/c, $|\eta_{\rm trig}|<0.35$, in central (0-5\%) $d$$+$Au collisions at RHIC.  $v_2 (\pt)$ for identified $\pi^{\pm}$ $p^{\pm}$ by the standard reaction plane method (Eq.~\ref{eq:siginv2}) for (b) central $d$$+$Au collisions at RHIC (\sqsn=200 GeV) and (c) central $p$$+$Pb collisions at LHC (\sqsn=5.02 TeV).
\label{fig:dAuv2pid}}\vspace*{-1.0pc}
\end{figure}

Figure~\ref{fig:dAuv2pid}a~\cite{PXppg161} shows the two-particle correlation function in $d$$+$Au (Eq.~\ref{eq:2partdphi}) fit with terms from $c_1$ to $c_4$ where the solid line is the fit and only the $c_1$ (dashes) and $c_2$ (dots) make significant contribution. 
\begin{equation} \frac{dN_{ta}}{d\phi_t d\phi_a}=C(\phi_t-\phi_a)\propto \left[1+\sum_n 2\ c_n  \cos n(\phi_t-\phi_a) \right] , \qquad c_n\equiv (v_n)_t (v_n)_a
\label{eq:2partdphi}\end{equation}
The $c_1$ comes into play because the trigger particle $t$ is a charged track at mid-rapidity while the associated particle $a$ is a count in an MPC tower ($\delta \eta \times \delta \phi\approx 0.12 \times 0.18$) from $\pi^0$ or $\eta$ meson decay photons at $-3.7\leq\eta_{\rm tower}\leq -3.1$. Also, there is no evidence of a di-jet contribution because the large pseudorapidity gap between $t$ and $a$ is beyond that of a di-jet. Thus, the long-range correlation in Fig.~\ref{fig:dAuv2pid}a which is not seen in $p$$+$$p$ comparison data but has the same properties as collective flow in Au$+$Au collisions is consistent with hydrodynamic collective flow in $d$$+$Au. Perhaps more convincing evidence for hydrodynamic flow is given in Fig.~\ref{fig:dAuv2pid}b,c where both at RHIC in $d$$+$Au (b) and LHC in $p$$+$Pb (c), the characteristic  $\pi$, $p$ mass splitting for $v_2(\pt)$ seen in Au$+$Au is observed~\cite{PXppg161}. The splitting occurs because, for a given transverse collective expansion velocity $\beta$, protons have a larger $p_T=\gamma\beta m$ than pions. \vspace*{-1.0pc}
\subsection{$\mathbf{v_2}$ in U$+$U collisions and constituent-quark participants} 
\begin{figure}[!thb]
   \begin{center}
a)\raisebox{0.4pc}{\includegraphics[width=0.38\linewidth]{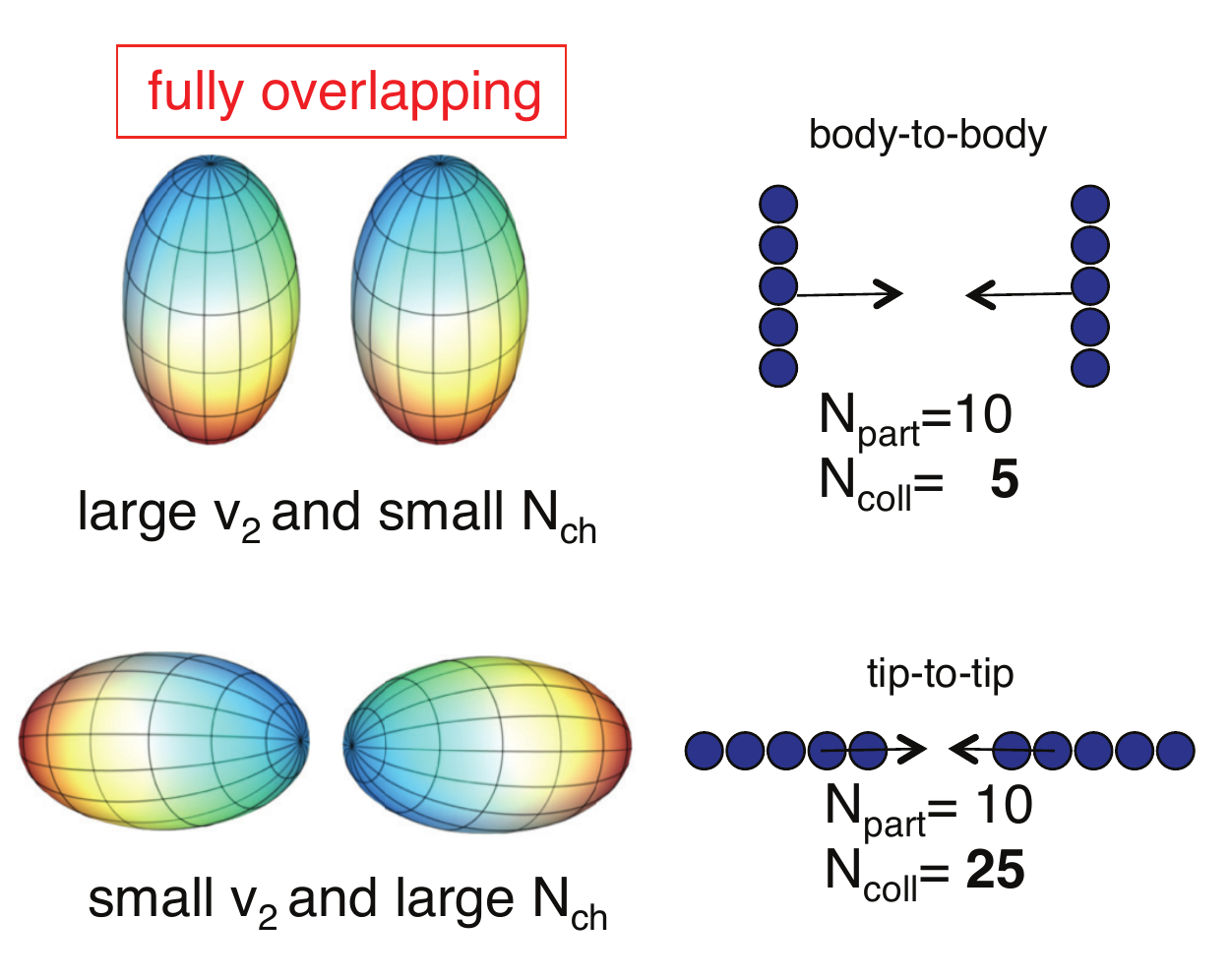}}\hspace*{1.0pc}
b)\raisebox{0.0pc}{\includegraphics[width=0.54\linewidth]{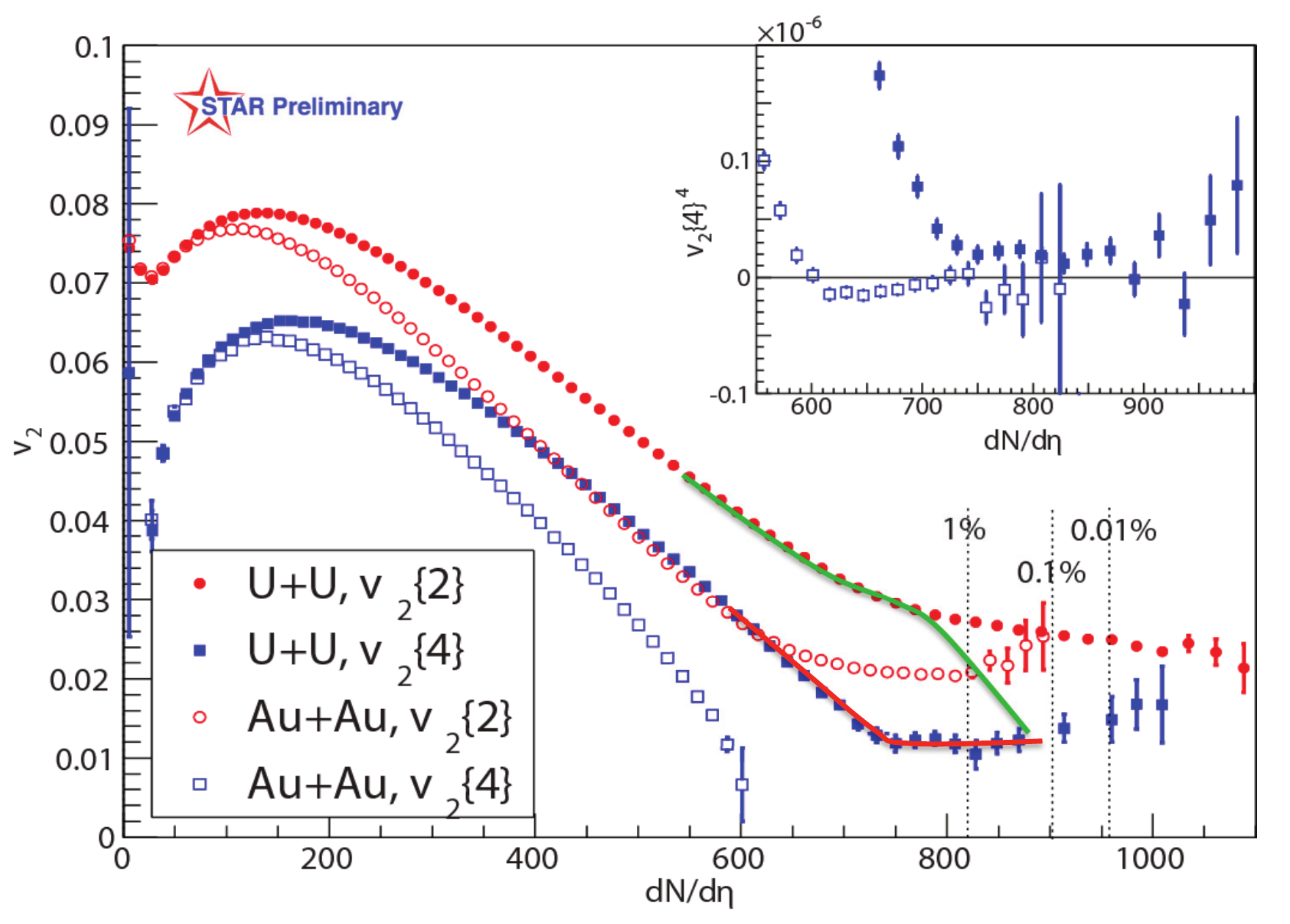}}
\end{center}\vspace*{-1.0pc}
\caption[]{\footnotesize (a) Body-to-body and tip-to-tip configurations in U+U collisions with zero impact parameter. The different relation of \Npart to \Ncoll is sketched next to each configuration. (Modified drawing from Ref.~\cite{STARv2UU}).
(b) STAR measurements of $v_2$ in Au$+$Au and U$+$U at $\sqsn\approx$ 200 GeV as a function of $d\Nch/d\eta$ with upper percentiles of centrality for U$+$U indicated by vertical dashed lines~\cite{STARv2UU}.
\label{fig:UUv2}}
\end{figure}

Because Uranium nuclei are prolate spheroids, there is the interesting possibility of large $v_2$ in body-to-body central collisions which have a significant eccentricity and almond shape (Fig.~\ref{fig:UUv2}a).  
Based on the assumption that the \Npart, \Ncoll ansatz (Eq.~\ref{eq:ansatz}) would describe the $d\Nch/d\eta$ distribution in U$+$U collisions, it was predicted that for the highest $d\Nch/d\eta$ (the most central collisions) the tip-to-tip configuration with much larger \Ncoll and small eccentricity (small $v_2$) would overtake the body-to-body configuration with large eccentricity corresponding to large $v_2$.  

This led to two predictions: i) the tip-to-tip configuration would be selected by the most central collisions~\cite{KuhlmanHeinzPRC72}; ii) these most central collisons would see a sharp decrease in $v_2$ with increasing $d\Nch/d\eta$~\cite{Filip--NuXuPRC80,VoloshinPRL105} called a cusp. This sharp decrease---represented by the bent line on the topmost U$+$U data (filled circles) in Fig.~\ref{fig:UUv2}b (called a knee in Ref.~\cite{STARv2UU})---is not observed. As discussed previously, this is because the \Ncoll term is not relevant for $d\Nch/d\eta$ distributions, which also argues against the method proposed in Ref.~\cite{KuhlmanHeinzPRC72} to select the tip-to-tip configuration. 

\section{RHIC Beam Energy Scan (BES)---in search of the critical point}
\begin{figure}[!bht]
\begin{center}
a) \includegraphics[width=0.40\textwidth]{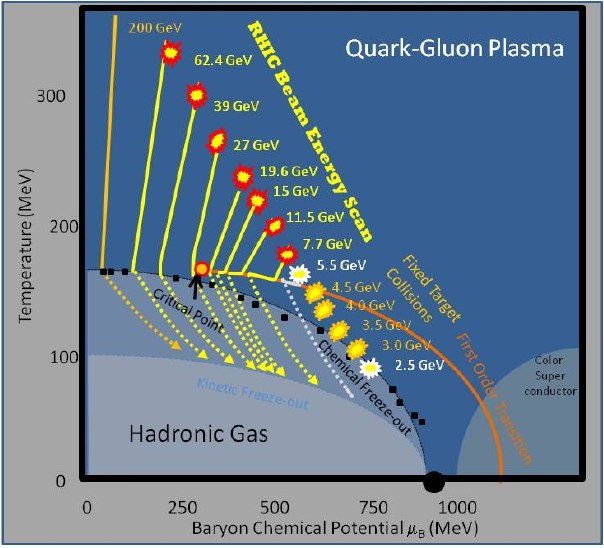}\hspace*{1.0pc}
b) \includegraphics[width=0.37\textwidth]{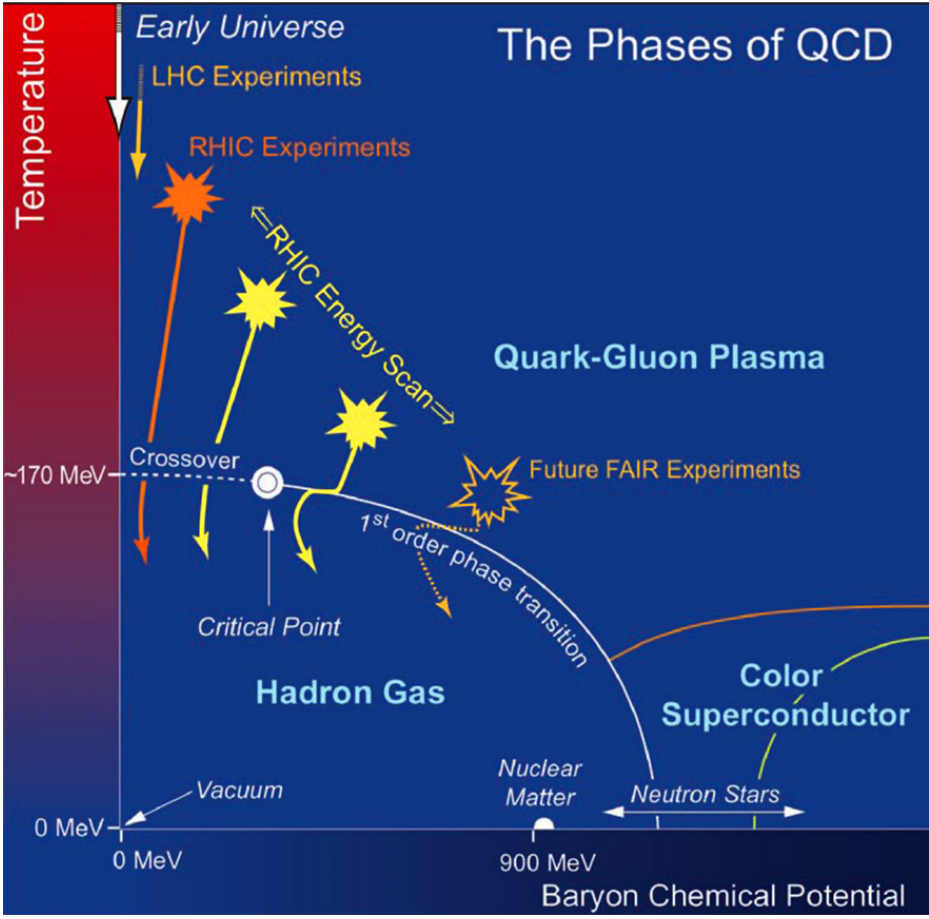}
\end{center}\vspace*{-1.0pc}
\caption[]{\footnotesize Proposed phase diagrams for nuclear matter:  Temperature,  $T$,  vs Baryon Chemical Potential, $\mu_B$. a)STAR's idea in 2013~\cite{STARBURfor2014-15}; b) STAR's more cautious idea in 2014~\cite{STARWPBES2014}. \label{fig:phase_boundary}}
\end{figure}

   In addition to discovering the \QGP\ and measuring its properties, another objective of the RHIC physics program is to measure the phase diagram of nuclear matter and to determine the equation of state in the various phases and the characteristics of the phase transitions. Two of the many proposed phase diagrams of nuclear matter (e.g. see Ref.~\cite{PawlowskiQM2014}) are shown in Fig.~\ref{fig:phase_boundary} together with the idealized trajectories of the 
evolution of the medium for Au+Au collisions at the \sqsn proposed for the Beam Energy Scan at RHIC to search for a \QCD\ critical point. The bursts represent the hottest and densest stage of the medium when thermal equilibrium is  reached shortly after the collision. The axes are the temperature $T$ vs. the baryon chemical potential $\mu_B$.      
The temperature for the transition from the Quark Gluon Plasma (\QGP) to a hadron gas is taken as 170 
MeV for $\mu_B=0$ and the phase boundary is predicted to be a smooth crossover down to a critical point below which the phase boundary becomes a first order phase transition.  

\begin{figure}[!bht]
\begin{center}
{\footnotesize a)} \includegraphics[width=0.45\textwidth]{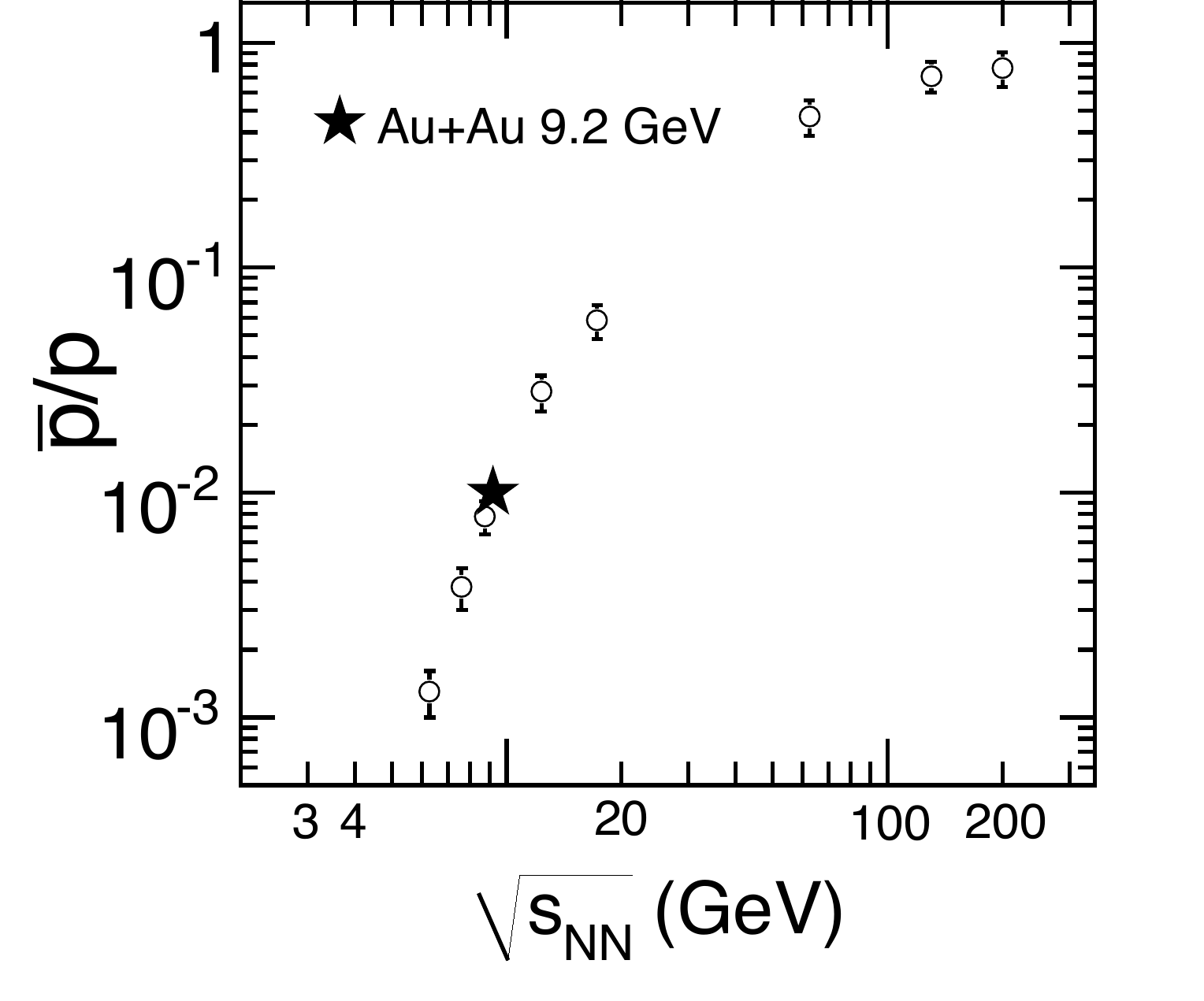}\hspace*{1.0pc}
{\footnotesize b)} \includegraphics[width=0.35\textwidth]{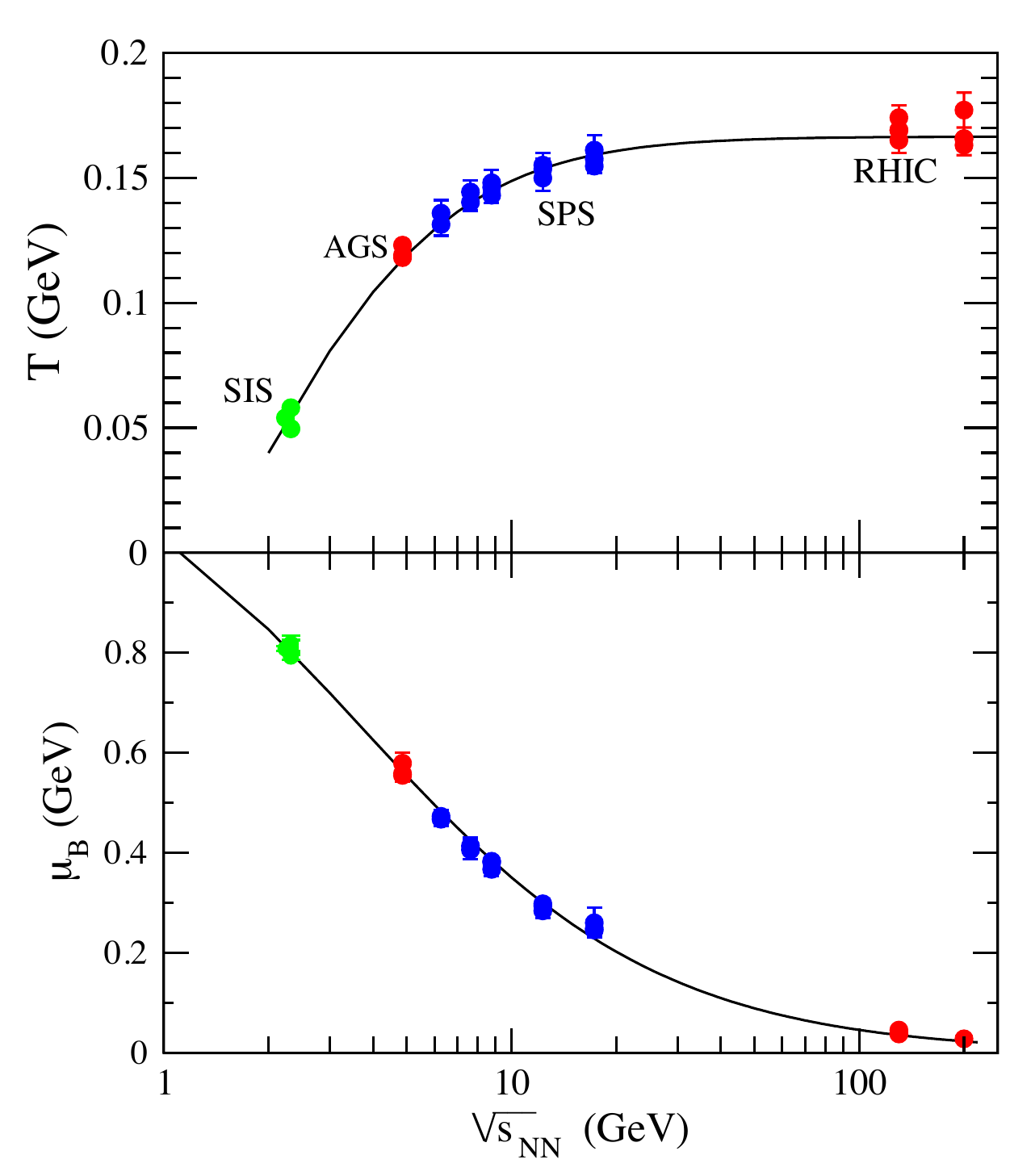}
\end{center}\vspace*{-1.0pc}
\caption[]{\footnotesize (a) STAR measurements of $\bar{p}/p$ vs. \sqsn~\cite{STARpidPRC81}; b) Best accepted analysis of $T$ and $\mu_B$ vs \sqsn~\cite{CleymansOeschler}. \label{fig:pbar-p}}
\end{figure}

In an equilibrated thermal medium, particles should follow a Boltzmann distribution in the local rest frame~\cite{CooperFrye}
\begin{equation}
{{d^2\sigma} \over {dp_L p_T dp_T}}={{d^2\sigma} \over {dp_L m_T dm_T}} \propto {1\over {e^{(E-\mu)/T} \pm 1}}\sim e^{-(E-\mu)/T} \qquad ,
\label{eq:boltz}
\end{equation}
where $m_T=\sqrt{p_T^2+m^2}$ and $\mu$ is a chemical potential. In fact,   the ratios of particle abundances (which are dominated by low $p_T$ particles) for central Au+Au collisions at RHIC, even for strange and multi-strange particles, 
are well described~\cite{STWP} by fits to a thermal distribution, 
          \begin{equation}
{{d^2\sigma} \over {dp_L p_T dp_T}}\sim e^{-(E-\mu)/T} \rightarrow {\bar{p} \over p}=\frac{e^{-(E+\mu_B)/T}}{e^{-(E-\mu_B)/T}}=e^{-(2\mu_B)/T} \qquad ,
\label{eq:boltz2}
\end{equation}
 with similar expressions for strange particles. $\mu_B$ (and $\mu_S$) are chemical potentials associated with each conserved quantity: baryon number, $\mu_B$,  (and strangeness, $\mu_S$). Thus it is simple and instructive to estimate the $\bar{p}/p$ ratio from Fig.~\ref{fig:phase_boundary}a, near the arrow, where I read $T=160$ MeV, $\mu_B=300$ MeV, $\sqsn\approx 30$ GeV, which gives $\bar{p}/p\approx 0.02$. Since the $\bar{p}/p$ ratio vs \sqsn will be an important issue later, it is not a good idea to get this important information from a sketch in a proposal but from measurements and the best analysis (Fig.~\ref{fig:pbar-p}).    
The results are: i) the correct $\bar{p}/p$ ratio at $\sqsn= 30$ GeV is $\sim 0.12$ from Fig.~\ref{fig:pbar-p}a; ii) this ratio also corresponds to the correct $\mu_b\approx 170$ MeV at $\sqsn= 30$ GeV from Fig.~\ref{fig:pbar-p}b. The lesson is: if it looks more like art than like science, be skeptical and look in refereed journals for the correct numbers.   

\subsection{A press release during ISSP 2011}
On June 23, 2011, shortly before I was to present my 2011 lectures, a press release from LBL arrived claiming that ``By comparing theory with data from STAR, Berkeley Lab scientists and their colleagues map phase changes in the QGP''~\cite{LBLJune23}. Since I was going to criticize in my lectures what I considered to be a particularly egregious case of ``physics by press release'' in the year 2000 by CERN (see Ref.~\cite{MJTIJMPA2014}), I felt that I was obliged to quickly absorb and present in my talk the physics behind this latest press release, hopefully a ``Highlight from RHIC''.  

The subject is ``Fluctuations of conserved quantities'', in this case the net baryon distribution taken as $p-\bar{p}$. Since there can be no fluctuations of conserved quantities such as net charge or net baryon number in the full phase space, one has go to  ``locally conserved quantities''~\cite{AsakawaHMPRL85} in small rapidity intervals to detect a small fraction of the protons and anti-protons which then fluctuates, i.e. varies from event to event. The argument is that, e.g. the fluctuation of one charged particle in or out of the considered interval produces a larger mean square fluctuation of the net electric charge if the system is in the hadron gas phase with integral charges than for the \QGP\ phase with fractional charges. 

However, while there are excellent statistical mechanical arguments about the utility of fluctuations of conserved quantities   such as net baryon number as a probe of a critical point~\cite{KochCFRNC06}, there were, in 2011, no adequate treatments of the mathematical statistics of the experimental measurements. There are also additional problems such as short-range rapidity correlations in A$+$A collisions between like-particles induced by Fermi or Bose quantum statistics that must be reckoned with (e.g. see Refs.~\cite{ZajcPRC29,MJTPLB347}). 

Theoretical analyses tend to be made by a Taylor expansion of the free energy \hbox{$F=-T\ln Z$} around the critical temperature $T_c$ where $Z$ is the partition function, or sum over states, which is of the form \vspace*{-0.5pc}
\begin{equation}\large
Z\propto e^{-(E-\sum_i \mu_i Q_i)/kT} \label{eq:partitionfn}  \end{equation} \normalsize
 and $\mu_i$ are chemical potentials associated with conserved charges $Q_i$~\cite{KochCFRNC06}.  The terms of the Taylor expansion, which are obtained by differentiation, are called susceptibilities, denoted $\chi$. The only connection of this method to mathematical statistics is that the Cumulant generating function in mathematical statistics for a random variable $x$ is also a Taylor expansion of the $\ln$ of an exponential:
\begin{equation}
g_x (t)=\ln\mean{e^{tx}}=\sum_{n=1}^\infty \kappa_n \frac{t^n}{n!} \qquad \kappa_m =\left.\frac{d^m g_x (t)}{dt^m}\right|_{t=0} \qquad .
\label{eq:cumgenfn}
\end{equation}
Thus, the susceptibilities are Cumulants in mathematical statistics terms, where, in general, the Cumulant $\kappa_m$ represents the $m^{\rm th}$ central moment, $\mu_m\equiv \mean{(x-\mu)^m}$, with all $m$-fold combinations of the lower order moments subtracted, where $\mu\equiv\mean{x}$. For instance, 
$\kappa_2=\mean{(x-\mu)^2}\equiv \sigma^2$, $\kappa_3=\mean{(x-\mu)^3}$, $\kappa_4=\mean{(x-\mu)^4}-3\kappa_2^2$, $\kappa_5=\mean{(x-\mu)^5}-10\kappa_3 \kappa_2$. Two so-called normalized or standardized Cumulants are common in this field, the skewness, $S\equiv\kappa_3/\sigma^3$ and the kurtosis, $\kappa\equiv\kappa_4/\sigma^4=\mean{(x-\mu)^4}/\sigma^4-3$.

\begin{figure}[!t]
   \begin{center}
\includegraphics[width=0.46\linewidth]{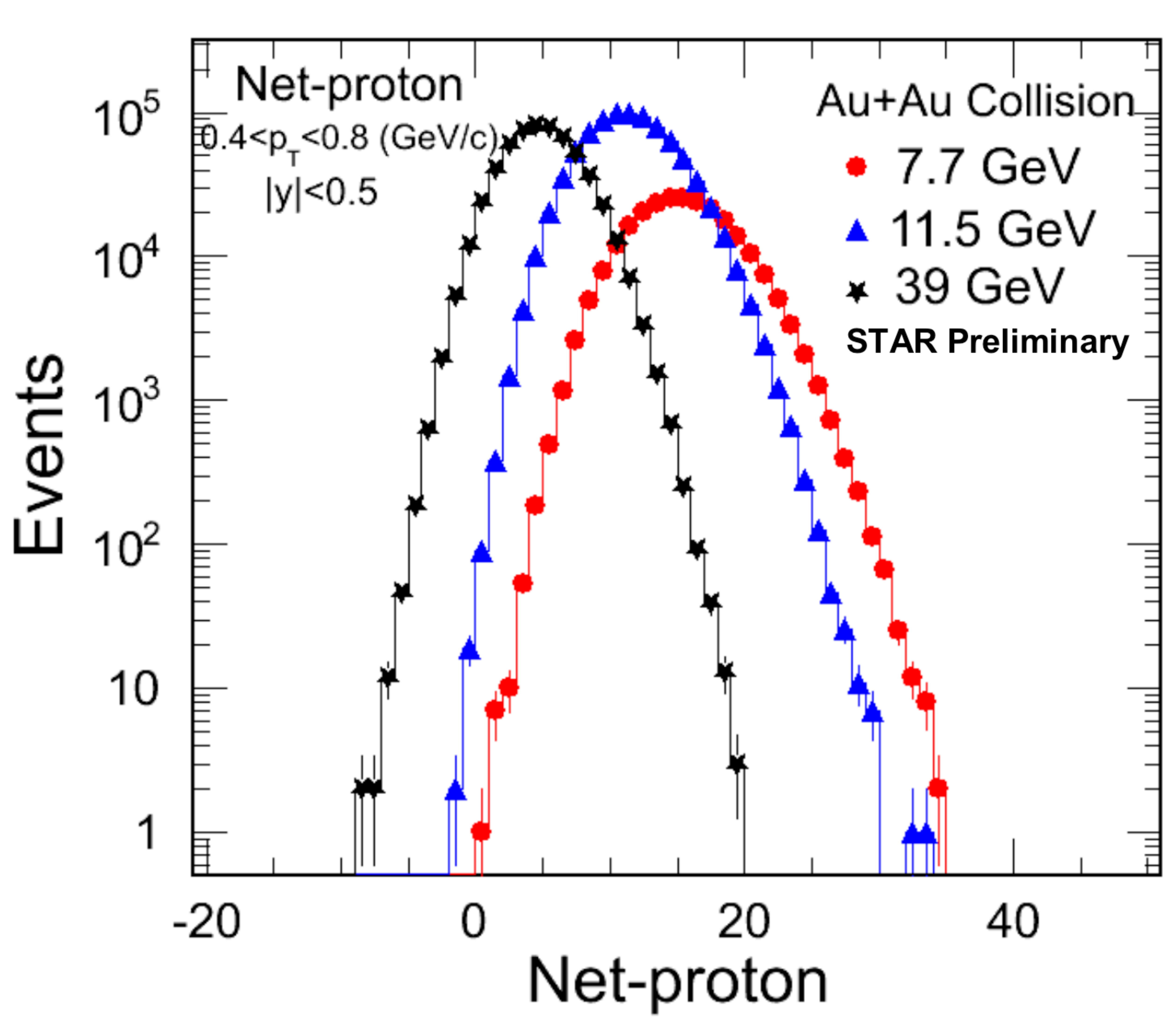}
\includegraphics[width=0.53\linewidth]{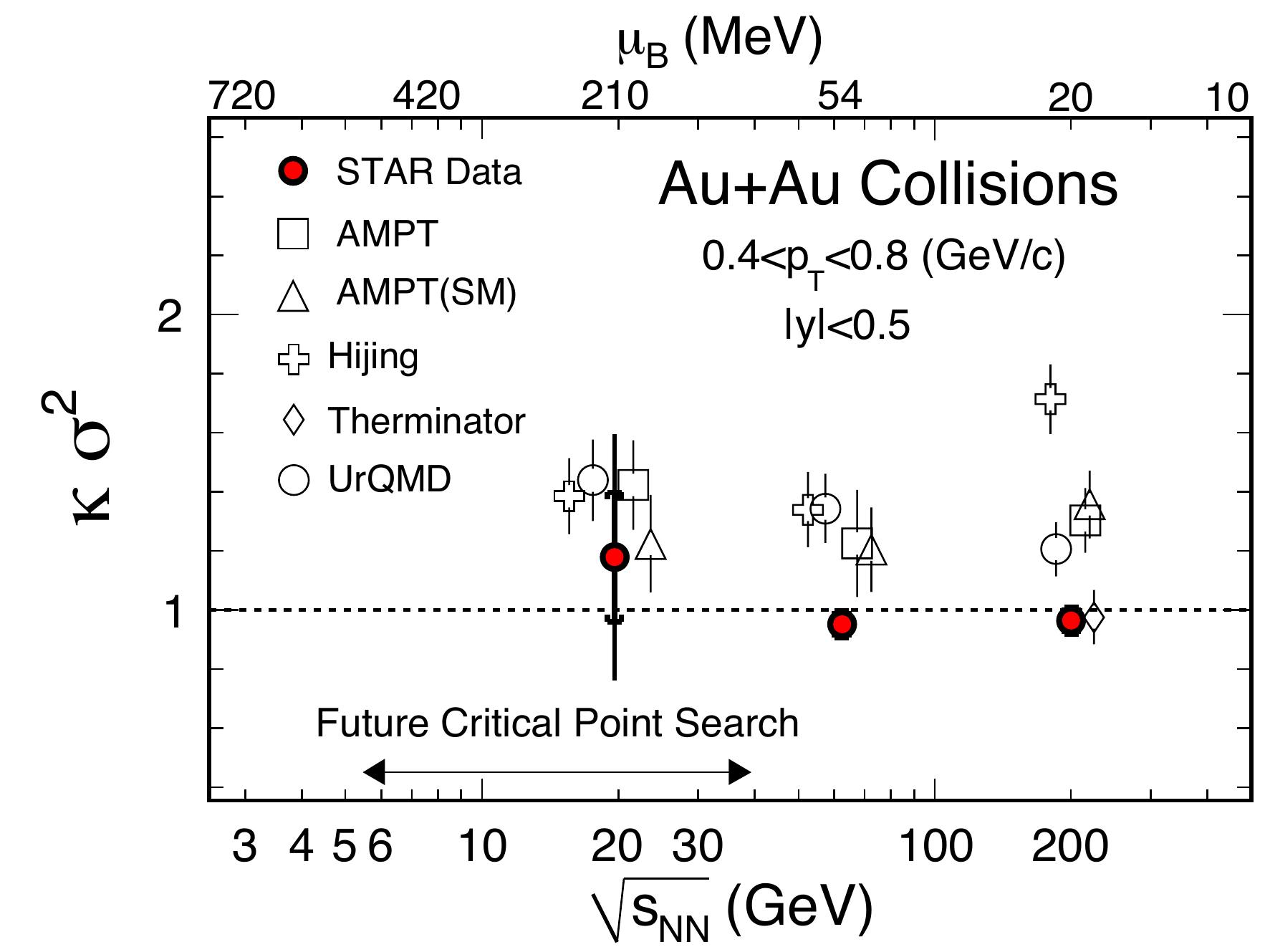}
\includegraphics[width=0.48\linewidth]{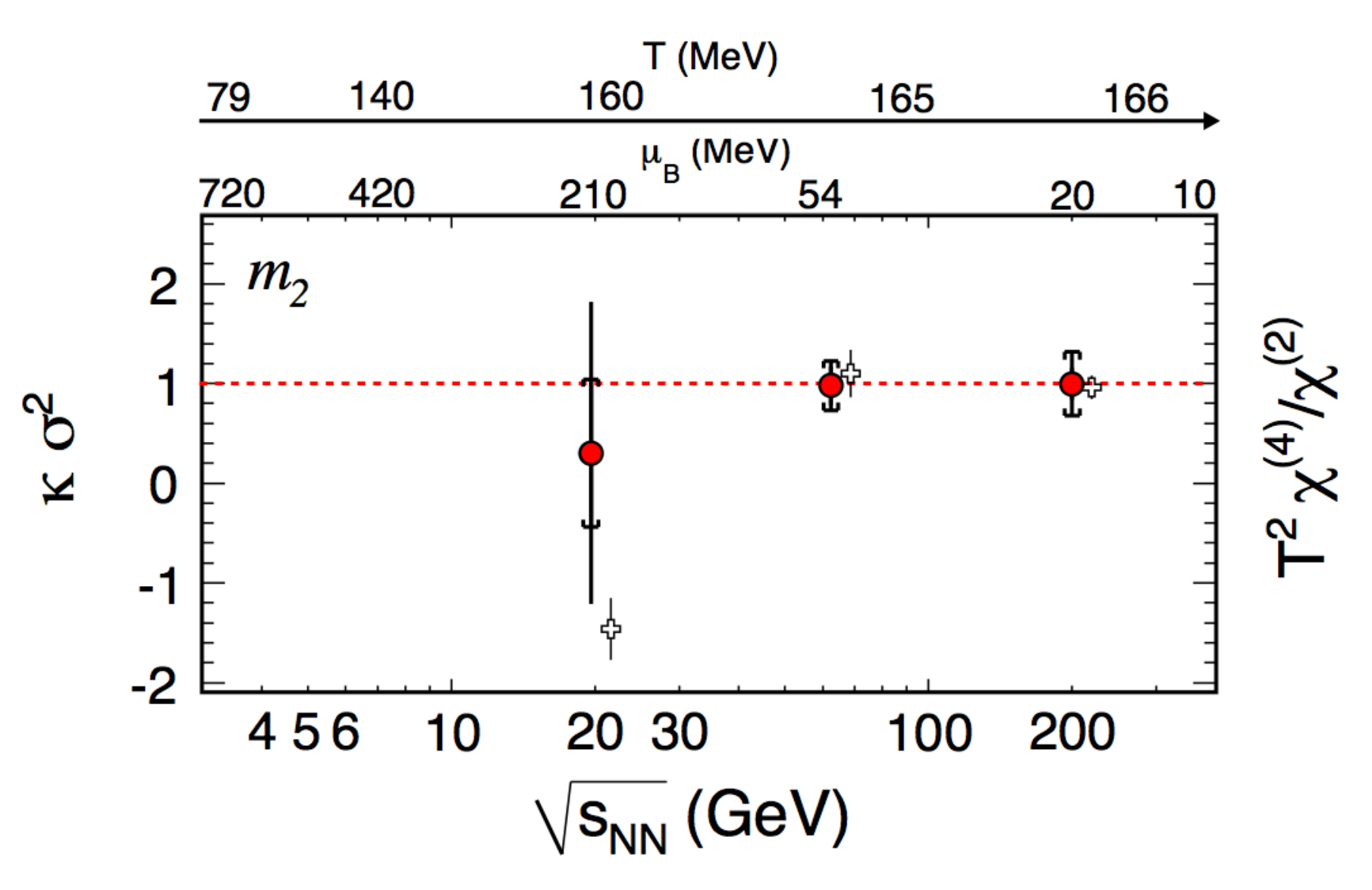}
\includegraphics[width=0.50\linewidth]{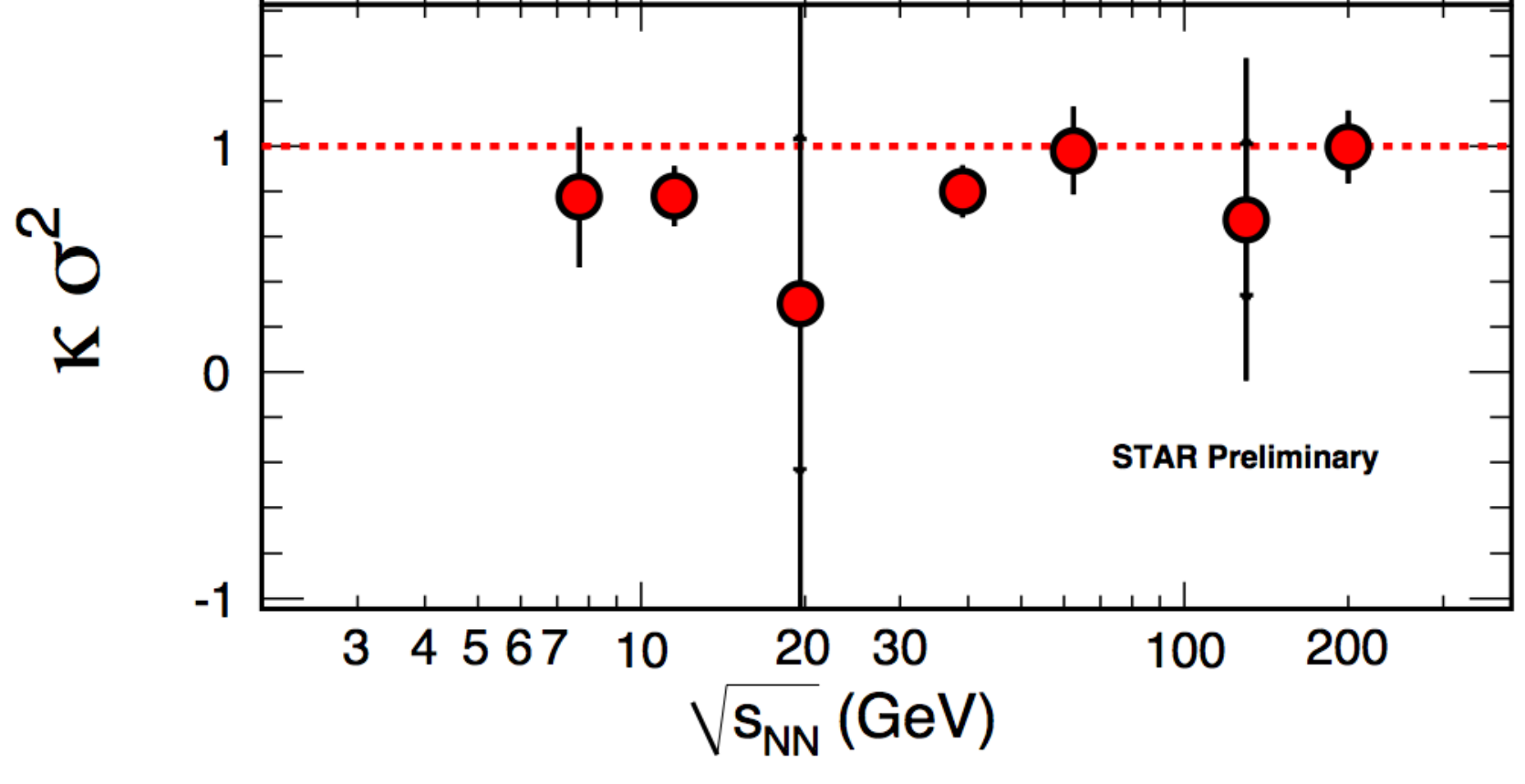}
\end{center}\vspace*{-2.0pc}
\caption[]{\footnotesize a) (top-left) STAR~\cite{TarnowskyQM2011} distribution of event-by-event $p-\bar{p}$ at 3 values of $\sqrt{s_{NN}}$; b) (top-right) STAR published~\cite{STARnetPPRL105} measurements of $\kappa\sigma^2$; c) (bottom-left) Measurements from (b) as shown in Ref.~\cite {Science332} compared to the predicted ratio of susceptibilities (open crosses); d) (bottom-right) compilation~\cite{TarnowskyQM2011} of STAR measurements of $\kappa\sigma^2$ for $p-\bar{p}$.   }
\label{fig:STAR-NetP}\vspace*{-1.0pc}
\end{figure}

A sample~\cite{TarnowskyQM2011} of STAR measurements of the distribution of net-protons in Au+Au collisions in the small interval $0.4\leq p_T\leq 0.8$ GeV/c, $|y|<0.5$ for different $\sqrt{s_{NN}}$ is shown in Fig.~\ref{fig:STAR-NetP}a. 
The moments in the form $\kappa\sigma^2=\kappa_4/\kappa_2$ are shown from a previous STAR publication~\cite{STARnetPPRL105} in  Fig.~\ref{fig:STAR-NetP}b while a plot, alleged to be of this same data, presented in the Lattice \QCD\ theory publication that generated the press-release, is shown in Fig.~\ref{fig:STAR-NetP}c~\cite{Science332}; and a plot of the $\kappa\sigma^2$ from the data of Fig.~\ref{fig:STAR-NetP}c, combined with the results from Fig.~\ref{fig:STAR-NetP}b, is shown in Fig.~\ref{fig:STAR-NetP}d~\cite{TarnowskyQM2011}. 
There are many interesting issues to be gleaned from Fig.~\ref{fig:STAR-NetP}. 

The data point at 20 GeV in Fig.~\ref{fig:STAR-NetP}c is not the published one from (b), as stated in the caption~\cite{Science332}, but the one from (d), which is different and with a much larger error. This, in my opinion, makes the data point look better compared to the predicted discontinuous value of $\kappa\sigma^2=-1.5$ for the critical point at 20 GeV (open crosses) in contrast to the predictions of 1.0 for both 62.4 and 200 GeV. The published measurements in (b) together with the newer measurements in (d) are all consistent with $\kappa\sigma^2=1$; but clearly indicate the need for a better measurement at $\sqrt{s_{NN}}=20$ GeV. Apart from these issues, the main problem of comparing Lattice \QCD\ ``data'' to experimental measurements is that it is like comparing peaches to a fish, since the prediction is the result of derivatives of the log of the calculated partition function of an idealized system, which may have little bearing on what is measured using finite sized nuclei in an experiment with severe kinematic cuts. Maybe this is too harsh a judgement; but since this is the first such comparison (hence the press release), perhaps the situation will improve in the future. If a future measurement would show a significant huge discontinuity of $\kappa\sigma^2$ similar to the theoretical prediction at $\sqrt{s_{NN}}=20$ GeV, then even I would admit that such a discovery would deserve a press release, maybe more! 
\subsubsection{If you know the distribution, you know all the moments and cumulants}
When I first saw the measured distributions in Fig.~\ref{fig:STAR-NetP}a in 2011, my immediate reaction was that STAR should fit them to Negative Binomial distributions (NBD) so that they would know all the Cumulants. However, I subsequently realized that my favorite 3 distributions for integer random variables, namely, Poisson, Binomial, and Negative Binomial, are all defined only for positive integers (e.g. see Ref.~\cite{RATCUP} for details), while the number of net-protons on an event can be negative as well as positive, especially at higher c.m. energies. Thanks to Gary Westfall of STAR, in a paper presented at the Erice School of Nuclear Physics in 2012~\cite{WestfallErice2012}, who found out that these three distributions fall into the class of ``integer valued L\'evy processes~\cite{Barndorff2013}'' for which the Cumulants $\kappa_j$ for the distribution $P(n-m)$ of the difference of samples from two such distributions, $P^{+}(n)$ and $P^{-}(m)$, with Cumulants $\kappa_j^+$ and $\kappa_j^-$, respectively, are~\cite{Westfall2013,Barndorff2013}:
\begin{equation}
\kappa_j=\kappa_j^+ +(-1)^j \kappa_j^- \qquad, \label{eq:difCum}
\end{equation} 
so long as the distributions are not 100\% correlated.
This result is the same as if the distributions $P^{+}(n)$ and  $P^{-}(m)$ were statistically independent. The first four Cumulants of the Poisson, Binomial and Negative Binomial distributions  are given in Table~\ref{tab:Cumulant}. 
\begin{table}[!h]\vspace*{-1.0pc}
\centering
\caption[]{Cumulants for Poisson, Binomial and Negative Binomial Distributions}
{\begin{tabular}{llll} 
\hline
Cumulant & Poisson & Binomial & Negative Binomial \\
\hline
$\kappa_1=\mu$ & $\mu$ & $np$ &$\mu$\\
$\kappa_2=\mu_2=\sigma^2$ &$\mu$& $\mu (1-p)$ & $\mu (1+{\mu}/{k})$\\[0.2pc]
$\kappa_3=\mu_3$ &$\mu$& $\sigma^2 (1-2p)$ & $\sigma^2 (1+2{\mu}/{k})$\\[0.2pc] 
$\kappa_4=\mu_4-3\kappa_2^2$\hspace*{1pc}&$\mu$& $\sigma^2 (1-6p+6p^2)$ & $\sigma^2 (1+6{\mu}/{k}+6{\mu^2}/{k^2})$\\[0.2pc]
\hline\\[-0.6pc]
$S\equiv{\kappa_3}/{\sigma^3}$& $ {1}/{\sqrt{\mu}}$ &${(1-2p)}/{\sigma}$ & $(1+2{\mu}/{k})/{\sigma}$\\[0.2pc]
$\kappa\equiv{\kappa_4}/{\kappa_2^2}$ & ${1}/{\mu}$ & ${(1-6p+6p^2)}/{\sigma^2}$\hspace*{1pc}& $(1+6{\mu}/{k}+6{\mu^2}/{k^2})/{\sigma^2}$\\[0.2pc]
$S\sigma=\kappa_3/\kappa_2$ & $1$ & $(1-2p)$ & $(1+2{\mu}/{k})$\\[0.2pc]
$\kappa\sigma^2=\kappa_4/\kappa_2$ & $1$ & $(1-6p+6p^2)$ & $(1+6{\mu}/{k}+6{\mu^2}/{k^2})$\\[0.2pc] 
\hline
\end{tabular}} \label{tab:Cumulant}\vspace*{-1.0pc}
\end{table}

\subsection{The latest measurements have appeared without a press release.}
\begin{figure}[!h] 
      \centering
\raisebox{0pc}{\begin{minipage}[b]{0.44\linewidth} \includegraphics[width=\linewidth]{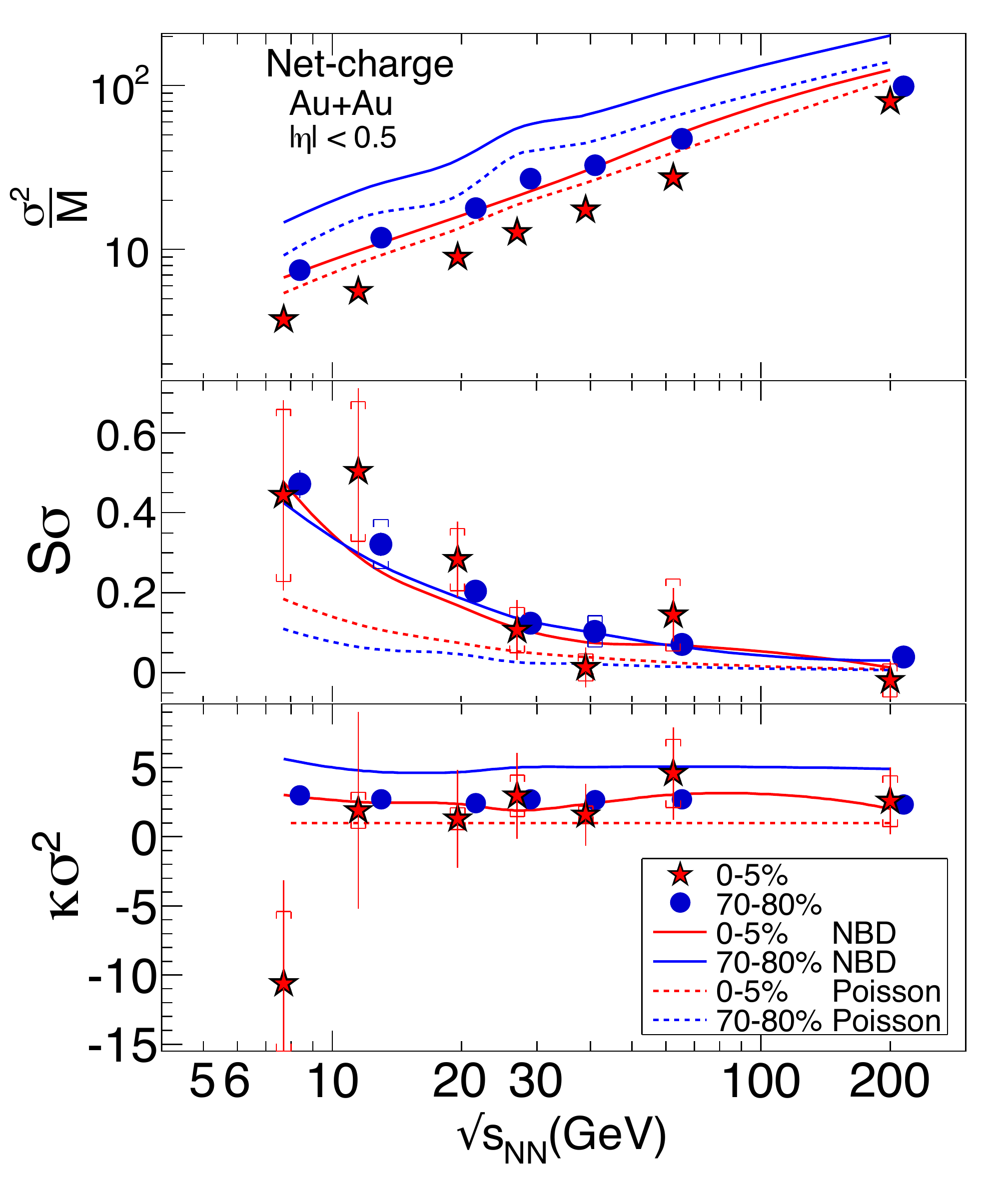} 
 \end{minipage}} 
\hspace*{1.0pc}\raisebox{1.0pc}{\begin{minipage}[b]{0.44\linewidth}  
\includegraphics[width=\linewidth]{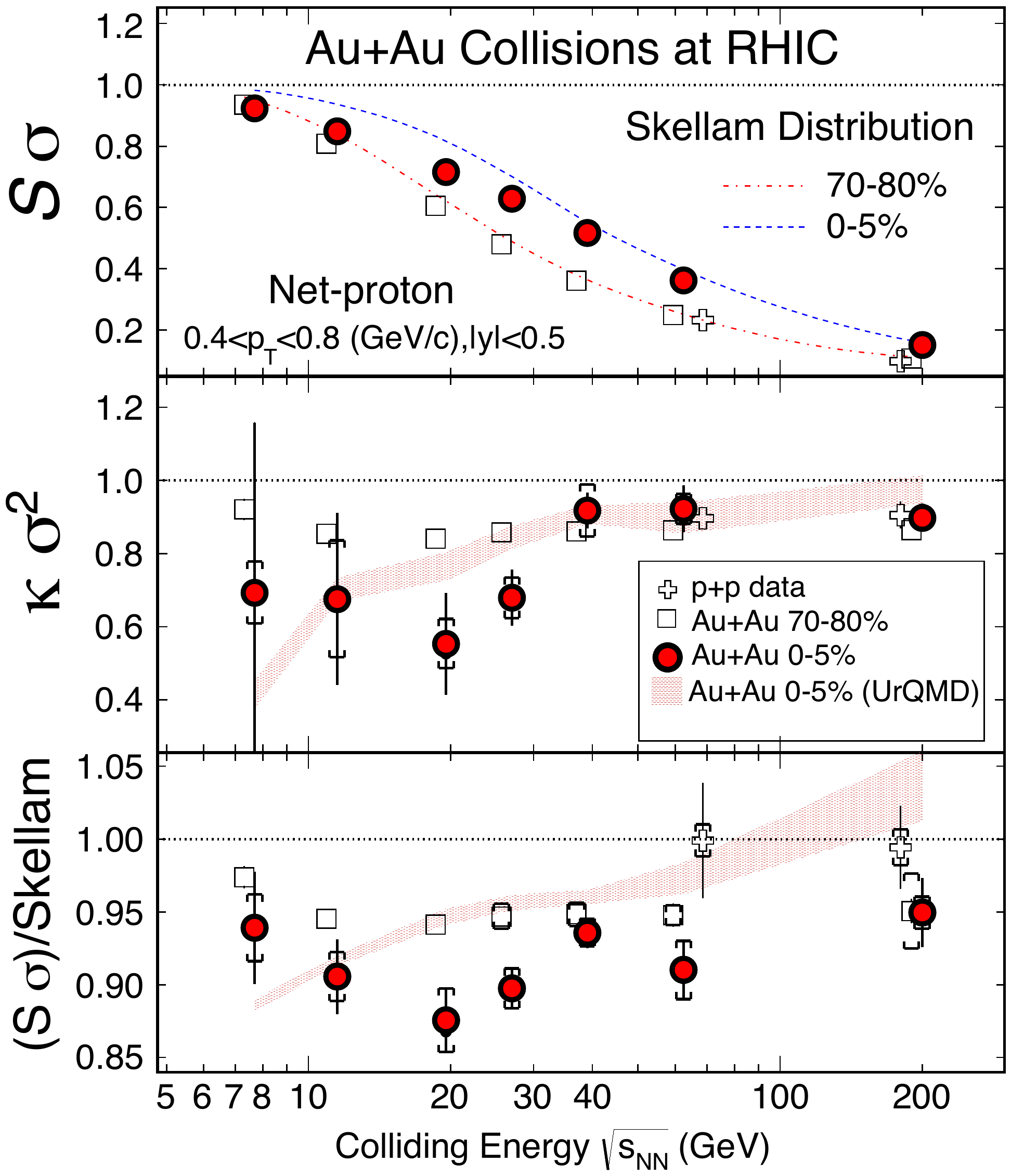} \end{minipage}}\vspace*{-1.0pc}
\caption[]{\footnotesize a) $\sqrt{s_{NN}}$ dependence of combinations of Cumulants in Au$+$Au (and $p$$+$$p$) from STAR: a) (left) net-charge Cumulants~\cite{STAR14021558}, where $M$ is used to represent the mean, $\mu$. b) (right) Cumulants of the $N_p-N_{\bar{p}}$ distributions~\cite{STARPRL112NetP}, {\bf where the error bars are statistical and the caps systematic errors}.}
\label{fig:STARCumpub}\vspace*{-1.0pc}
   \end{figure}

In the intervening period since 2011, the STAR collaboration has improved the preliminary measurements to publications and has improved the analysis by comparing to both Poisson and Negative Binomial distributions.  Figure~\ref{fig:STARCumpub}a~\cite{STAR14021558} shows the STAR measurements of Cumulants of the net charge $(N^+ -N^-)$ distributions      
from the ``number of positive ($N^+$) and negative ($N^-$) charged particles within $|\eta|<0.5$ and $0.2< p_T< 2.0$ GeV/c on each event (after removing protons and antiprotons with $p_T<400$ MeV/c)~\cite{STAR14021558}''. The corresponding Poisson and NBD Cumulants were calculated from the measured mean, $\mu$, and variance, $\sigma^2$, of the $N^+$ and $N^-$ distributions, respectively, and then calculated using Eq.~\ref{eq:difCum}.  
In contrast to Fig.~\ref{fig:STAR-NetP}, no non-monotonic behavior with $\sqrt{s_{NN}}$ is observed (or claimed) and the measurements of $S\sigma$ and $\kappa\sigma^2$ are all above the Poisson baseline. The $S\sigma$ measurements clearly favor the NBD. 

\begin{figure}[!h] 
\begin{center}
{\footnotesize a)} \raisebox{0.0pc}{\includegraphics[width=0.48\textwidth]{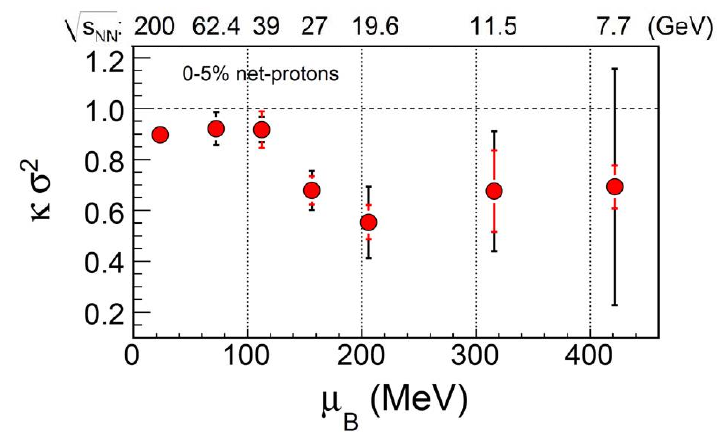}}\hspace*{1.0pc}
{\footnotesize b)} \raisebox{0.4pc}{\includegraphics[width=0.43\textwidth]{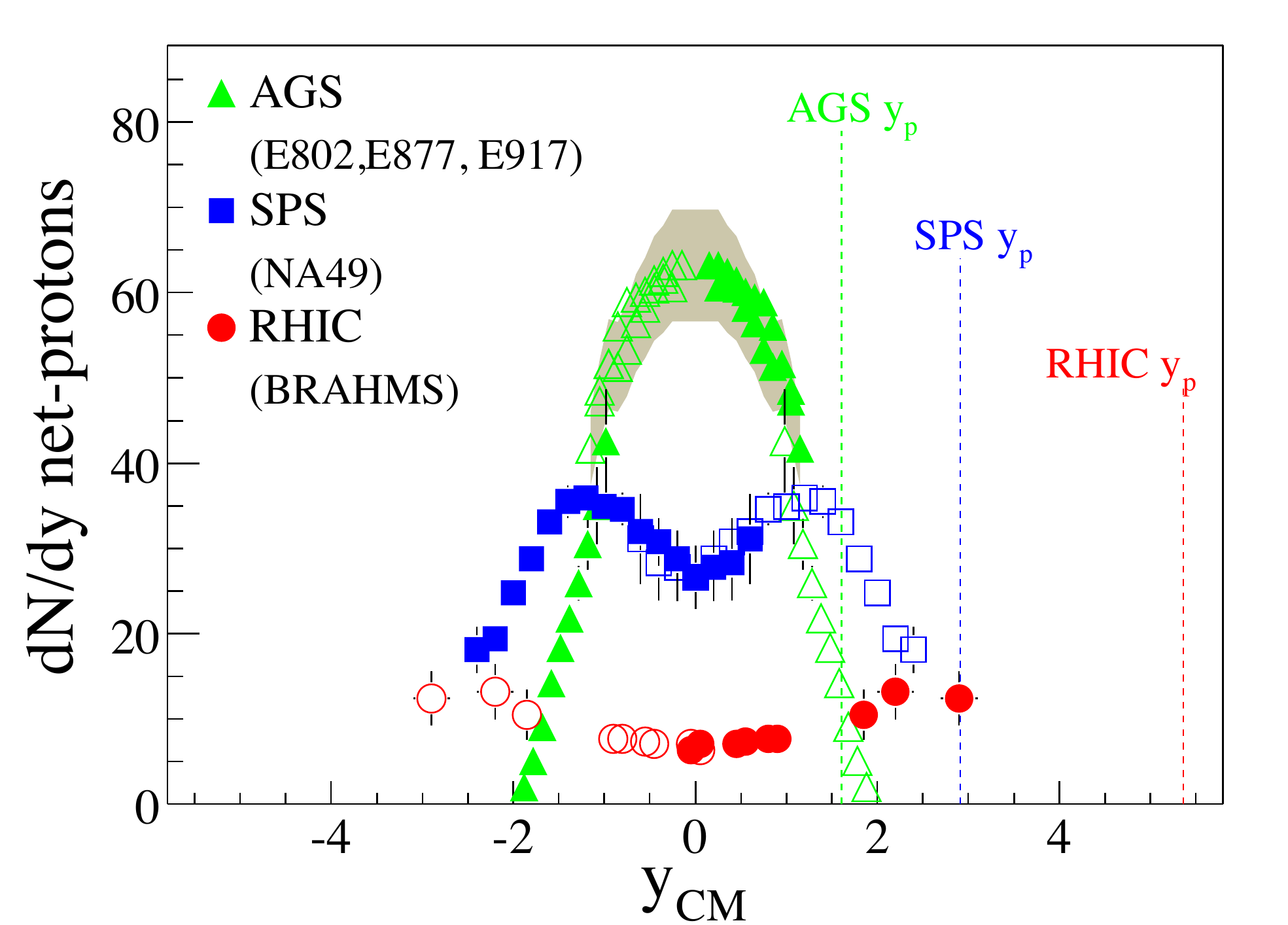}}
\end{center}\vspace*{-1.0pc}
\caption[]{\footnotesize (a)(left) $\kappa\sigma^2$ vs. \sqsn and $\mu_B$~\cite{STARWPBES2014}; b) (right) ($dN_p/dy-dN_{\bar{p}}/dy$) for top 5\% centrality at AGS (Au+Au, \sqsn=4.9 GeV), SPS (Pb+Pb, \sqsn=17.2 GeV) and RHIC(Au+Au, \sqsn=200 GeV), beam rapidity in c.m. system $y_{\rm p}=1.6,2.9,5.4$~\cite{BRAHMSNetPPRL93}.\label{fig:backtoAGS}}\vspace*{-1.0pc}
\end{figure}

The situation is quite different for the net-proton ($N_p-N_{\bar{p}}$) Cumulants (Fig.~\ref{fig:STARCumpub}b)~\cite{STARPRL112NetP} measured within $|y|<0.5$ over the range $0.4< p_T< 0.8$ GeV/c which covers roughly half the $p_T$ spectrum.   Here the measurements of $S\sigma$  and $\kappa\sigma^2$ are all below the Poisson baseline, denoted Skellam, which is the distribution of the difference between two Poissons and reflects ``a system of totally uncorrelated, statistically random particle production''\cite{STARPRL112NetP}.  From Eq.~\ref{eq:difCum} and Table~\ref{tab:Cumulant} for a Poisson one can see that, for a Skellam, $S\sigma=\kappa_3/\kappa_2=(\mu_p-\mu_{\bar{p}})/(\mu_p+\mu_{\bar{p}})$ which increases with decreasing \sqsn because the $\bar{p}$ vanish ($\mu_{\bar{p}}=\mean{N_{\bar{p}}}\ll \mu_{p}$) so that the shape of the net distribution becomes dominated by the protons. This is easier to see in a plot of $\kappa\sigma^2$ vs $\mu_B$ with \sqsn indicated (Fig.~\ref{fig:backtoAGS}a)~\cite{STARWPBES2014} which shows clearly that $\kappa\sigma^2$ starts dropping for $\sqsn<39$ GeV where the $\bar{p}/p$ ratio drops below $\sim 0.3$ (recall Fig.~\ref{fig:pbar-p}). 

The errors are still too large to determine whether or not $\kappa\sigma^2$ stays constant below \sqsn=19.6 GeV but are sufficient to clearly rule out the value of $\kappa\sigma^2=-1.5$ at $\sqsn\approx 20$ GeV predicted in Fig.~\ref{fig:STAR-NetP}~\cite{Science332} which created the fuss in 2011.      
It is also important to point out that in addition to the vanishing of the anti-protons, the physics of the protons at mid-rapidity changes dramatically---the protons are no longer produced particles, which would conserve $N_p-N_{\bar{p}}$, but are the participants and fragments from the colliding nuclei which move to mid-rapidity and eventially stop as \sqsn is reduced from 200, to 17.2 to 4.9 GeV (Fig.~\ref{fig:backtoAGS}b)~\cite{BRAHMSNetPPRL93}. 
All these results indicate that the search for a \QCD\  critical point at RHIC in the Beam Energy Scan (BES) in 2018-19 may not be as straightforward as originally assumed.\vspace*{-1.0pc} 
\section{Jet quenching, RHIC's main claim to fame}\vspace*{-0.5pc}
The gold-plated signature for the \QGP\ since 1986~\cite{MatsuiSatz86} has been the suppression of $J/\Psi$ because the color potential between $c,\bar{c}$ quarks would be screened (Debye screening) by all the free color charges in the medium so that the $c,\bar{c}$ would not be able to bind to form the $J/\Psi$. In fact the PHENIX experiment at RHIC was specifically designed to detect the $J/\Psi$ at mid-rapidity at rest or with very low $p_T$ (where the screening effect would be the largest) via the decay $J/\Psi\rightarrow e^+ +e^-$. $J/\Psi$ suppression was reportedly observed several times at the CERN SpS fixed target heavy ion program starting with NA38 in O$+$U collisions in 1989~\cite{NA38PLB220} but  was plagued with many problems. The principal physics problem is that the $J/\Psi$ does not follow the standard hard-scattering pointlike scaling in A$+$B collisions, $\sigma^{J/\Psi}_{AB}=(A\cdot B)^\alpha \cdot \sigma^{J/\Psi}_{NN}$ with $\alpha\equiv 1$, but is suppressed in cold nuclear matter (CNM) in $p$$+$A and A$+$B collisions, with $\alpha\approx 0.91$. Thus, the ultimate discovery by NA50 in Pb$+$Pb collisions at \sqsn=17.2 GeV~\cite{GoninQM96} in 1996--1998 was called ``anomalous suppression'' because it was below the CNM cross section dependence which was itself well below the hard-scattering pointlike scaling.  

In 1998 at the \QCD\ workshop in Paris~\cite{4thQCDWks}, I found what I thought was a cleaner signal of the \QGP\ when Rolf Baier asked me whether jets could be measured in Au$+$Au collisions because he had made studies in p\QCD~\cite{BDMPS} of the energy loss of partons, produced by hard-scattering ``with their color charge fully exposed'',  in traversing a medium ``with a large density of similarly exposed color charges''. The conclusion was that ``Numerical estimates of the loss suggest that it may be significantly greater in hot matter than in cold. {\em This makes the magnitude of the radiative energy loss a remarkable signal for \QGP\  formation}''~\cite{BDMPS}. In addition to being a probe of the \QGP\ the fully exposed color charges allow the study of parton-scattering with $Q^2 \ll 1-5$ (GeV/c)$^2$ in the medium where new collective \QCD\ effects may possibly be observed.

Because the expected energy in a typical jet cone $R=\sqrt{(\Delta\eta)^2+ (\Delta\phi)^2}$ in central Au$+$Au collisions at \sqsn=200 GeV would be $\pi R^2\times1/2\pi \times d\Et/d\eta=R^2/2 \times d\Et/d\eta~\sim 350$ GeV for $R=1$, where the kinematic limit is 100 GeV, I said (and wrote~\cite{4thQCDWks}) that jets can not be reconstructed in Au$+$Au central collisions at RHIC---still correct after 16 years. On the other hand, hard-scattering was discovered in $p+p$ collisions at the CERN-ISR in 1972 with single particle and two-particle correlations, while jets had a long learning curve from 1977--1982 with a notorious false claim (e.g. see Refs.~\cite{RATCUP,MJTIJMPA2014}), so I said (and wrote~\cite{4thQCDWks}) that we should use single and two-particle measurements---which we did and it WORKED! The present solution for jets in A$+$A collisions (LHC 2010 and RHIC c.2014) is to take smaller cones, with 56 GeV in $R=0.4$, 32 GeV in $R=0.3$, 14 GeV in $R=0.2$ at RHIC.  
\subsection{Jet quenching at RHIC --- Suppression of high $p_T$ particles }
\begin{figure}[!h]
\vspace*{-0.3pc}        \centering
\includegraphics[height=0.27\textheight]{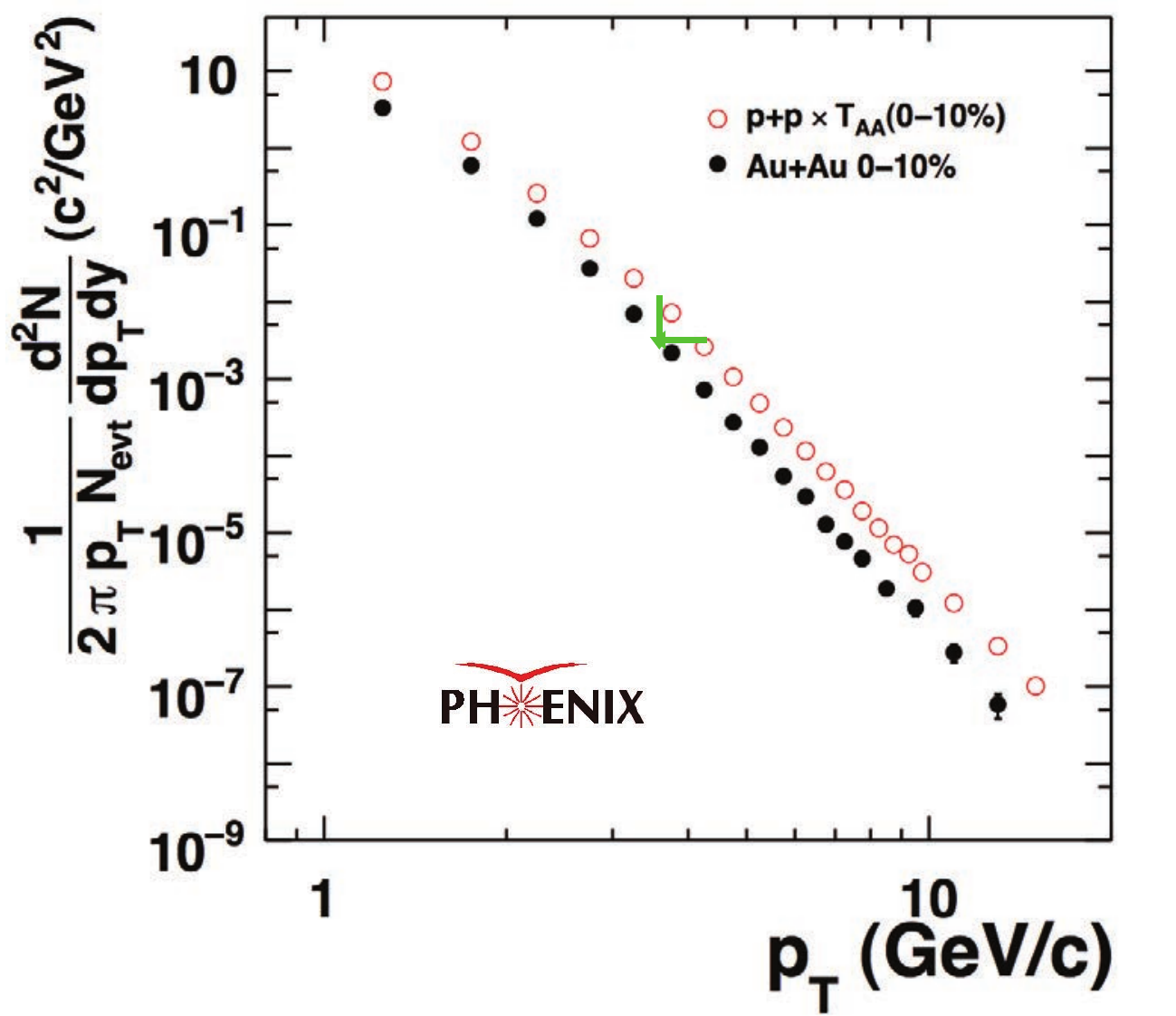}
\hspace*{-0.011\textwidth} \includegraphics[height=0.27\textheight]{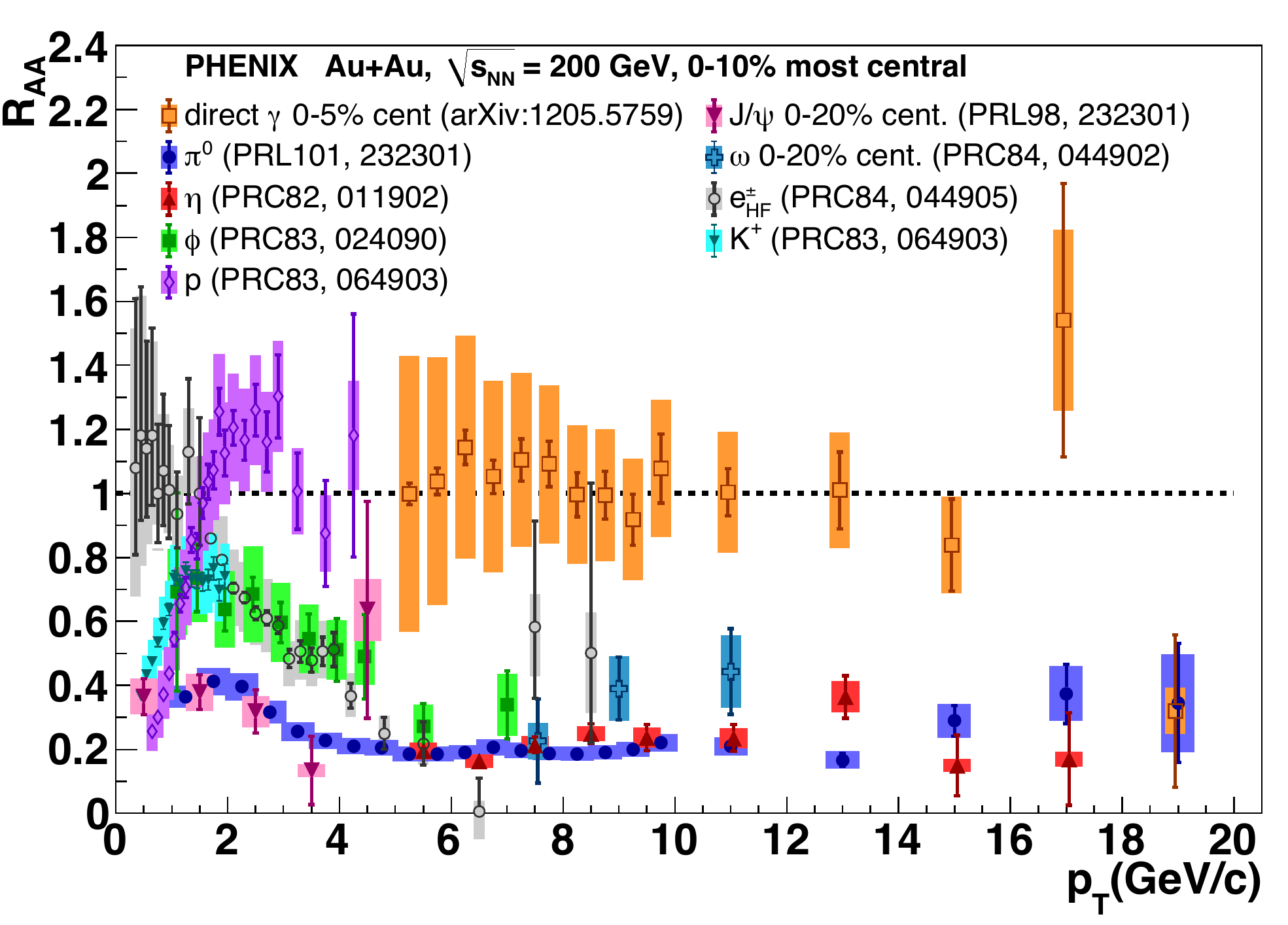}\vspace*{-0.5pc}
\caption{\footnotesize a) (left) Log-log plot of invariant yield of $\pi^0$ at $\sqrt{s_{NN}}=200$ GeV as a function of transverse momentum $p_T$ in $p$$+$$p$ collisions multiplied by $\mean{T_{AA}}$ for Au+Au central (0--10\%) collisions compared to the Au+Au measurement~\cite{ppg054}. Vertical arrow is for $R_{AA}(p_T)$, horizontal arrow for $\delta p_T'$. b) (right) $R_{AA}(p_T)$ for all identified particles so far measured by PHENIX in Au+Au central collisions at $\sqrt{s_{NN}}=200$ GeV.}
\label{fig:Tshirt}
\end{figure}

   The discovery at RHIC~\cite{ppg003} that $\pi^0$'s produced at large transverse momenta are suppressed in central Au+Au collisions by a factor of $\sim5$ compared to pointlike scaling from $p$$+$$p$ collisions is arguably {\em the}  major discovery in Relativistic Heavy Ion Physics. For $\pi^0$ (Fig.~\ref{fig:Tshirt}a)~\cite{ppg054} the hard-scattering in $p$$+$$p$ collisions is indicated by the power law behavior $p_T^{-n}$ for the invariant cross section, $E d^3\sigma/dp^3$, with $n=8.1\pm 0.1$ for $p_T\geq 3$ GeV/c.  The Au+Au data at a given $p_T$ can be characterized either as shifted lower in \pt by $\delta p_T'$ from the pointlike scaled $p$$+$$p$ data at $p'_T=p_T+\delta p_T'$, or shifted down in magnitude, i.e. suppressed. In Fig.~\ref{fig:Tshirt}b, the suppression of the many identified particles measured by PHENIX at RHIC is presented as the Nuclear Modification Factor, 
$R_{AA}(p_T)$, the ratio of the yield of e.g. $\pi$ per central Au+Au collision (upper 10\%-ile of observed multiplicity)  to the pointlike-scaled $p$$+$$p$ cross section at the same $p_T$, where $\mean{T_{AA}}$ is the average overlap integral of the nuclear thickness functions: 
   \begin{equation}
  R_{AA}(p_T)=\frac{(1/N_{AA})\;{d^2N^{\pi}_{AA}/dp_T dy}} { \mean{T_{AA}}\;\, d^2\sigma^{\pi}_{pp}/dp_T dy} \quad . 
  \label{eq:RAA}
  \end{equation}

The striking differences of $R_{AA}(p_T)$ in central Au+Au collisions for the many particles measured by PHENIX  (Fig.~\ref{fig:Tshirt}b) illustrates the importance of particle identification for understanding the physics of the medium produced at RHIC. Most notable are: the equal suppression of $\pi^0$ and $\eta$ mesons by a constant factor of 5 ($R_{AA}=0.2$) for $4\leq p_T \leq 15$ GeV/c, with suggestion of an increase in $R_{AA}$ for $p_T > 15$ GeV/c; the equality of suppression of direct-single $e^{\pm}$ (from heavy quark ($c$, $b$) decay) and $\pi^0$ at $p_T\gsim 5$ GeV/c; the non-suppression of direct-$\gamma$ for $p_T\geq 4$ GeV/c; the exponential rise of $R_{AA}$ of direct-$\gamma$ for $p_T<2$ GeV/c~\cite{ppg086}, which is totally and dramatically different from all other particles and attributed to thermal photon production by many authors (e.g. see citations in Ref.~\cite {ppg086}). For $p_T\gsim 4$ GeV/c, the hard-scattering region,  the fact that all hadrons are suppressed, but direct-$\gamma$ are not suppressed, indicates that suppression is a medium effect on outgoing color-charged partons likely due to energy loss by coherent Landau-Pomeranchuk-Migdal radiation of gluons, predicted in p\QCD~\cite{BDMPS}, which is sensitive to properties of the medium. 

One nice advantage that hard-scattering and high $p_T$ suppression have as a \QGP\ probe compared to $J/\Psi$ suppression is that although there is a CNM effect, it is an enhancment rather than a suppression; and as far as is known, the enhancement, historically called the Cronin effect~\cite{CroninPRD11}, only occurs for baryons and not mesons at RHIC energies. 
         \begin{figure}[!h]
\vspace*{-0.3pc}   \begin{center}
\includegraphics[width=0.45\textwidth]{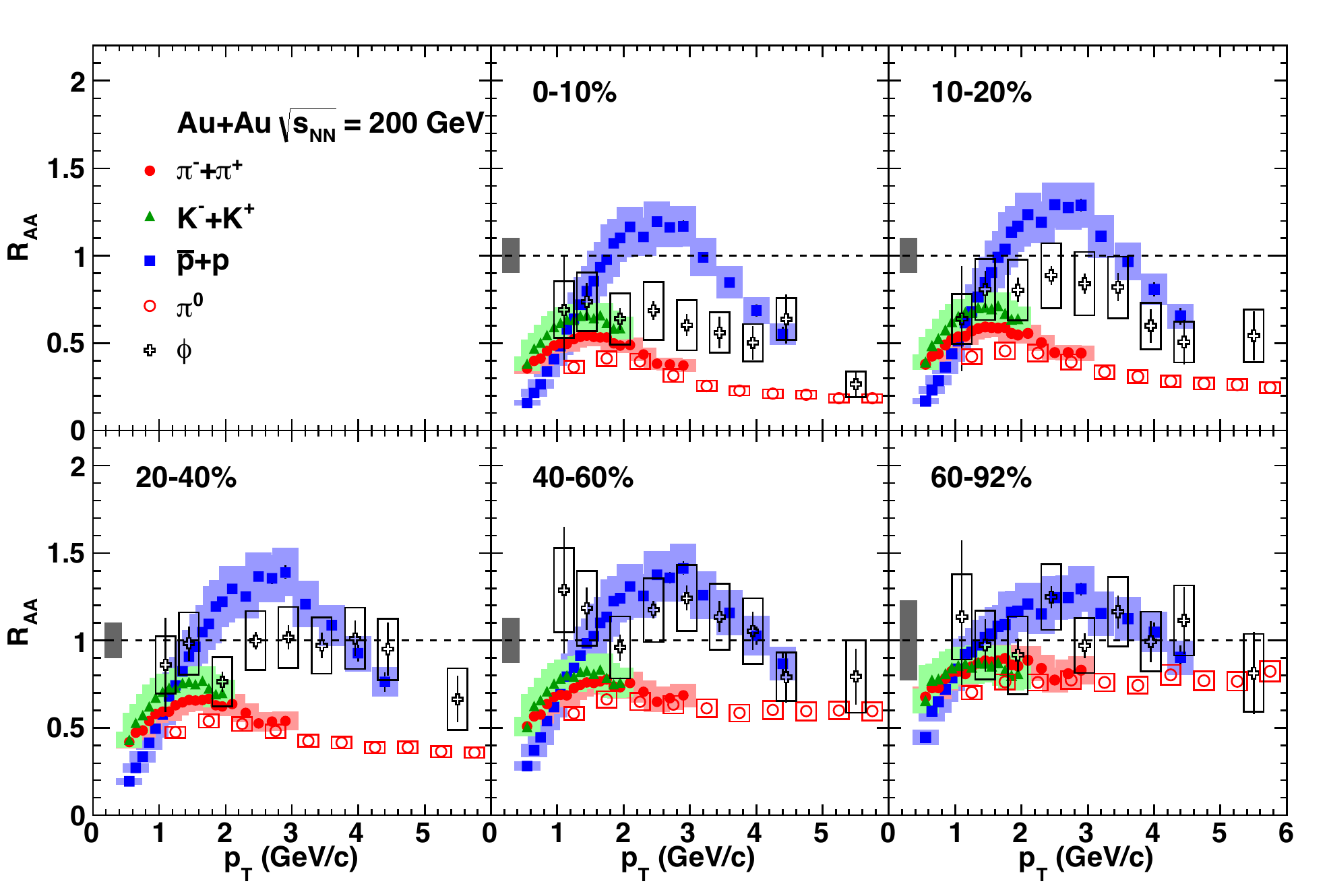}
\includegraphics[width=0.45\textwidth]{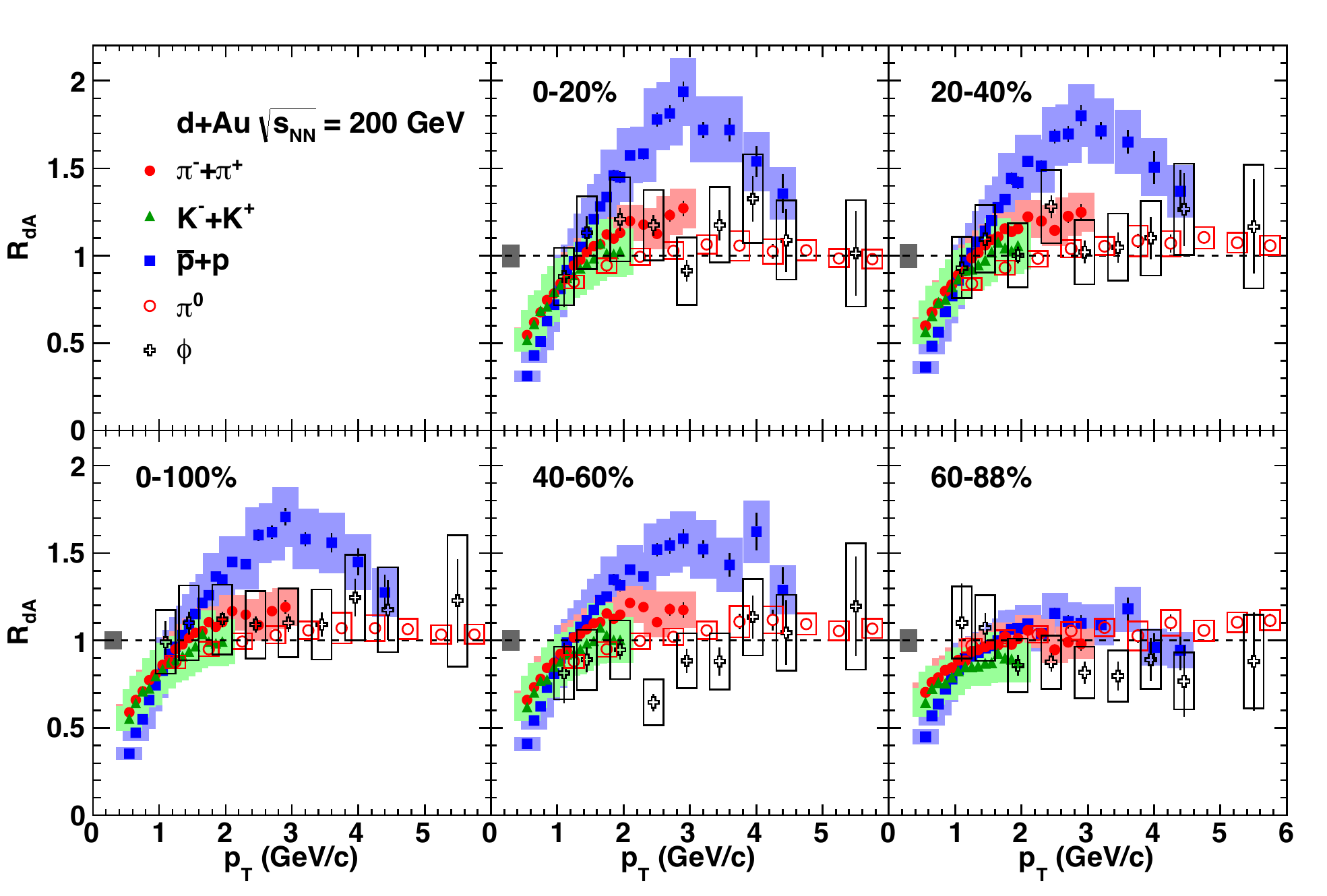}
\end{center}\vspace*{-1.5pc}
\caption[]{\footnotesize Measurements of $R_{AA}$ of identified particles $\pi^\pm, K^\pm, p^\pm, \pi^0,\phi$ as a function $p_T$ and centrality at \mbox{$\sqrt{s_{NN}}=200$ GeV~\cite{PXPRC88}:} a) (left) Au+Au; b) (right) d+Au. 
\label{fig:RAAdAuAuAu}}\vspace*{-0.2pc}
\end{figure}
Figure~\ref{fig:RAAdAuAuAu}a shows $R_{AA}$ in Au+Au for protons and mesons in the range $0.5<p_T<6.0$ GeV/c, where, in central collisions (0-10\%), all the mesons are suppressed for {$p_T>2$ GeV/c}  while the protons are enhanced for $2<p_T<4$ GeV/c and then become suppressed at larger $p_T$. The d+Au results in Fig.~\ref{fig:RAAdAuAuAu}b show no CNM effect for the mesons, $R_{AA}\approx 1$ out to $p_T=6$ GeV/c; while the protons show a huge enhancement (Cronin effect) in all centralities except for the most peripheral (60-88\%).  At present, there is no explanation of the proton enhancement in either Au$+$Au or $d$$+$Au collisions, so $\pi^0$ and $\eta$ are the favored hard-probes.  \vspace*{-0.1pc} 
\subsection{$\mathbf{\delta p_T'/p_T'}$, the fractional shift in the $\mathbf{p_T'}$ spectrum}
\begin{figure}[!h]
   \begin{center}
\includegraphics[width=0.49\textwidth]{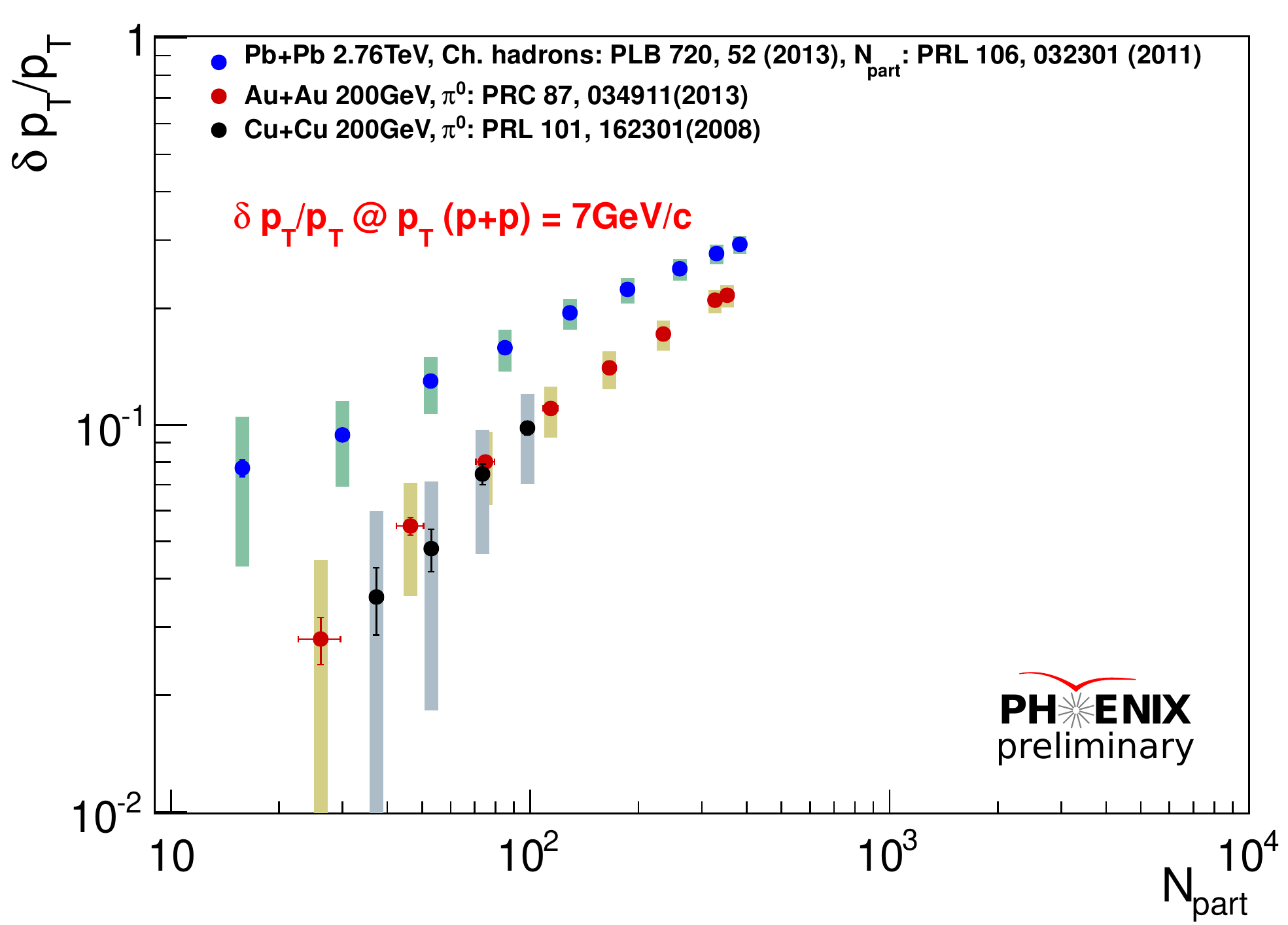}
\includegraphics[width=0.49\textwidth]{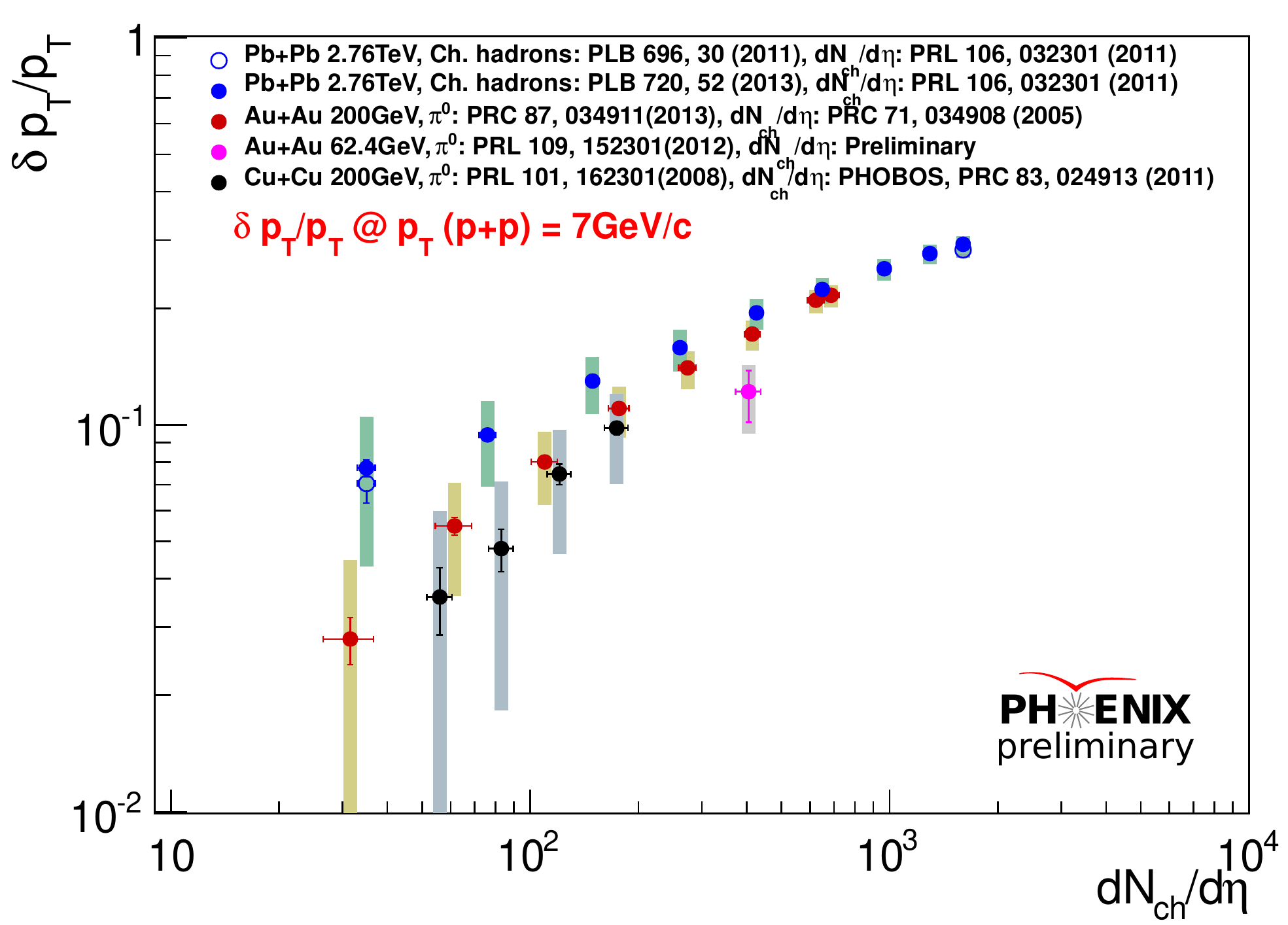}
\end{center}\vspace*{-1.0pc}
\caption[]{\footnotesize Plots from PHENIX~\cite{TakaoNantes} of $\delta p_T'/p_T'$ at $p_T'\equiv p_T(p+p)=7$ GeV for $\pi^0$ (RHIC) and charged hadrons (LHC): a) as a function of centrality (\Npart), b) as a function of $d\Nch/d\eta$.  
\label{fig:dpTpT}}\vspace*{-0.1pc}
\end{figure}

After more than a decade of using the ratio $R_{AA}$, we are now paying more attention to $\delta p_T'/p_T'$, the fractional shift of the $p_T'$ spectrum, as an indicator of energy loss in the \QGP\ Fig.~\ref{fig:dpTpT}~\cite{TakaoNantes}. 
For a constant fractional energy loss, which is true at RHIC in the range $6<p_T<12$ GeV/c (as shown in Fig.~\ref{fig:Tshirt}a where the $p$$+$$p$ reference and Au$+$Au measurement are parallel on a log-log plot) there is a simple relationship between $R_{AA}$, $\delta p_T'/p_T'$ and $n$,  the power in the invariant $p_T$ spectra:
\begin{equation}
R_{AA}(p_T')=R_{AA}(p_T)=(1-\delta p_T/p_T)^{n-2}\qquad . \label{eq:RAAdelta}
\end{equation}
Using  $\delta p_T/p_T$ is important for comparison to the LHC measurements where the power is $n\approx 6$ compared to $n=8.1$ at RHIC, so that the same $R_{AA}$ does not mean the same $\delta p_T/p_T$. Strictly $\delta p_T/p_T$ is not a measure of the parton energy loss in the \QGP\ but is used as a proxy.   
Figure \ref{fig:dpTpT}a shows that $\delta p_T/p_T$ at $p_T=7$ GeV/c for RHIC and LHC both increase monotonically with centrality (\Npart) but is a factor of 2 to 1.4 larger at LHC, depending on centrality, a likely indication of a hotter and/or denser medium. Figure \ref{fig:dpTpT}b attempts to determine whether $\delta p_T/p_T$ is a universal function of the charged particle density, $d\Nch/d\eta$ at both RHIC and LHC. The dependence is not quite universal. A fit of $\delta p_T/p_T\propto (d\Nch/d\eta)^\alpha$ gives $\alpha\approx 0.35$ at LHC and 0.55 at RHIC, although the data at $\sqsn=200$ GeV and 2.76 TeV do appear to merge for $(d\Nch/d\eta)\geq 300$. Hopefully, measurements of $\delta p_T/p_T$ will eventually lead to the determination of $dE/dx$ of partons in the \QGP. 
\subsection{At last: jet measurements in Au$+$Au at RHIC in 2014?} 
\begin{figure}[!htb] 
\begin{center}
 \raisebox{0.0pc}{\includegraphics[width=0.9\textwidth]{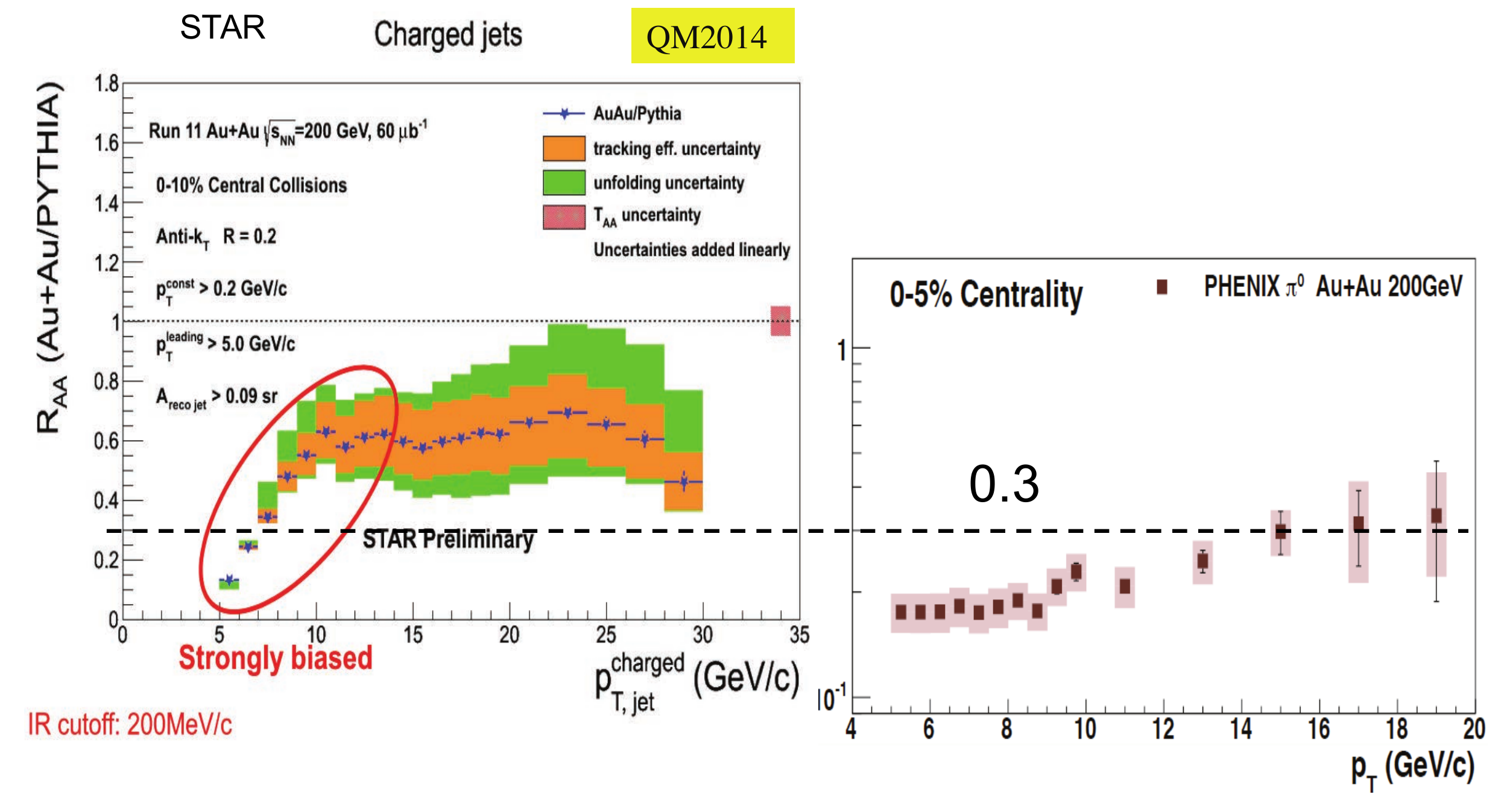}}
\end{center}\vspace*{-1.0pc}
\caption[]{\footnotesize a) (left) STAR $R_{AA}$ for charged jets at \sqsn=200 GeV in central Au$+$Au collisions (see details in legend) compared to b) $R_{AA}$ for PHENIX $\pi^0$. The dashed line at 0.3 is the maximum $R_{AA}$ for $\pi^0$ in this $p_T$ range.\label{fig:STARjetvssingle}}
\end{figure}

Some interesting new jet measurements in A$+$A collisions at RHIC were presented at Quark Matter 2014 in a plenary review talk on jets by Yen-Jie Lee who works on CMS~\cite{YJLeeQM2014}. Figure~\ref{fig:STARjetvssingle} shows that the STAR charged jets in a cone with $R=0.2$ have much less suppression ($R_{AA}\gg 0.3$) than $\pi^0$ ($0.2\leq R_{AA}\leq 0.3$) in the range $10< p_T<20$ GeV. 

\begin{figure}[!htb] 
\begin{center}
 \raisebox{0.0pc}{\includegraphics[width=0.9\textwidth]{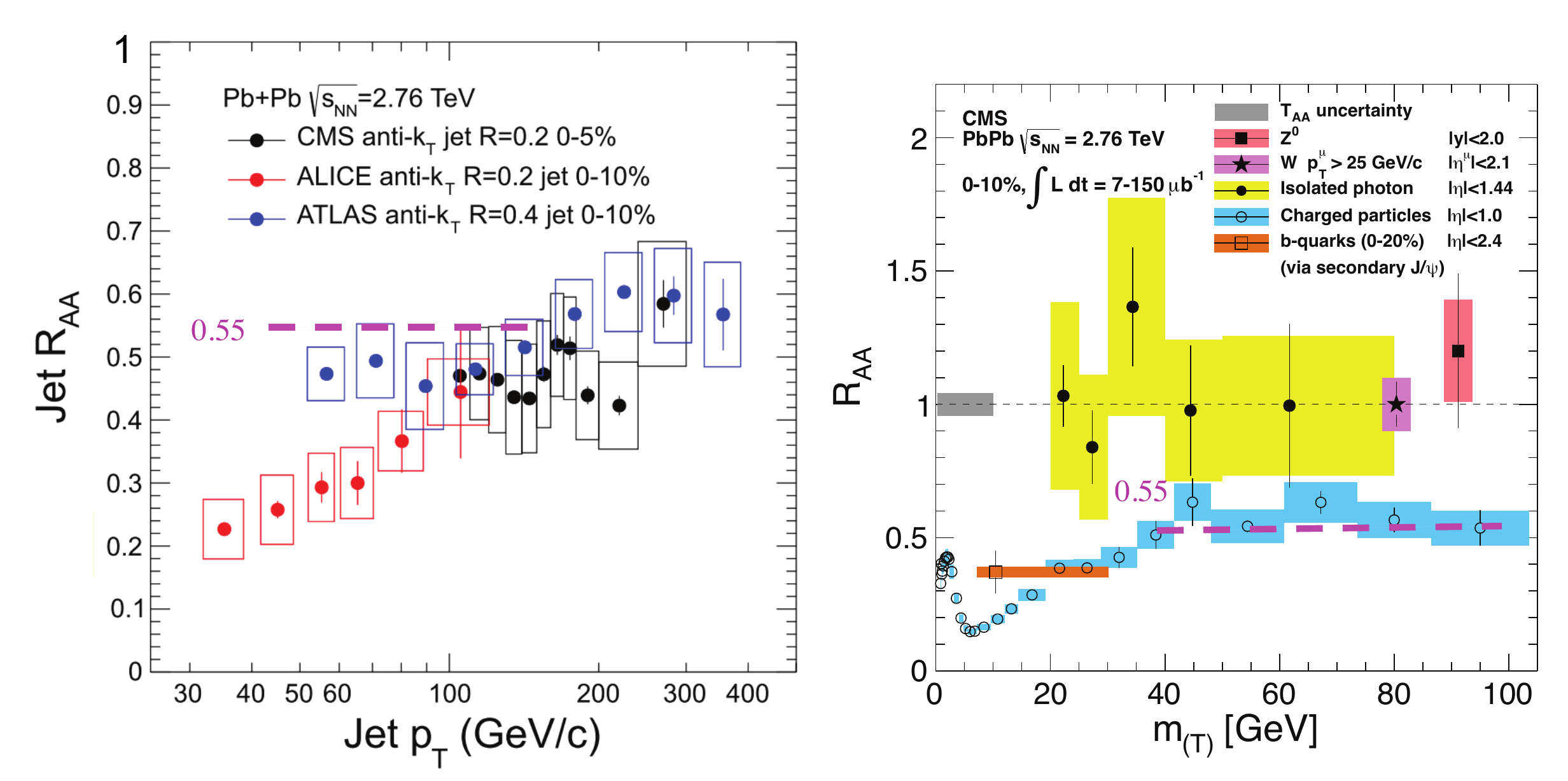}}
\end{center}\vspace*{-1.0pc}
\caption[]{\footnotesize a) (left) $R_{AA}$ for jets at \sqsn=2.76 GeV by CMS and ALICE  compared to b) CMS $R_{AA}$ for charged hadrons ($R_{AA}\approx 0.55$), $b$-quarks and 3 favorite Electro-Weak Bosons~\cite{YJLeeQM2014}.\label{fig:CMSjetvssingle}}
\end{figure}
\noindent This is quite different from jets at the LHC (Fig.~\ref{fig:CMSjetvssingle}) which have comparable or smaller $R_{AA}$ than charged particles from jet fragmentation in the range $30<p_T<100$ GeV. Note that the $\gamma$, $W$ and $Z^0$ bosons in Fig.~\ref{fig:CMSjetvssingle}b which are not coupled to color are not suppressed.

\begin{figure}[!htb]\vspace*{-0.8pc} 
\begin{center}
 \raisebox{0.0pc}{\includegraphics[width=0.9\textwidth]{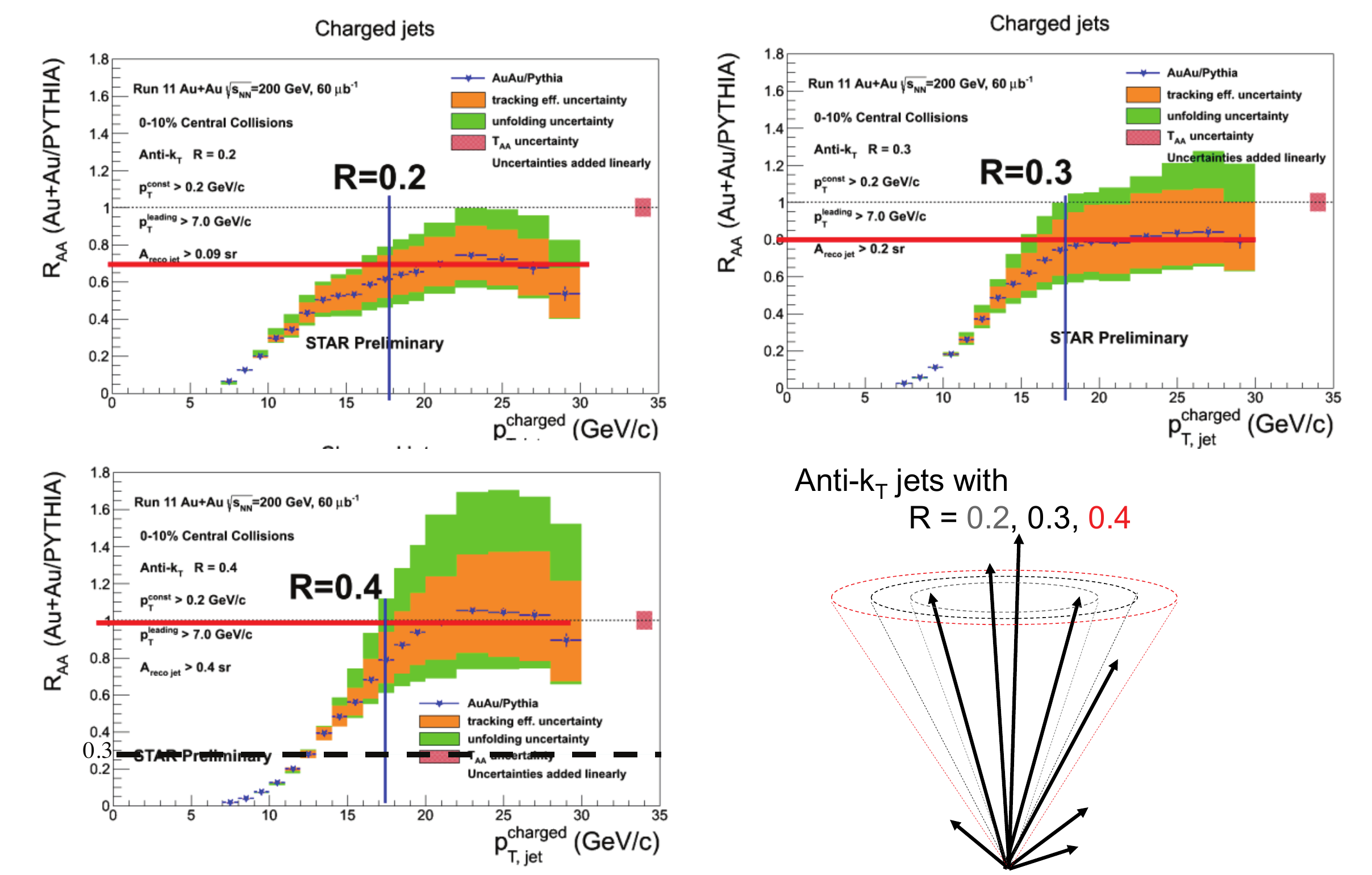}}
\end{center}\vspace*{-1.5pc}
\caption[]{\footnotesize STAR $R_{AA}$ for charged jets at \sqsn=200 GeV in central Au$+$Au collisions for 3 different jet cones with $R=0.2,0.3,0.4$ (see details in legend and sketch)~\cite{YJLeeQM2014}. \label{fig:STAR3Rs}}\vspace*{-0.5pc}
\end{figure}

For STAR, the disagreement of the jet and single particle $R_{AA}$ gets worse as the jet cone is increased from $R$=0.2 to 0.3 to 0.4 (Fig.~\ref{fig:STAR3Rs}). Some people would say that this is great because all the jet fragments and/or any energy lost in the \QGP\ by the originating parton have been captured in the $R$=0.4 cone. Skeptics like myself can hardly wait to see what happens when the jet cone is further increased. After 14 runs at RHIC, the jet learning curve in Au$+$Au central collisions still has a way to go.

The good news for the future is that a new detector, now called sPHENIX, to find jets by the more traditional method using hadron calorimetry has been proposed, is moving along on the approval process and is on the schedule at RHIC (Fig.~\ref{fig:RHICperf}b) 
         \begin{figure}[!h]
   \begin{center}
\includegraphics[width=0.43\textwidth]{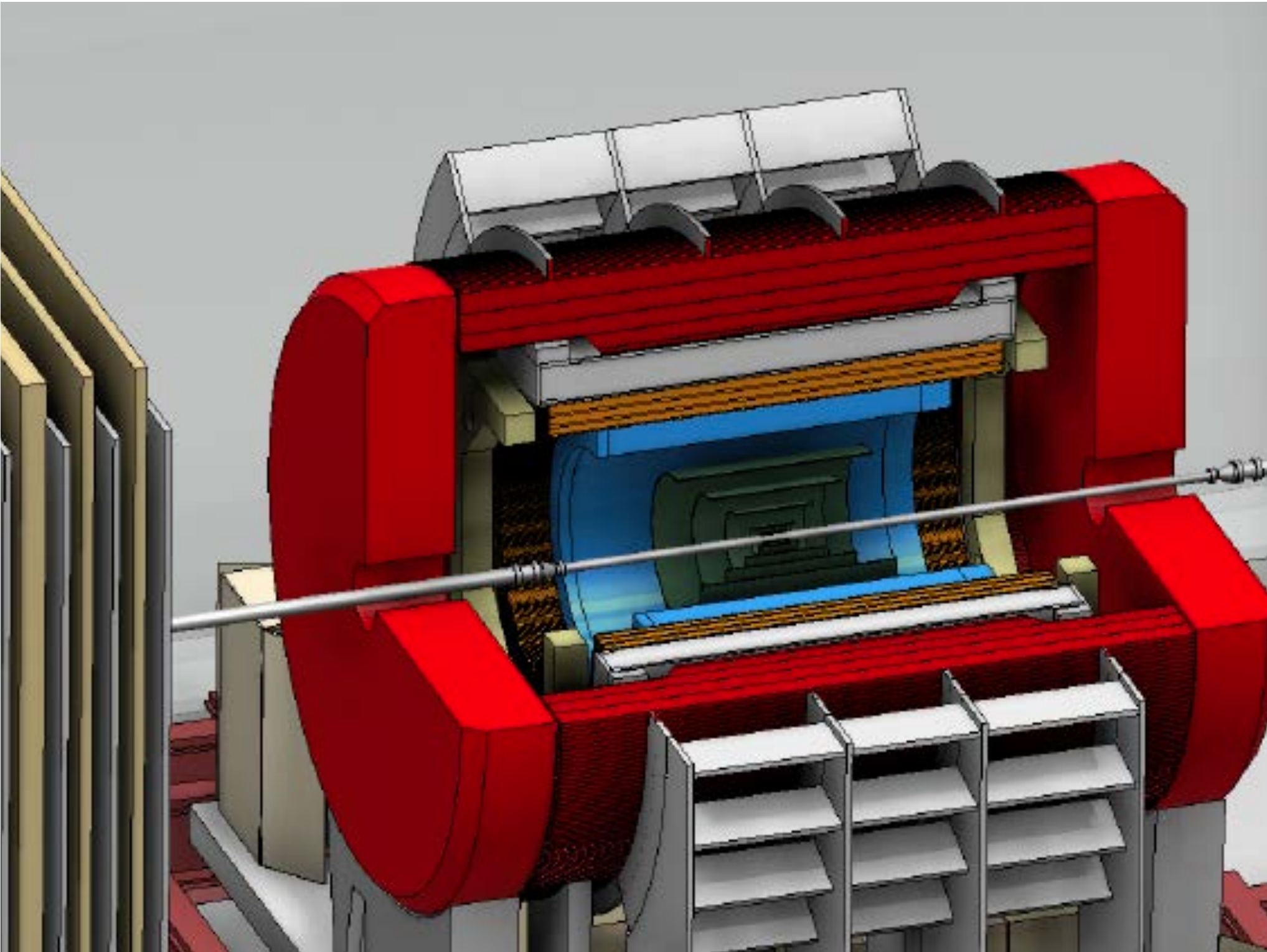}
\includegraphics[width=0.49\textwidth]{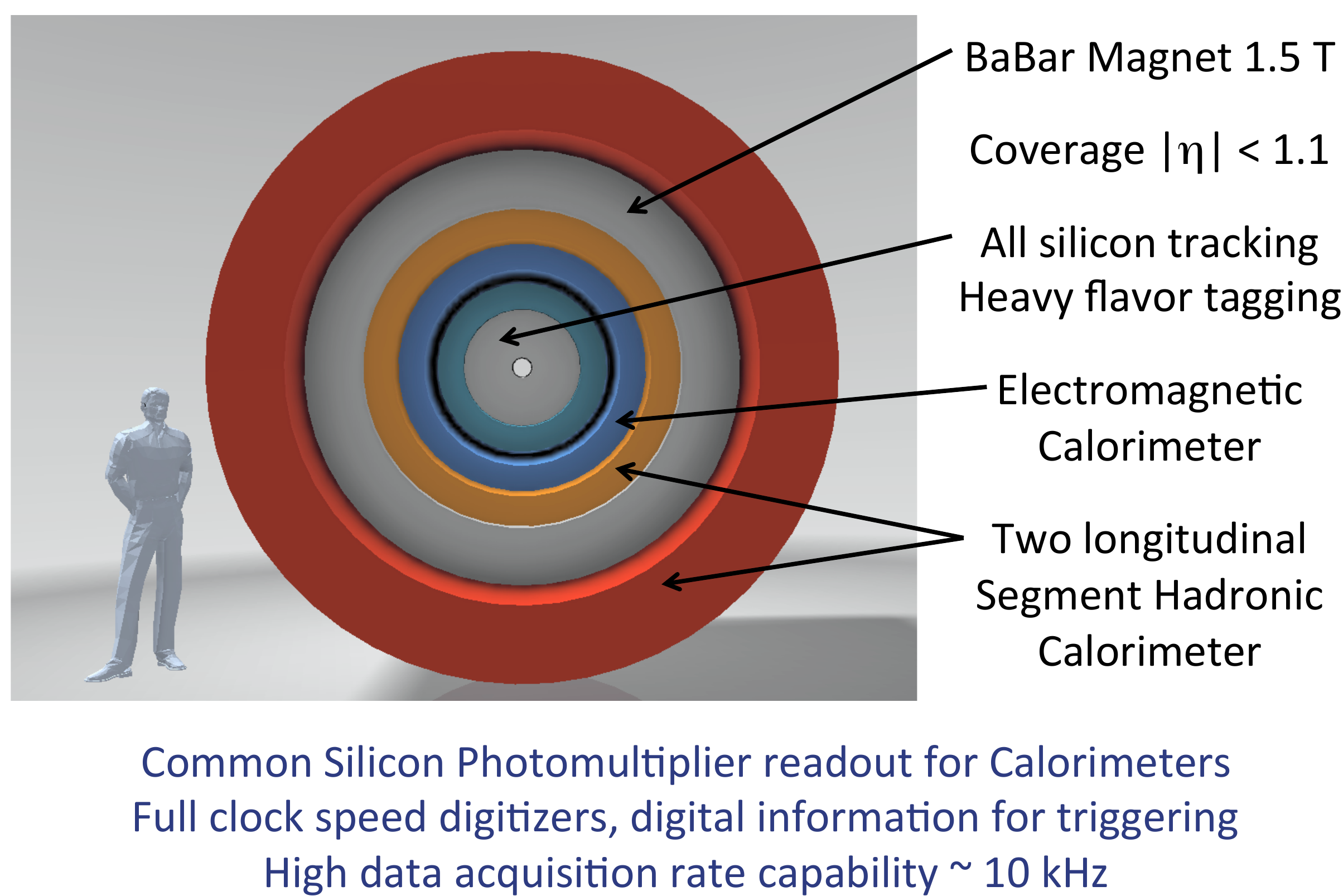}
\end{center}\vspace*{-1.0pc}
\caption[]{\footnotesize a)(left) new detector in place of PHENIX; b) (right) beam's eye view with details~\cite{JNDoE714}. 
\label{fig:sPHENIX}}\vspace*{-0.1pc}
\end{figure}
for comissioning and full installation in the PHENIX IR in 2019--20. It is based on the (made in Italy) BABAR superconducting solenoid from SLAC which became available when the B-factory in Italy was unfortunately cancelled. Design work is moving along quickly (Fig.~\ref{fig:sPHENIX}) so now is the optimum time for new collaborators to join.

\end{document}